\documentclass[3p]{elsarticle}
\usepackage{graphicx}
\usepackage{amsmath,amsfonts,amssymb,amsthm,stmaryrd}
\usepackage{bm}
\usepackage{tabularx}
\usepackage{booktabs}
\usepackage{comment}
\usepackage[lined]{algorithm2e}
\usepackage{ulem}
\usepackage[hidelinks]{hyperref}
\usepackage{float,caption,subcaption}
\usepackage{color, xcolor}
\usepackage{enumerate}
\usepackage{pifont}
\usepackage{array}
\usepackage{multirow}

\usepackage{placeins}

\newcommand{\jump}[1]{\llbracket #1 \rrbracket}

\begin{document}

\begin{frontmatter}
\title{On phase-field regularization in dynamic fracture\\with brittle and cohesive formulations}
\journal{arXiv}

\author[eth]{J. Heinzmann}
\author[ut]{F. Vicentini}
\author[eth]{P. Carrara}
\author[eth]{L. De Lorenzis}

\address[eth]{Computational Mechanics Group, ETH Zürich, Tannenstrasse 3, 8092 Zürich, Switzerland}
\address[ut]{Department of Aerospace Engineering \& Engineering Mechanics, The University of Texas at Austin, 2617 Wichita Street, Austin, TX 78712, USA}

\begin{abstract}
    Phase-field models of fracture are widely used for simulating crack nucleation and propagation, yet the role of the phase-field regularization in the dynamic regime is not fully understood and depends critically on how the damage variable is coupled to the displacement field.
    In this paper, we analyze three alternative formulations: the brittle model with stiffness degradation, its variant with stiffness+density degradation, and our recently proposed phase-field regularization of cohesive fracture, which we extend to elastodynamics.
    By studying the interaction of a tensile and a compressive elastic wave with a phase-field crack in a one-dimensional bar, we determine for which models and under which conditions the phase-field regularization preserves the features of the wave-crack interaction expected for a sharp crack, and we theoretically explain which variables control the behavior.
    For the new cohesive model extended to dynamics, we further derive an analytical dynamic cohesive opening law.
    Finally, we study the dynamic behavior including branching of a two-dimensional notched plate at two loading intensities.
\end{abstract}


\end{frontmatter}

\section{Introduction}\label{sec:introduction}
Many real-world structures undergo fracture under dynamic loading, either deliberately (e.g.\ mining, fracking) or accidentally (e.g.\ crashes).
For fracture under high strain rates, complex mechanisms emerge, including limiting crack-tip speeds \cite{Ravichandar_2004,Ravi_experimental_1984a}, crack-tip instabilities \cite{Buehler_2006,Vasudevan_oscillatory_2021,wang_tensile_2023,Fineberg_1991}, and (micro-)branching phenomena \cite{Sharon_1995,Sharon_1996,Livine_2005}.
The phase-field approach naturally lends itself to capturing intricate crack patterns without explicit branching or propagation criteria, tracking algorithms, or re-meshing.
Originally proposed in the quasi-static regime as a regularization of Griffith's crack propagation criterion \cite{griffith_phenomena_1921} and later shown to amount to a gradient damage model \cite{pham_gradient_2011}, the phase-field approach to brittle fracture \cite{bourdin_numerical_2000} formulates crack evolution as the minimization of a total energy functional \cite{francfort_revisiting_1998}.
Extensions to dynamics \cite{Bourdin_time_2011,Schlueter_phase_2014,Hofacker_phase_2013} retain the same regularization and supplement it with inertia effects; their governing equations can be derived from an extension of Hamilton's principle to systems with dissipation \cite{li_phd_2016,li_cracktip_2017,fischer_gradient_2019}.

Dynamic phase-field fracture models have reproduced at least qualitatively a wide range of experimentally observed phenomena, including the dependence of the energy release rate on crack-tip velocity \cite{Agraval_dependence_2017}, velocity toughening \cite{Bleyer_dynamic_2012}, a limiting mode-I crack speed of approximately 60\% of the Rayleigh wave speed \cite{Borden_2012_phase}, inter- and transonic propagation \cite{Schlueter_investigation_2016}, oscillatory crack-tip instabilities \cite{chen_instability_2017}, micro- and macro-branching \cite{Bleyer_microbrancing_2017}, and fragmentation \cite{fischer_gradient_2019b,Mandal_evaluation_2020}.
Substantial work has addressed mathematical aspects such as existence of solutions and convergence \cite{Larsen_existence_2010}, stability \cite{fischer_gradient_2019}, as well as numerical aspects \cite{Ren_explicit_2019,Schlueter_numerical_2017,Nagaraja_phase_2019,Gupta_adaptive_2022,ZiaeiRad_massive_2016,Schlueter_gpu_2013}.

However, several works have reported anomalous behavior in the interaction of elastic waves with regularized cracks. Steinke et al.~\cite{Steinke_comparative_2016} observe high-frequency oscillations following the interaction of an elastic wave with a pre-existing phase-field crack in a quasi-1D setting. They interpret these oscillations as a numerical artifact which can be alleviated through dissipative time integration and a crack region spanning at least one fully damaged element.
Schlüter et al.~\cite{Schlueter_numerical_2017} report similar oscillations in a 1D bar and note that, while some numerical solution approaches yield \textit{`no or minor oscillations'} \cite[p.~41]{Schlueter_numerical_2017}, with all tested approaches the shape of the elastic wave is altered by the interaction with the pre-existing phase-field crack.
More recently, Durussel et al.~\cite{durussel_origins_2025} interpret the phenomenon as a \textit{`trapping of elastic waves'} within the damaged zone, leading to spurious effects and a widening of the phase-field profile; they argue that these effects vanish as the phase-field regularization length goes to zero and classify them as numerical artifacts.

In classical models, the coupling between the displacement and the phase-field (or damage) variable is realized by multiplying the stiffness of the material by a degradation function depending on the phase-field variable. In the following, we refer to these models as phase-field models of brittle fracture \textit{with stiffness degradation}. In dynamics, other models have been proposed which degrade both the stiffness and the mass density \cite{chen_instability_2017}. Their motivation is that the mismatch between degraded stiffness and undegraded density spuriously reduces the local wave speed within the damaged zone.
Subsequent works in the large-strain setting~\cite{Lubomirsky_2018,Tian_dynamic_2020,Vasudevan_oscillatory_2021} adopt this variant and report on its ability to capture higher crack-tip velocities and the experimentally observed oscillatory crack-tip instability.
Tian et al.~\cite{Tian_dynamic_2020} further argue that, without density degradation, energy accumulates at the crack tip which may yield unrealistic crack-tip splitting. In this paper, we refer to these models as phase-field models of brittle fracture \textit{with stiffness+density degradation}.

The phase-field models of cohesive fracture recently proposed in~\cite{bourdin_variational_2025,Vicentini_variational_2026}  preserve the elastic stiffness and degrade only the material strength; hence, we denote them as phase-field models of cohesive fracture \textit{with strength degradation}.
These models have been proposed for quasi-static fracture and have not yet been extended to the dynamic regime.
In this paper, we perform this extension and show that these models, besides their advantages in the quasi-static framework, are also promising alternatives for dynamics.
In summary, this work makes the following contributions:
\begin{itemize}
    \item For the classical model with stiffness degradation, we show that the high-frequency oscillations and the widening of the phase-field profile reported in the literature are not purely numerical artifacts, but consequences of the phase-field regularization in the dynamic setting.
    By a scattering analysis and computations with the transfer matrix method, we show that the wave-crack interaction is governed by the ratio of the regularization length to the wavelength. By finite element simulations, we explain the widening of the phase-field profile.
    We demonstrate that degrading the mass density in addition to the stiffness does not resolve the above issues.
    \item We extend the recently proposed phase-field model of cohesive fracture~\cite{bourdin_variational_2025,Vicentini_variational_2026} to the dynamic setting.
    We study under which conditions this new formulation solves the above issues and recovers the desired elastodynamic response.
    Under these conditions, we further derive analytically a dynamic cohesive zone law and identify the ratio of the wave speed to the Irwin cohesive length as parameter governing the temporal response.
\end{itemize}

The remainder of this work is organized as follows.
Section~\ref{sec:phase_field_model_brittle} analyzes the phase-field models of dynamic brittle fracture in both variants with stiffness and stiffness+density degradation.
Section~\ref{sec:phase_field_model_cohesive} extends the cohesive phase-field formulation with strength degradation in \cite{Vicentini_variational_2026} to the dynamic setting.
In both sections, after presenting the governing equations in the multidimensional setting, we analyze the 1D problem of the interaction between an elastic wave and a phase-field crack both analytically and numerically. 
Section~\ref{sec:model_comparison} compares the numerical results from all studied formulations on a 2D benchmark of dynamic fracture.
Section~\ref{sec:conclusions} concludes with a summary and outlook.

\FloatBarrier
\section{Phase-field modeling of dynamic brittle fracture with stiffness and stiffness+density degradation}
\label{sec:phase_field_model_brittle}
We start by analyzing phase-field models of dynamic brittle fracture based on stiffness degradation \cite{Li_numerical_2016,li_cracktip_2017,fischer_gradient_2019}.
Then, we shift our attention to the model proposed by \cite{chen_instability_2017,Tian_dynamic_2020} with stiffness and density degradation.

\subsection{Brittle model with stiffness degradation}\label{sec:stiffdeg}
In this section, we first recapitulate the main aspects of phase-field models of dynamic brittle fracture with stiffness degradation.
Subsequently, we analyze the interaction of elastic waves with a pre-existing phase-field crack in a 1D bar.

\subsubsection{Energetic quantities}\label{sec:model_formulation_brittle}
We consider a homogeneous, isotropic, linear elastic body occupying a portion of a $d$-dimensional  space $\Omega \subset \mathbb{R}^d$ with boundary $\partial\Omega$.
The state at a point $\boldsymbol{x} \in \Omega$ and time $t \in [0,T]$ is described by the displacement field $\boldsymbol{u}(\boldsymbol{x}, t) : \Omega \times [0,T] \rightarrow \mathbb{R}^d$ and the phase (or damage) field $\alpha(\boldsymbol{x}, t) : \Omega \times [0,T] \rightarrow [0,1]$, where $\alpha=0$ denotes the pristine material and $\alpha=1$ the fully damaged material.
The infinitesimal strain tensor is defined as $\boldsymbol{\varepsilon}(\boldsymbol{x}, t) = \nabla_{\text{sym}} \boldsymbol{u}(\boldsymbol{x}, t)$, with $\nabla_{\text{sym}}$ as the symmetric spatial gradient operator.
We denote with  $\boldsymbol{\bar u}(\boldsymbol{x},t)$ the prescribed displacement on the Dirichlet part of the boundary $\partial\Omega_D$  and with  $\boldsymbol{f}(\boldsymbol{x},t)$ the prescribed traction on the Neumann part of the boundary $\partial\Omega_N$.
Unless stated otherwise, we assume the domain to be initially undamaged and at rest, namely $\alpha(\boldsymbol{x}, 0)=0$ and $\dot {\boldsymbol{u}}(\boldsymbol{x}, 0)=\boldsymbol{0}$ for all $\boldsymbol{x} \in \Omega$, where $\dot{(\bullet)}$ stands for the time derivative of $(\bullet)$.

The total potential energy of the body reads
\begin{equation}\label{eq:P}
    \mathcal{P} (\boldsymbol{u}, \alpha)
    = \mathcal{P}^\text{e} (\boldsymbol{u},\alpha) + \mathcal{P}^\text{f} (\alpha) - \mathcal{W} (\boldsymbol{u})
    = \int_{\Omega} \psi (\boldsymbol{\varepsilon}(\boldsymbol{u}),\alpha) \,\mathrm{d}\boldsymbol{x} + \int_{\Omega} \frac{G_{\text{c}}}{c_w} \left( \frac{w(\alpha)}{\ell} + \ell |\nabla \alpha|^2 \right) \mathrm{d}\boldsymbol{x} - \int_{\partial \Omega_{\text{N}}} \boldsymbol{f} \cdot \boldsymbol{u} \,\mathrm{d}\boldsymbol{x}
\end{equation}
comprising the elastic energy $\mathcal{P}^\text{e}$, the fracture energy $\mathcal{P}^\text{f}$, and the work of the external forces $\mathcal{W}$.

Following \cite{amor_regularized_2009,miehe_2010_thermodynamically,freddi_regularized_2010,de_lorenzis_nucleation_2022,Vicentini_energy_2024}, the \textit{elastic strain energy density} $\psi$ is expressed as
\begin{equation}\label{eq:psi_brittle}
    \psi(\boldsymbol{\varepsilon}, \alpha) = g(\alpha) \psi_{\text{D}} (\boldsymbol{\varepsilon}) + \psi_{\text{R}} (\boldsymbol{\varepsilon}) \quad\text{,}
\end{equation}
where $\psi_{\text{D}}$ is the degraded contribution driving damage evolution and $\psi_{\text{R}}$ is the residual part.
In this work, we employ the volumetric-deviatoric split \cite{amor_regularized_2009},
\begin{equation}\label{eq:vol_dev_decomposition}
    \psi_{\text{D}} (\boldsymbol{\varepsilon}) = \frac{\kappa_0}{2} \langle \text{tr} (\boldsymbol{\varepsilon}) \rangle_+^2 + \mu_0 \left|\boldsymbol{\varepsilon}_{\text{dev}}\right|^2
    \quad\text{,}\quad
    \psi_{\text{R}} (\boldsymbol{\varepsilon}) = \frac{\kappa_0}{2} \langle \text{tr} (\boldsymbol{\varepsilon}) \rangle_-^2
    \quad\text{, with}\quad
    \psi_{\text{D}}(\boldsymbol{\varepsilon}) + \psi_{\text{R}}(\boldsymbol{\varepsilon}) = \psi_0 (\boldsymbol{\varepsilon})\quad\text{,}\quad
\end{equation}
where  $\kappa_0$ and $\mu_0$ are the undamaged bulk and shear moduli, while $\psi_0 (\boldsymbol{\varepsilon}) = \frac{\kappa_0}{2} \text{tr}(\boldsymbol{\varepsilon})^2 + \mu_0 \left|\boldsymbol{\varepsilon}_{\text{dev}}\right|^2$ is the undamaged elastic strain energy density.
Also, $\text{tr} (\boldsymbol{\varepsilon})$ is the trace of $\boldsymbol{\varepsilon}$ and $\boldsymbol{\varepsilon}_{\text{dev}}=\boldsymbol{\varepsilon}-\frac{1}{d}\text{tr} (\boldsymbol{\varepsilon})\boldsymbol{I}$ is its deviatoric part with $\boldsymbol{I}$ as the second-order identity tensor, while we use $\langle(\bullet)\rangle_\pm = \tfrac{1}{2} ((\bullet) \pm |(\bullet)|)$.
The \textit{degradation function} $g(\alpha) = (1-\alpha)^2 + g_0$ couples the phase field and the displacement field and contains the constant $0 < g_0 \ll 1$ ($g_0=o(\ell)$) leading to a small residual stiffness in fully damaged conditions. In the quasi-static case, the residual stiffness is used to ensure strict convexity of the total energy functional in $\boldsymbol{u}$. In the dynamic case, we retain it for the reasons clarified in Section~\ref{sec:interaction_wave_brittle_crack}.

The second contribution in \eqref{eq:P} is the Ambrosio-Tortorelli (\texttt{AT}) regularization of the surface energy \cite{ambrosio_approximation_1992}, with fracture toughness $G_{\text{c}}$, regularization length $\ell$, and normalization constant $c_w = 4 \int_0^1 \sqrt{w(\beta)} \mathrm{d}\beta$ depending on the choice of $w(\alpha)$.
The two most common choices are the \texttt{AT1} model with $w(\alpha) = \alpha$ ($c_w = 8/3$) and the \texttt{AT2} model with $w(\alpha) = \alpha^2$ ($c_w=2$).
In this paper, for the brittle case we adopt the \texttt{AT1} model, since \texttt{AT2} leads to a vanishing elastic domain. The implications of adopting the \texttt{AT2} model in the dynamic setting are briefly discussed in Section~\ref{sec:summary_brittle}.

Finally, the kinetic energy reads
\begin{equation}\label{eq:K}
    \mathcal{K} (\dot{\boldsymbol{u}}) = \int_\Omega \frac{1}{2} \rho_0 \left|\dot{\boldsymbol{u}}\right|^2 \mathrm{d}\boldsymbol{x}
\end{equation}
with the mass density $\rho_0$.

\subsubsection{Governing equations}\label{sec:governing_equations_brittle}
Following \cite{Li_numerical_2016,li_cracktip_2017,fischer_gradient_2019}, the governing equations are obtained imposing the \textit{principle of stationary action}, the \textit{energy balance}, and the \textit{irreversibility condition}. The latter reads
\begin{equation}\label{eq:irr}
    \dot{\alpha} \geq 0 \qquad\text{.}
\end{equation}
Along with the initial condition $\alpha(\boldsymbol{x}, 0) \geq 0$, \eqref{eq:irr} ensures the non-negativity of the damage parameter $\alpha \geq 0$.
The \textit{space-time action functional} between two time instants $t_2>t_1$ is defined as
\begin{equation}
    \mathcal{A}(\boldsymbol{u}, \alpha) = \int_{t_1}^{t_2} \mathcal{L} (\boldsymbol{u}, \dot{\boldsymbol{u}}, \alpha) \,\mathrm{d}t \qquad\text{with}\qquad \mathcal{L} (\boldsymbol{u}, \dot{\boldsymbol{u}}, \alpha) = \mathcal{P} (\boldsymbol{u}, \alpha) - \mathcal{K} (\dot{\boldsymbol{u}})\,,
\end{equation}
where $ \mathcal{L}$ is the Lagrangian of the system.
The principle of stationary action requires the first variation of $\mathcal{A}$ to be non-negative for all admissible variations of the state variables over any interval $[t_1, t_2]$ \cite{li_phd_2016,gray_when_2007}, namely
\begin{equation}\label{eq:PSA}
    \mathcal{A}^\prime (\boldsymbol{z}) (\hat{\boldsymbol{z}} - \boldsymbol{z}) \geq 0\,, \qquad \forall \hat{\boldsymbol{z}} \in \mathcal{Z}\,,
\end{equation}
where $\boldsymbol{z} = (\boldsymbol{u}, \alpha)$ is the state vector, $\mathcal{A}^\prime (\boldsymbol{z}) (\hat{\boldsymbol{z}} - \boldsymbol{z})$ denotes the Gâteaux derivative of $\mathcal{A}$ in the direction $\hat{\boldsymbol{z}} - \boldsymbol{z}$, and $\mathcal{Z}$ is a sufficiently regular functional space incorporating the  Dirichlet boundary conditions.
Finally, the energy balance requires conservation of the total energy,
\begin{equation}\label{eq:EB}
    \frac{\mathrm{d}}{\mathrm{d}t} \mathcal{E} (\boldsymbol{u}, \dot{\boldsymbol{u}}, \alpha) = - \int_{\partial \Omega_{\text{N}}}\!\! \dot{\boldsymbol{f}} \cdot \boldsymbol{u} \,\mathrm{d}\boldsymbol{x} + \int_{\partial \Omega_{\text{D}}}\!\! \boldsymbol{\sigma} \boldsymbol{n} \cdot \dot{\bar{\boldsymbol{u}}} \,\mathrm{d}\boldsymbol{x} \qquad\text{with}\qquad \mathcal{E} (\boldsymbol{u}, \dot{\boldsymbol{u}}, \alpha) = \mathcal{P} (\boldsymbol{u},\alpha) + \mathcal{K} (\dot{\boldsymbol{u}}) \qquad\text{.}
\end{equation}

Defining the stress
\begin{equation}\label{eq:sigma_brittle}
    \boldsymbol{\sigma} (\boldsymbol{\varepsilon}, \alpha)
    = \frac{\partial \psi (\boldsymbol{\varepsilon}, \alpha)}{\partial \boldsymbol{\varepsilon}}
    = g(\alpha) \left[ \kappa_0 \langle \text{tr} (\boldsymbol{\varepsilon}) \rangle_+ \boldsymbol{I} + 2 \mu_0 \boldsymbol{\varepsilon}_{\text{dev}} \right] + \kappa_0 \langle \text{tr} (\boldsymbol{\varepsilon}) \rangle_- \boldsymbol{I}
\end{equation}
and the \textit{damage energy release rate} 
\begin{equation}\label{eq:standard_err}
    Y (\boldsymbol{\varepsilon}, \alpha) = - \frac{\partial \psi (\boldsymbol{\varepsilon}, \alpha)}{\partial \alpha} = - g^\prime(\alpha) \psi_{\text{D}} (\boldsymbol{\varepsilon})\qquad\text{,}
\end{equation}
\eqref{eq:irr}, \eqref{eq:PSA} and \eqref{eq:EB} yield the dynamic equilibrium equations
\begin{subequations}\label{eq:u_strongform}
    \begin{equation}\label{eq:u_strongform_bulk}
        - \nabla \cdot \boldsymbol{\sigma} + \rho_0 \ddot{\boldsymbol{u}} = \boldsymbol{0}\,, \qquad \forall (\boldsymbol{x},t) \in \Omega \times [0,T]\,,
    \end{equation}
    \begin{equation}\label{eq:NeumannBC}
        \boldsymbol{\sigma} \boldsymbol{n} = \boldsymbol{f}\,, \qquad \forall (\boldsymbol{x},t) \in \partial \Omega_{\text{N}} \times [0,T]\,,\\
    \end{equation}
\end{subequations}
which differ from those of the quasi-static case only by the presence of the inertia term $\rho_0 \ddot{\boldsymbol{u}}$, and the Karush-Kuhn-Tucker (KKT) conditions for damage evolution
\begin{subequations}\label{eq:alpha_strongform}
    \begin{equation}
            - Y (\boldsymbol{z}) + \frac{G_{\text{c}}}{c_w} \left( \frac{w^\prime(\alpha)}{\ell} - 2 \ell \Delta \alpha \right)\geq 0\,,
            \dot{\alpha} \geq 0\,,
            \left[ - Y (\boldsymbol{z}) + \frac{G_{\text{c}}}{c_w} \left( \frac{w^\prime(\alpha)}{\ell} - 2 \ell \Delta \alpha \right) \right] \dot{\alpha} = 0
        \,\,\forall (\boldsymbol{x},t) \in \Omega \times [0,T]
        \label{KKT_bulk}
    \end{equation}
    \begin{equation}
        \nabla \alpha \cdot \boldsymbol{n} \geq 0 \qquad \dot{\alpha} \geq 0 \qquad (\nabla \alpha \cdot \boldsymbol{n}) \dot{\alpha} = 0 \qquad \forall (\boldsymbol{x},t) \in \partial \Omega \times [0,T] \qquad\text{,}
        \label{KKT_boundary}
    \end{equation}
\end{subequations}
which coincide with those of the quasi-static case.

In the quasi-static case, the 1D counterpart of \eqref{KKT_bulk} written for the local damage model (i.e., for $\alpha^\prime \equiv 0$) delivers the 1D elastic domain \cite{pham_gradient_2011} which, for the \texttt{AT1} model, reads 
\begin{equation}\label{eq:elastic_domain_brittle_1D}
    \mathcal{S} (\alpha) = (-\infty, \hat{\sigma}_{\text{c}}(\alpha)]
    \qquad\text{with}\qquad
    \hat{\sigma}_{\text{c}}(\alpha) = \left( (1-\alpha)^2 +g_0 \right)\sqrt{\frac{3 G_{\text{c}} E_0}{8 \ell (1-\alpha)}}
    \qquad\text{.}
\end{equation}
The critical stress at the initial undamaged state
\begin{equation}\label{eq:tens_str}
    \hat{\sigma}_{\text{c}}(0) = \left(1+g_0 \right)\sqrt{\frac{3 G_{\text{c}} E_0}{8 \ell}}
\end{equation}
coincides with the tensile strength of the material.

\subsubsection{Interaction of a tensile wave with a fixed phase-field crack in a 1D bar} \label{1D_stiffdeg}
We now consider a 1D bar which contains a phase-field crack at its center $x=0$, represented by the optimal \texttt{AT1} phase-field profile \cite{gerasimov_2019}
\begin{equation}\label{eq:profile_AT1}
    \alpha (x) = \begin{cases}
        \left( 1 - \frac{|x|}{2\ell} \right)^2 &x \in (-2\ell, 2\ell)\\
        0 &\text{else}\\
    \end{cases} \qquad\text{.}
\end{equation}
In this section, this phase-field profile is held constant in time and denoted as a \textit{fixed} phase-field crack (whereas in later sections we will let it evolve and denote it as an \textit{evolving} phase-field crack).
Hence, the damage field is known and only the elastodynamic problem \eqref{eq:u_strongform} with the appropriate boundary and initial conditions needs to be solved.
For the purpose of the analytical study in this section, we ignore the volumetric-deviatoric energy decomposition and apply the degradation function to the whole elastic energy density, i.e. we adopt $\psi_{\text{D}} (\varepsilon) = \psi(\varepsilon)$ and $\psi_{\text{R}} (\varepsilon) = 0$. 
Combining the kinematics $\varepsilon = u^\prime$ and the constitutive law $\sigma = g(\alpha) E_0 \varepsilon$ with the balance of linear momentum~\eqref{eq:u_strongform}, we obtain the governing equation
\begin{equation}\label{eq:wave_equation_standard_1D_nosplit}
    - \frac{\partial}{\partial x} \left[ g(\alpha) E_0 u^\prime \right] + \rho_0 \ddot{u} = 0
\end{equation}
which is the wave equation in a 1D medium with smoothly varying material properties. 

\begin{figure}[t!]
    \centering
    \includegraphics{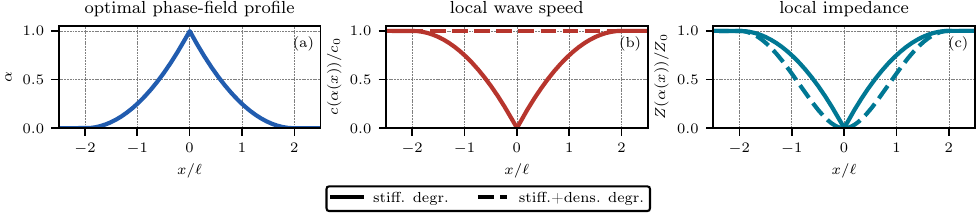}
    \caption{Optimal phase-field profile (a), resulting local wave speed (b) and local acoustic impedance (c) for the \texttt{AT1} model.}
    \label{fig:phasefield_regularization}
\end{figure}

The so-called scattering analysis, i.e. the analysis of the frequency-dependent reflection and transmission of an elastic wave in a heterogeneous elastic medium, is a classical subject \cite{bremmer_1951, brekhovskikh_1980} and its main concepts and results are summarized in Appendix~\ref{scatt}. As noted in \cite{chen_instability_2017,Tian_dynamic_2020,durussel_origins_2025}, the spatially variable stiffness leads to a wave speed $c(\alpha)= \sqrt{g(\alpha)E_0/\rho_0}=\sqrt{g(\alpha)}c_0$ which decreases from $c_0$ in the intact material to a very small value (depending on $g_0$) for $\alpha=1$ (Fig.~\ref{fig:phasefield_regularization}).
As a consequence, in absence of a residual stiffness no mechanical signal reaching the crack can propagate away from it.
For this reason, we argue that $g_0 > 0$ is needed not only in the quasi-static but also in the dynamic brittle case.
However, in contrast to the claims in literature \cite{chen_instability_2017,Tian_dynamic_2020,durussel_origins_2025}, the reduction of the local wave speed alone does not characterize the wave-crack interaction.
As shown in Appendix~\ref{scatt}, the full characterization of the scattering behavior is given by two profiles \cite{brekhovskikh_1980,brekhovskikh_1990}: the local wave speed $c(\alpha)$, and the acoustic impedance
\begin{equation}\label{eq:impedance}
    Z(\alpha) = \sqrt{g(\alpha)E_0\,\rho_0} = \sqrt{g(\alpha)}\,Z_0 \quad\text{with}\quad Z_0 = \sqrt{E_0\rho_0} \quad\text{,}
\end{equation}
both reported in Fig.~\ref{fig:phasefield_regularization}. In the model with stiffness degradation, both profiles degrade simultaneously, so that the regularized crack acts on the wave not only as a region of reduced speed, but also as a smooth impedance well. According to the Wentzel–Kramers–Brillouin (WKB) criterion for smoothly graded media \cite{brekhovskikh_1980,brekhovskikh_1990},
a harmonic wave of angular frequency $\omega$ traverses a heterogeneity without appreciable reflection wherever the relative impedance variation per local wavelength is small, i.e., wherever
\begin{equation}\label{eq:WKB}
    \delta(x) = \frac{|Z'(x)|}{Z(x)} \frac{c(x)}{\omega} \ll 1 \quad\text{,}
\end{equation}
while values of $\delta$ of order one mark the onset of reflection. Note that, with a little abuse of notation, we are denoting $c(x)=c(\alpha(x))$ as the composition of $c(\alpha)$ with the phase-field profile \eqref{eq:profile_AT1}, and similarly for $Z(x)$.
For the optimal \texttt{AT1} profile \eqref{eq:profile_AT1} we have $1-\alpha(x) \simeq |x|/\ell$, 
therefore $\sqrt{g(\alpha(x))} \simeq |x|/\ell$ to leading order near the center of the crack.
The relative impedance gradient then diverges toward the center of the crack, $|Z'|/Z \simeq 1/|x|$, but the local wavelength shrinks at exactly the compensating rate, $c(x)/\omega \simeq (\lambda/2\pi)\,|x|/\ell$, so that
\begin{equation}\label{eq:WKB-AT1}
    \delta \simeq \frac{\lambda}{2\pi\ell}
\end{equation}
to leading order near the crack center.
Hence, the wave-crack interaction is controlled by the single global parameter $\ell/\lambda$, to which $\delta$ is inversely proportional: a small $\ell/\lambda$ corresponds to the non-adiabatic regime $\delta \gg 1$, in which the wave is reflected, while a large $\ell/\lambda$ corresponds to the adiabatic regime $\delta \ll 1$, in which the wave is transmitted as through a homogeneous medium; a transition region separates the two. Note that this analysis identifies the governing parameter and the two limit regimes, but not the quantitative location of the transition. 

A quantitative analysis can be performed using the transfer matrix method (TMM), which approximates a heterogeneous medium with a stack of homogeneous layers with constant material properties (Fig.~\ref{fig:tmm_standard}). Its main features are recalled in Appendix~\ref{app:tmm}.
The TMM makes the separation of roles between wave speed and impedance explicit. The global transfer matrix \eqref{global_transfer_matrix} is assembled from two types of factors with disjoint dependencies: the discontinuity matrices \eqref{discont_matrix} depend only on the impedance ratios of adjacent layers, while the propagation matrices \eqref{propag_matrix} depend only on the local wavenumber $k(\alpha) = \omega/c(\alpha)$, i.e., on the local wave speed, and accumulate phase without generating reflection.
Consequently, the reflected power is produced solely by the impedance profile, whereas the wave speed enters the scattering problem only indirectly, by dictating the local wavelength against which the impedance variation is measured.
The wave speed reduction within the phase-field support is thus not \emph{per se} the origin of the anomalous reflection and transmission; it is the accompanying impedance variation that governs them.
Fig.~\ref{fig:tmm_standard}a shows the obtained reflection and transmission  coefficients in dependence of $\ell/\lambda$ for the optimal \texttt{AT1} profile with $g_0 = 10^{-6}$.
Three distinct regimes can be distinguished.
If the wavelength is sufficiently larger than the phase-field support, in our case for $\ell/\lambda \lessapprox 0.08$ (at this ratio, the reflection power coefficient first drops below $99$\%), the quantitative behavior resembles that of a sharp crack, leading to a full reflection.
The opposite happens when the wavelength is smaller than (or comparable to) the phase-field support, in our setup for $\ell/\lambda \gtrapprox 0.22$\footnote{Recall from \eqref{eq:profile_AT1} that the length of the phase-field support is $4\ell$.} (from then onwards, the reflection coefficient is below $1$\%).
In this case the behavior is similar to that of a homogeneous material and the wave is fully transmitted.
In between these two extremes there is a regime where the wave is partially transmitted and partially reflected.
The TMM results thus suggest that the brittle model with stiffness degradation yields the intended wave-crack interaction only for $\lambda \gtrapprox \ell/0.08$. This condition is difficult to enforce in realistic scenarios where crack propagation can induce high-frequency waves. Moreover, as shown by \eqref{eq:tens_str}, in the present model the regularization length plays the role of a material parameter if the tensile strength of the material has to be quantitatively reproduced. 

\begin{figure}[t!]
    \centering
    \includegraphics{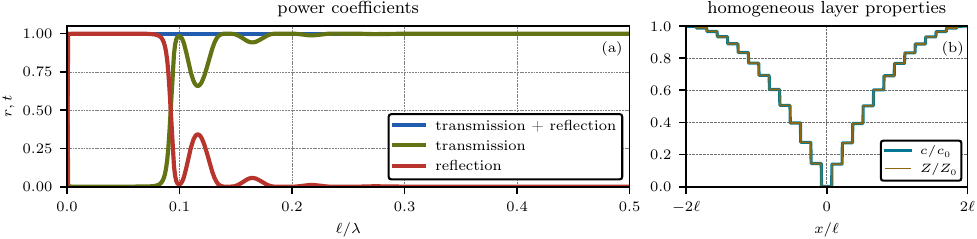}
    \caption{Reflected (r) and transmitted (t) fractions of the incident power depending on the $\ell/\lambda$ ratio for a tensile stress wave propagating through a \textit{fixed} phase-field crack using the brittle model with stiffness degradation (a). Normalized profile of the wave speed $c=\sqrt{E/\rho}$ and acoustic impedance $Z=\sqrt{\rho E}$ of the homogeneous layers used for the TMM computations (b). Note that for the actual computation, significantly more layers are used.}
    \label{fig:tmm_standard}
\end{figure}

\subsubsection{Interaction of a compressive wave with a fixed phase-field crack in a 1D bar}\label{sec:compr_stiffdeg}
For a compressive wave, the expected full transmission (with both crack faces in contact) is recovered in the 1D case through the strain energy decomposition.
With $\psi_{\text{D}} (\varepsilon) = \tfrac{1}{2} E_0 \langle \varepsilon \rangle_+^2$ and $\psi_{\text{R}} (\varepsilon) = \tfrac{1}{2} E_0 \langle \varepsilon \rangle_-^2$, the balance of linear momentum becomes
\begin{equation}\label{eq:wave_equation_standard_1D_split}
    - g^\prime(\alpha) \alpha^\prime E_0 \langle u^\prime \rangle_+ - g(\alpha) E_0 H (u^\prime) u^{\prime\prime} - E_0 H (- u^\prime) u^{\prime\prime} + \rho_0 \ddot{u} = 0
\end{equation}
which collapses to the bulk wave equation for $u^\prime < 0$. In \eqref{eq:wave_equation_standard_1D_split}, $H(\bullet)$ is the Heaviside function that takes the value $1$ if $(\bullet) \geq 0$ and $0$ if $(\bullet) < 0$.
\subsubsection{Numerical examples in 1D}\label{sec:interaction_wave_brittle_crack}
\begin{figure}[t!]
    \centering
    \includegraphics{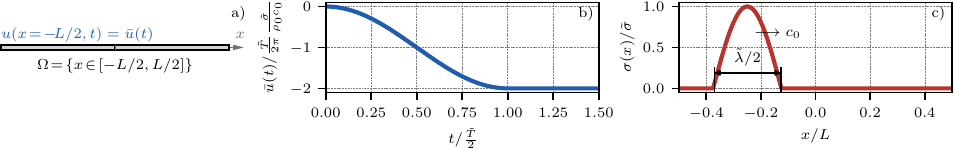}
    \caption{1D setup for the study of an elastic wave interacting with a pre-existing crack.}
    \label{fig:setup_1D_brittle}
\end{figure}

\begin{figure}[t!]
    \centering
    \includegraphics{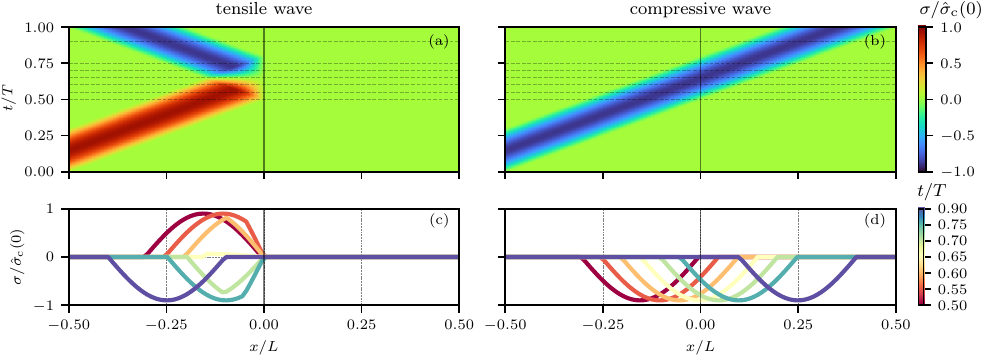}
    \caption{Interaction of a sinusoidal stress wave with a sharp crack in terms of $x$-$t$ diagram (a,c), and snapshots of the stress field (b,d).
    The dashed horizontal lines in the first row represent the time instants for the snapshots of the plots in the rows below.}
    \label{fig:interaction_sharp}
\end{figure}

As follows, we report the results of a numerical investigation in which a sample pulse interacts with a phase-field crack. 
\paragraph{Problem setup} 
To study the interaction between phase-field regularization and elastodynamics, we consider a 1D bar of length $L$ (Fig.~\ref{fig:setup_1D_brittle}a) subjected to an imposed displacement
\begin{equation}\label{eq:1D_pulse_loading}
    \bar{u} (t) =
    \begin{cases}
        \frac{\tilde{\sigma}}{\rho_0 c_0} \frac{\tilde{T}}{2\pi} \left[ \cos \left( 2 \pi \frac{t}{\tilde{T}} \right) - 1 \right] &t \in \left[0, \tfrac{\tilde{T}}{2} \right]\\
        - \frac{\tilde{\sigma}}{\rho_0 c_0} \frac{\tilde{T}}{\pi} &\text{else}\\
    \end{cases}
    \qquad\text{with}\qquad
    \tilde{T} = \frac{\tilde{\lambda}}{c_0}
\end{equation}
at $x=-L/2$ (Fig.~\ref{fig:setup_1D_brittle}b). 
This generates a sinusoidal tensile stress half-wave with wavelength $\tilde{\lambda}$, full period  $\tilde{T}$ (or angular frequency $\tilde{\omega}=2\pi/\tilde{T}$) and amplitude $\tilde{\sigma}$ traveling at a wave speed $c_0 = \sqrt{E_0 / \rho_0}$ in the undamaged material, with $E_0$ as the undamaged Young's modulus (Fig.~\ref{fig:setup_1D_brittle}c).
Note that, as a truncated sinusoid, the pulse is not monochromatic; we therefore distinguish its nominal wavelength $\tilde \lambda$ from the wavelength $\lambda$ of a generic harmonic component.
The results expected from the interaction of this stress pulse with a sharp crack are shown in Fig.~\ref{fig:interaction_sharp} by means of space-time diagrams (\ref{fig:interaction_sharp}a,c) and snapshots of the stress field at various time instants (\ref{fig:interaction_sharp}b,d).
The first column shows the result for a tensile wave, where the crack acts as a free boundary, reflecting the wave as undistorted compressive wave without energy transmission to the right part of the bar.
The second column shows the results for a compressive wave, which is transmitted without any distortion to the other side of the crack, as the crack lips are in contact.

In the numerical simulations, the problem is solved by the finite element method (FEM) with linear elements and a mesh size $\Delta x \approx \ell/5$ (unless specified otherwise), known to be sufficiently fine in both quasi-static \cite{Tanne_2018,Levy_2024} and dynamic \cite{Borden_2012_phase,Mandal_evaluation_2020} cases.
For time integration, we use the implicit Newmark-$\beta$ scheme \cite{newmark_method_1959} with $c_0\Delta t/\Delta x = 0.1$, $\gamma = 1/2$, and $\beta = 1/4$, providing unconditional stability, second-order accuracy, and no numerical dissipation in the undamaged elastodynamic case. We stop the computations as soon as the wave has reached the free end of the bar $x= L/2$, hence there are no boundary reflections.
Details of our numerical implementation are reported in Appendix~\ref{app:fem}.

The FEM computations in this section employ the constitutive model with the strain energy decomposition \eqref{eq:vol_dev_decomposition}, which in 1D reduces to the tension–compression split in \eqref{eq:wave_equation_standard_1D_split}: the stiffness degradation is applied or not depending on the local sign of the strain. This is in contrast to the scattering analysis and the TMM of Section \ref{1D_stiffdeg}, which presuppose a fixed form of the wave equation, with degradation applied irrespective of the strain sign. The two descriptions coincide wherever the strain field within the phase-field support is uniformly tensile (or, for the compressive case, uniformly compressive); once both signs coexist — as in the superposition of incident and reflected waves and in the high-frequency oscillations developing during the interaction — the FEM problem becomes nonlinear, with coefficients depending on the solution through the sign of the strain. The comparison between FEM and TMM results is therefore only qualitative.

In the following analyses, we adopt for the bar $L=1000$~mm, for the material $E_0 = 30\,000$~MPa, $G_{\text{c}} = 0.1$~N/mm, and $\rho_0 = 2\,400$~kg/m$^3$, and for the imposed stress wave $\tilde{\sigma} = 4.5$~MPa and $\tilde{\lambda}=600$~mm.
We test two values of the regularization length, namely $\ell = 15$~mm and $\ell = 45$~mm; both ensure a  small phase-field support compared to domain size ($\ell/L=0.015$ and $0.045$) and wavelength of the stress wave ($\ell/\tilde{\lambda} = 0.025$ and $0.075$). With the above values, it is $\hat{\sigma}_{\text{c}}(0)=8.7$~MPa and $5$~MPa, hence $\tilde{\sigma} / \hat{\sigma}_{\text{c}}(0) = 0.52$ and $0.9$.
Finally, we set $g_0 = 10^{-6}$.

\begin{figure}[t!]
    \centering
    \includegraphics{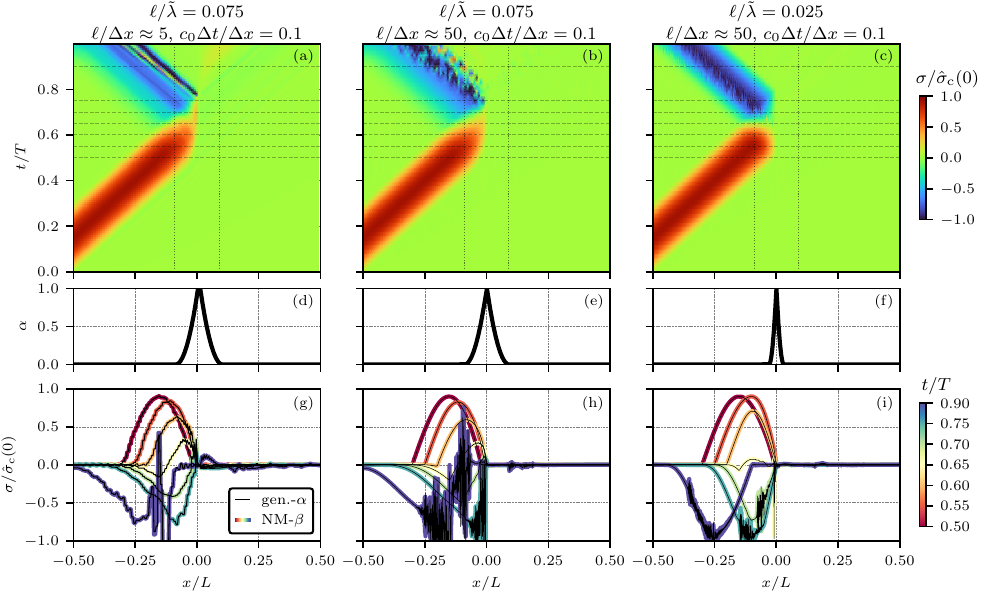}
    \caption{Interaction of a sinusoidal stress wave with a \textit{fixed} phase-field crack using the brittle model with stiffness degradation (middle and right columns) in terms of $x$-$t$ diagram (a,b,c), phase field (d,e,f) and snapshots of the stress field (g,h,i).
    In the left column (a,d,g), a standard spatial discretization with element size $\Delta x \approx \ell/5$ is used, while in the middle column (b,e,h), a significantly finer discretization with $\Delta x \approx \ell/50$ is employed.
    The right column (c,f,i) has a different ratio of $\ell/\tilde{\lambda} = 0.025$.
    These results are obtained with the Newmark-$\beta$ integration scheme, while we additionally show the snapshots from the generalized-$\alpha$ scheme with thin black lines in the last row.
    The dashed horizontal lines in the first row represent the time instants for the snapshots of the plots in the rows below.}
    \label{fig:interaction_standard}
\end{figure}

\paragraph{Numerical results for a tensile wave and a fixed phase-field crack} 
Fig.~\ref{fig:interaction_standard} reports a space-time diagram of the stress (Fig.~\ref{fig:interaction_standard}a-c), the fixed phase-field profile (Fig.~\ref{fig:interaction_standard}d-f), and snapshots of the stress field at selected time steps (Fig.~\ref{fig:interaction_standard}g-i).
The left column (Fig.~\ref{fig:interaction_standard}b,d,g) shows the results for $\ell/\tilde{\lambda} = 0.075$.
Once the wave enters the region of the phase-field crack, its speed is reduced and its shape is distorted.
Upon arriving at the center of the crack at $x=0$, part of the wave is reflected with an altered shape and high-frequency oscillations, while the remaining portion is transmitted to the right part of the bar.
In \cite{Steinke_comparative_2016,Schlueter_numerical_2017} this behavior is interpreted as a numerical artifact due to a coarse spatial and temporal discretization.
To show that this is not the only reason, we illustrate in the middle column (Fig.~\ref{fig:interaction_standard}b,e,h) the results obtained with a significantly finer discretization of $\Delta x \approx \ell/50$ (keeping $c_0\Delta t/\Delta x = 0.1$).
There, the high-frequency distortions of the wave are even more pronounced, and part of the wave energy is still transmitted through the crack.
In the last row, we overlay the results with further ones obtained using the generalized-$\alpha$ time integration scheme with parameters $\alpha_m = 0.2$, $\alpha_f = 0.4$, $\gamma = \tfrac{1}{2} - \alpha_m + \alpha_f$ and $\beta= \tfrac{1}{4} (1 - \alpha_m + \alpha_f)^2$.
These parameters correspond to a high-frequency spectral radius $\rho_\infty = 2/3$, so as to introduce controlled numerical dissipation of high-frequency modes while retaining second-order accuracy and unconditional stability.
While the numerical damping slightly lowers the amplitudes of the oscillations, the same behavior obtained with Newmark-$\beta$ persists.
From now on, we exclusively employ the generalized-$\alpha$ scheme with the mentioned parameters.
Finally, the last column of Fig.~\ref{fig:interaction_standard} (c,f,i) shows the results for a smaller ratio $\ell/\tilde{\lambda} = 0.025$, but keeping $\Delta x \approx \ell/50$.
Confirming the theoretical results, the interaction of the elastic wave and the phase-field crack resembles more closely that of the sharp crack (Fig.~\ref{fig:interaction_sharp}), while the shape of the reflected wave is still significantly altered.

\begin{figure}[t!]
    \centering
    \includegraphics{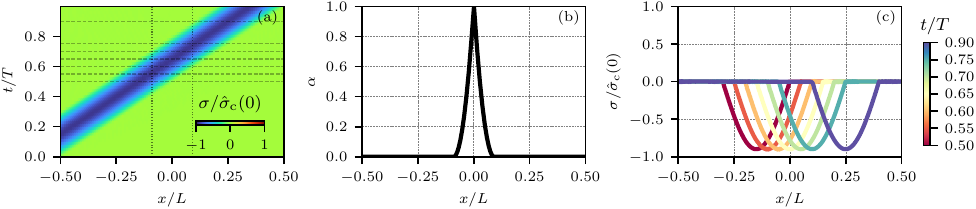}
    \caption{Interaction of a sinusoidal, compressive stress wave with a \textit{fixed} phase-field crack using the brittle model with stiffness degradation in terms of $x$-$t$ diagram (a), phase field (b) and snapshots of the stress field (c).
    A spatial discretization with element size $\Delta x \approx \ell/50$ and the generalized-$\alpha$ integration scheme with $c_0 \Delta t / \Delta x = 0.1$ are used.
    The dashed horizontal lines in (a) represent the time instants for the snapshots of the plot in (c).}
    \label{fig:interaction_standard_compressive}
\end{figure}

\paragraph{Numerical results for a compressive wave and a fixed phase-field crack}
In the FEM simulations a compressive wave is transmitted with no distortion, as shown in Fig.~\ref{fig:interaction_standard_compressive}, matching the expected behavior (Fig.~\ref{fig:interaction_sharp}b,d).

\begin{figure}[t!]
    \centering
    \includegraphics{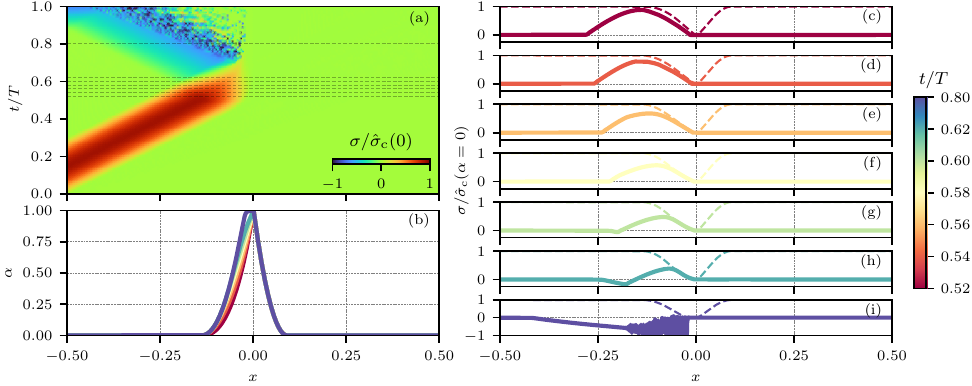}
    \caption{Interaction of a sinusoidal stress wave with an \textit{evolving} phase-field crack using the brittle model with stiffness degradation in terms of $x$-$t$ diagram (a) and snapshots of the phase field (b) and stress field (c-i).
    The material strength $\hat{\sigma}_{\text{c}}(0)$ in \eqref{eq:tens_str} is plotted with a dashed line. The dashed horizontal lines in (a) represent the time instants for the snapshots of the plots in the rows below.}
    \label{fig:widening_standard}
\end{figure}

\paragraph{Numerical results for a tensile wave and an evolving phase-field crack} So far we held the phase field fixed ($\dot{\alpha} = 0$). Next, we repeat the computation while letting both displacement and phase field evolve.
We retain the setup of Fig.~\ref{fig:setup_1D_brittle} with $\ell/\tilde{\lambda} = 0.075$, $\ell/\Delta x \approx 50$, and $c_0 \Delta t / \Delta x = 0.1$. The results, illustrated in Fig.~\ref{fig:widening_standard}, show a widening of the phase-field profile as soon as the wave interacts with the regularized crack, indicating the fulfillment of the damage evolution criterion \eqref{eq:alpha_strongform} and a spurious increase of the dissipated energy that no longer matches  the fracture toughness $G_{\text{c}}$. 

This phenomenon can be explained considering the 1D elastic domain \eqref{eq:elastic_domain_brittle_1D}, from which the profile of the material strength along the bar is reported in Fig.~\ref{fig:widening_standard}.
As the wave enters the phase-field support, damage evolution is triggered since the stress exceeds the local $\hat{\sigma}_{\text{c}}(\alpha)$, which becomes extremely small (depending on $g_0$) at the center of the phase-field crack (Fig.~\ref{fig:widening_standard}c,d).
As the damage reduces $\hat{\sigma}_{\text{c}}(\alpha)$, further points are exposed to damage evolution (Fig.~\ref{fig:widening_standard}e-h), which  widens the damaged band.
In addition, the local reduction of the wave speed  forces the incoming stresses to pile up as the wave stalls, further facilitating damage.
The phase-field evolution along with the wave--stiffness interaction discussed earlier involving partial wave reflection ultimately leads to high-frequency oscillations as illustrated in Fig.~\ref{fig:widening_standard}i.

Note that, for a fixed phase field, the reflection and transmission behavior quantified with the TMM is independent of the stress wave amplitude.
Instead, the widening behavior (as well as the accompanying reflection and transmission) does depend on the amplitude of the stress, as this amplitude directly influences the attainment of the local material strength.

\subsection{Brittle model with stiffness+density degradation}\label{sec:density_degradation}
To avoid a spatially varying wave speed, Chen et al.~\cite{chen_instability_2017} and Tian et al.~\cite{Tian_dynamic_2020} propose to modify the kinetic energy by degrading also the mass density.
In this section, we first briefly recapitulate the main equations of this model and then we again analyze the interaction of elastic waves with a pre-existing phase-field crack in a 1D bar as predicted by this model.

\subsubsection{Energetic quantities and governing equations}
With the brittle model with stiffness+density degradation, the kinetic energy reads
\begin{equation}\label{eq:kinetic_energy_degraded}
    \mathcal{K} (\dot{\boldsymbol{u}}, \alpha) = \int_\Omega \frac{1}{2} h(\alpha) \rho_0 \left|\dot{\boldsymbol{u}}\right|^2 \mathrm{d}\boldsymbol{x} \quad\text{,}
\end{equation}
where $h(\alpha)$ is a monotonically decreasing density degradation function, which, along with \eqref{eq:irr}, leads to the violation of the mass conservation condition, namely
\begin{equation}
    \dot{m} = \frac{\mathrm{d}}{\mathrm{d} t} \int_{\Omega} h(\alpha) \rho_0 \mathrm{d}\boldsymbol{x} = \int_{\Omega} h^\prime(\alpha) \dot{\alpha} \rho_0 \mathrm{d}\boldsymbol{x} \leq 0 \qquad\text{for}\qquad \dot{\alpha} \geq 0 \qquad\text{.}
\end{equation}

By reapplying the principles in Section \ref{sec:governing_equations_brittle} with \eqref{eq:kinetic_energy_degraded} in place of \eqref{eq:K}, the previous governing equation \eqref{eq:u_strongform} becomes
\begin{equation}\label{eq:BF_strong_form_u_degraded}
    - \nabla \cdot \boldsymbol{\sigma} + \underbrace{h^\prime(\alpha) \dot{\alpha} \rho_0 \dot{\boldsymbol{u}}}_{(\dagger)} + h(\alpha) \rho_0 \ddot{\boldsymbol{u}} = \boldsymbol{0} \qquad \forall (\boldsymbol{x},t) \in \Omega \times [0,T]\quad,
\end{equation}
still accompanied by boundary conditions \eqref{eq:NeumannBC}, where $(\dagger)$ is an additional term not present in the sharp crack case nor in the brittle model with stiffness degradation.
The KKT conditions~\eqref{eq:alpha_strongform} for the phase field remain unchanged, but with a modified damage energy release rate
\begin{equation}
    Y(\boldsymbol{\varepsilon}, \dot{\boldsymbol{u}}, \alpha) = - g^\prime(\alpha) \psi_{\text{D}} (\boldsymbol{\varepsilon}) + \underbrace{\frac{1}{2} h^\prime(\alpha) \rho_0 \left|\dot{\boldsymbol{u}}\right|^2}_{(\diamond)} \qquad\text{,}
\end{equation}
which now includes the contribution $(\diamond)$ depending on the velocity $\dot{\boldsymbol{u}}$.
Since $h^\prime(\alpha) \leq 0$ and $\rho_0 \left|\dot{\boldsymbol{u}}\right|^2 \geq 0$, this term leads to a lower energy release rate compared to \eqref{eq:standard_err}.

In the 1D case, the critical stress reads
\begin{equation}\label{eq:elastic_domain_brittle_1D_densdegrad}
    \hat{\sigma}_{\text{c}}(\alpha,\dot{u}) = \left( (1-\alpha)^2 +g_0 \right) \sqrt{\frac{3 G_{\text{c}}E_0}{8 \ell (1-\alpha)} + E_0 \rho_0 \dot{u}^2 }
    \qquad\text{,}
\end{equation}
i.e. it now depends also on $\dot{u}$ and grows for increasing velocities.

\subsubsection{Interaction of a tensile wave with a fixed phase-field crack in a 1D bar}\label{sec_interaction_tensile_fixed_pf}
\begin{figure}[t!]
    \centering
    \includegraphics{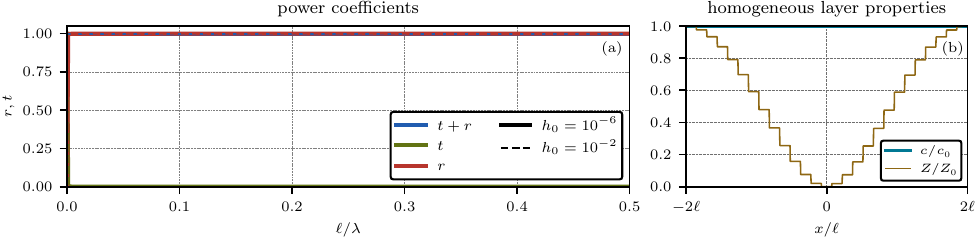}
    \caption{Reflected (r) and transmitted (t) fractions of the incident power depending on the $\ell/\lambda$ ratio for a tensile stress wave propagating through a \textit{fixed} phase-field crack using the brittle model with stiffness+density degradation (a). Normalized profile of the wave speed $c=\sqrt{E/\rho}$ and acoustic impedance $Z=\sqrt{\rho E}$ of the homogeneous layers used for the TMM computations (b). Note that for the actual computation, significantly more layers are used.}
    \label{fig:tmm_degraded}
\end{figure}

We now consider the same problem of Section \ref{1D_stiffdeg}, i.e. we want to study in 1D the interaction of a tensile wave with a phase-field crack represented by the optimal profile \eqref{eq:profile_AT1} which is held constant in time.
As in Section \ref{1D_stiffdeg}, we ignore the volumetric-deviatoric decomposition. 
The equation of motion becomes
\begin{equation}\label{eq:wave_equation_density}
    - \frac{\partial}{\partial x} \left[ g(\alpha) E_0 u^\prime \right] + \frac{\partial}{\partial t}\left[h(\alpha)\rho_0 \dot{u}\right] = 0\qquad .
\end{equation}
For a fixed phase field ($\dot\alpha = 0$), the elastodynamic problem reduces again to a heterogeneous wave equation, now with independently degraded stiffness $E(\alpha) = g(\alpha)E_0$ and density $\rho(\alpha) = h(\alpha)\rho_0$. Thus, the scattering analysis in Appendix~\ref{scatt} continues to apply, now with
\begin{equation}\label{eq:wavespeed_density_deg}
    c (\alpha) = \sqrt{\frac{g(\alpha)}{h(\alpha)}}c_0\qquad Z(\alpha) = \sqrt{g(\alpha)\,h(\alpha)}Z_0
\end{equation}
The choice $h(\alpha) = g(\alpha)$ leads to a constant wave speed $c(\alpha)=c_0$, whereas the acoustic impedance becomes
\begin{equation}
  Z(\alpha) \;=\; g(\alpha)\,Z_0 \quad\text{.}
  \label{eq:impedance-dd}
\end{equation}
Thus, the density degradation does not remove the impedance well accompanying the regularized crack, but rather makes it stronger.
Let us now analyze the adiabatic criterion \eqref{eq:WKB}.
With the wave speed equal to $c_0$, the local wavenumber is $k(\alpha) = \omega/c_0 = 2\pi/\lambda$. 
The local wavelength no longer contracts toward the crack center, and nothing compensates the diverging relative impedance gradient, hence
\begin{equation}\label{eq:WKB-dd}
  \delta(x) = \frac{|Z'(x)|}{Z(x)}\,\frac{c_0}{\omega} \simeq\; \frac{\lambda}{\pi\,|x|}\, \xrightarrow[\;x\to 0\;]{}\; \infty \quad\text{.}
\end{equation}
Thus, a non-adiabatic region of size $|x| \lesssim \lambda/\pi$ exists at every frequency and its extent is set by the wavelength, not by $\ell$.
Across this region the impedance drops by several orders of magnitude, down to its residual value at the crack center.
Total reflection therefore occurs independently of the $\ell/\lambda$ ratio. In the TMM, 
the propagation matrices \eqref{propag_matrix} accumulate a spatially uniform phase and the entire scattering response is produced by the discontinuity matrices \eqref{discont_matrix} through the impedance ratios $Z^{l+1}/Z^{l}$ (with $l$ denoting the layer index), whose profile is reported in Fig.~\ref{fig:tmm_degraded}b.
This behavior is shown in Fig.~\ref{fig:tmm_degraded}a, for which we set $h(\alpha) = (1-\alpha)^2 + h_0$ with a \textit{residual density} $h_0$ and investigate two values of the residual density $h_0$, one equal to $g_0=10^{-6}$ and a much larger one $h_0=10^{-2}$. While TMM results are unaffected by this choice, the reason for using a larger value will become apparent later.
Unlike in the stiffness-degraded model, no adiabatic (transmissive) regime can be reached by tuning $\ell/\lambda$.

\subsubsection{Interaction of a compressive wave with a fixed phase-field crack in a 1D bar}

To study the interaction of a compressive wave, we use the volumetric-deviatoric decomposition as in Section~\ref{sec:compr_stiffdeg}, which yields the governing equation
\begin{equation}\label{eq:wave_equation_stiffdens}
    \begin{aligned}
        &- g^\prime(\alpha) \alpha^\prime E_0 \langle u^\prime \rangle_+ - g(\alpha) E_0 H (u^\prime) u^{\prime\prime} - E_0 H (- u^\prime) u^{\prime\prime} +  h(\alpha) \rho_0 \ddot{u} = 0 \quad\text{.}
    \end{aligned}
\end{equation}
In compression, the stiffness is not degraded but the density is, leading to
\begin{equation}\label{eq:wavespeed_density_deg_compr}
    c (\alpha) =
        \frac{c_0}{\sqrt{h(\alpha)}}\qquad Z(\alpha) = \sqrt{h(\alpha)}Z_0 \quad\text{,}
\end{equation}
thus the impedance still decreases within the phase-field support while the local wavelength now \emph{grows} toward the crack center, giving $\delta(x) \simeq \lambda\ell/(2\pi x^2)$ and hence an even more strongly non-adiabatic region.

\subsubsection{Numerical results}
As follows we report numerical results obtained with the same problem setup of Section \ref{sec:interaction_wave_brittle_crack}, this time using the brittle model with stiffness and density degradation. Once again the FEM computations employ the constitutive model with the strain energy decomposition \eqref{eq:vol_dev_decomposition}, which in 1D now reduces to the tension–compression split in \eqref{eq:wave_equation_stiffdens}. This is in contrast to the scattering analysis and the TMM of Section \ref{sec_interaction_tensile_fixed_pf}, which presuppose a fixed form of the wave equation. Therefore, also in this case the comparison between FEM and TMM results is only qualitative.

\begin{figure}[t!]
    \centering
    \includegraphics{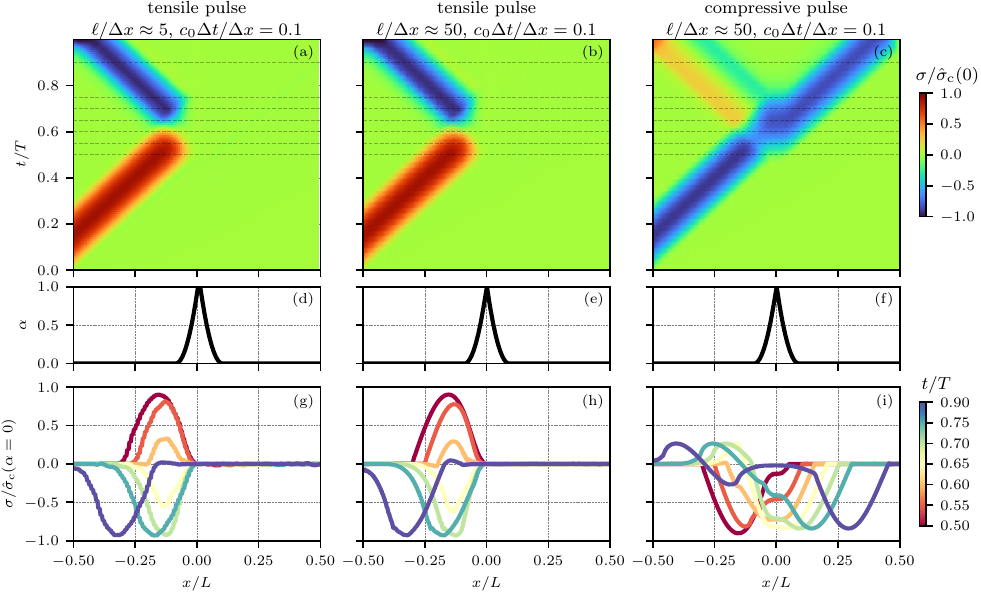}
    \caption{Interaction of a sinusoidal stress wave with a \textit{fixed} phase-field crack using the brittle model with stiffness+density degradation, in terms of $x$-$t$ diagram (a,b,c), phase field (d,e,f) and snapshots of the stress field (g,h,i). All results are obtained using the generalized-$\alpha$ time integration scheme. The dashed horizontal lines in the first row represent the time instants for the snapshots of the plots in the rows below.}
    \label{fig:interaction_degraded}
\end{figure}

\paragraph{Numerical results for a tensile wave and a fixed phase-field crack}
We repeat the numerical experiment of Fig.~\ref{fig:interaction_standard} for the model with stiffness+density degradation and report the results in Fig.~\ref{fig:interaction_degraded}.
We set $h_0 = 10^{-2}$, a value significantly larger than $g_0 = 10^{-6}$.
This choice stems from the fact that the degraded mass induces high frequency oscillations of the stress and numerical instabilities once the wave interacts with the phase-field profile.
To limit oscillations, alternative strategies are available; however, these modify the material behavior by introducing viscosity in the constitutive law \cite{chen_instability_2017,Tian_dynamic_2020} or give up variational consistency by neglecting the concave term in $Y$ as in \cite{durussel_origins_2025}.
Apart from the numerical difficulties, another issue with this model is that the velocity-dependent term $\tfrac{1}{2} h^\prime(\alpha) \rho_0 \left|\dot{\boldsymbol{u}}\right|^2$ in $Y$ potentially leads to a negative energy release rate in case of high velocities.
A negative energy release rate can lead to the loss of convexity of the damage subproblem as explained in \cite[Section 4.2]{heinzmann_exact_2026}.

Figs.~\ref{fig:interaction_degraded}a,d,g involve a tensile stress wave and use $\ell/\Delta x \approx 5$, while Figs.~\ref{fig:interaction_degraded}b,e,h report the results with  $\ell/\Delta x \approx 50$. The wave is fully reflected (apart from minor numerical artifacts for $\Delta x \approx \ell/5$ which are visible in Fig.~\ref{fig:interaction_degraded}g).
Unlike with the model with stiffness degradation, here the results are independent of the discretization.

\paragraph{Numerical results for a compressive wave and a fixed phase-field crack}
Let us now study the interaction of a compressive wave ($|\tilde{\sigma}| = 4.5$~MPa and remaining parameters same as before) with a phase-field crack, see Figs.~\ref{fig:interaction_degraded}c,f,i for $\ell/\Delta x \approx 50$. 
For a compressive wave, the combination of undegraded stiffness and degraded density results in an increase of the wave speed approaching $\alpha=1$ (diverging to $+\infty$ if $h_0$ is neglected), see \eqref{eq:wavespeed_density_deg_compr}.
In the results of Fig.~\ref{fig:interaction_degraded}c,i, the transmitted wave is reduced in amplitude and altered in shape, while a significant reflection occurs at the regularized crack, highlighting that the strain energy decomposition does not yield the expected behavior.
Moreover, in Fig.~\ref{fig:interaction_degraded}c a faster motion of the wave is visible  around $t/T \approx 0.6$ at the regularized crack, which reflects the increase of wave speed indicated by \eqref{eq:wavespeed_density_deg_compr} in the compressive case.
For the compressive wave, we obtained an even more strongly non-adiabatic region, consistent with the significant reflection and waveform alteration observed in Fig.~\ref{fig:interaction_degraded}c,i.

\begin{figure}[t!]
    \centering
    \includegraphics{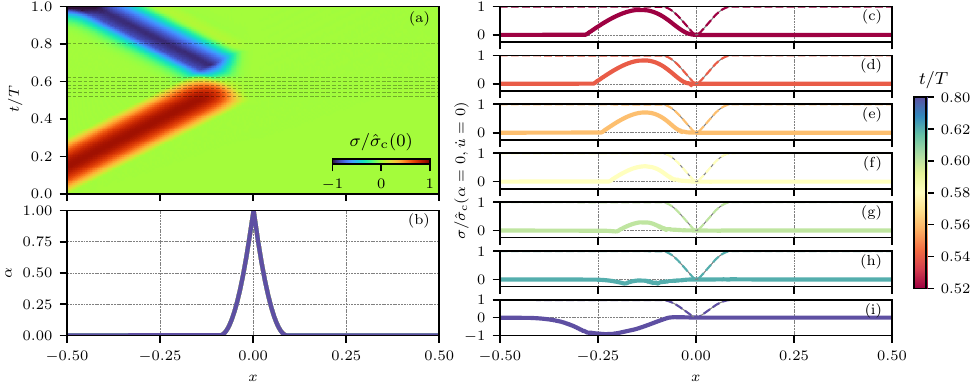}
    \caption{Interaction of a sinusoidal stress wave with an \textit{evolving} phase-field crack using the brittle model with stiffness+density degradation in terms of $x$-$t$ diagram (a), phase field (b) and snapshots of the stress field (c-i).
    The material strength $\hat{\sigma}_{\text{c}}(0, \dot{u})$ is plotted with a dashed line, whereas its counterpart without velocity dependence $\hat{\sigma}_{\text{c}}(0)$ in \eqref{eq:tens_str} is plotted for comparison as a gray line. The dashed horizontal lines in (a) represent the time instants for the snapshots of the plots in the rows below.}
    \label{fig:widening_densdegrad}
\end{figure}

\paragraph{Numerical results for a tensile wave and an evolving phase-field crack}
We now repeat the experiment of Fig.~\ref{fig:widening_standard}, where the phase field is allowed to evolve, for the model with stiffness+density degradation.
The results in terms of $x$-$t$ diagram are illustrated in Fig.~\ref{fig:widening_densdegrad}a, while the phase-field profile and the stress field along with the material strength are reported in Fig.~\ref{fig:widening_densdegrad}b and Fig.~\ref{fig:widening_densdegrad}c-i, respectively.
Unlike in Fig.~\ref{fig:widening_standard}, now the damage does not evolve, hence the phase-field profile does not widen and the tensile stress wave keeps being completely reflected.
In  Fig.~\ref{fig:widening_densdegrad}c-i we plot the material strength given by \eqref{eq:elastic_domain_brittle_1D_densdegrad} for $\alpha=0$ along with its counterpart without the velocity-dependent contribution, \eqref{eq:tens_str}, to show that, for the case analyzed here, the dependence of the strength on the velocity is negligible.

\subsection{Summary of the results with brittle models}\label{sec:summary_brittle}
From the results reported in Sections~\ref{sec:stiffdeg} and \ref{sec:density_degradation}, we conclude that phase-field formulations of dynamic brittle fracture cannot consistently represent the interaction of stress waves with a pre-existing crack, regardless of whether they involve degrading the stiffness or both the stiffness and the mass density.
Phase-field models with degradation functions different from the ones chosen here should display similar issues if the same degradation mechanisms and coupling between damage and mechanical fields are retained. This is the case for the models recovering cohesive-like behavior proposed in \cite{lorentz_2011,Lorentz_2012,Wu_2018} for the quasi-static case and extended to dynamics in \cite{Geelen_phase_2019,Nguyen_modeling_2018,Mandal_evaluation_2020}.
We therefore do not consider them separately here.
Concerning the \texttt{AT2} model, its vanishing elastic domain implies that any non-zero stress wave triggers local damage evolution and hence a local change of material properties, as well as a mix of reflections and transmissions, ultimately leading to a distorted wave regardless of the wave amplitude.

Hence, a different modeling approach is required. A potential candidate is the cohesive model proposed for the quasi-static case in \cite{Vicentini_variational_2026,bourdin_variational_2025} where the material stiffness is not degraded.
In the following section, we extend this model to dynamics and analyze its predictions.

\FloatBarrier
\section{Phase-field modeling of dynamic cohesive fracture with strength degradation}\label{sec:phase_field_model_cohesive}
In this section, we extend the recently proposed phase-field regularization of cohesive fracture \cite{Vicentini_variational_2026,bourdin_variational_2025} to dynamics.
Following the structure of the previous section on brittle models, we first derive the governing equations and subsequently analyze the model behavior in 1D, focusing on the interaction of a stress wave with a pre-existing crack.

\subsection{Energetic quantities and governing equations}\label{sec:cohesive_model_formulation}
The main modification introduced by the cohesive model in \cite{Vicentini_variational_2026,bourdin_variational_2025} lies in the elastic strain energy density and the regularity of the fields. 
In particular, the cohesive formulation introduces an additional reversible kinematic field, the \textit{eigenstrain} $\boldsymbol{\eta} : \Omega \times [0,T] \rightarrow \mathbb{M}^{d}_{\text{sym}}$, where $\mathbb{M}^{d}_{\text{sym}}$ is the set of symmetric second-order tensors.
Specifically, the elastic strain energy density is written as\footnote{Originally, the deviatoric contribution is written as $\mu_0 \left| \boldsymbol{\varepsilon}_{\text{dev}} - \boldsymbol{\eta}_{\text{dev}} \right|^2$.
However, since the elastic domain depends only on the spherical and deviatoric components of $\boldsymbol{\eta}$ and is axisymmetric about the purely volumetric direction, it can be replaced by its lower bound $\mu_0 \left( \left| \boldsymbol{\varepsilon}_{\text{dev}} \right| - \left| \boldsymbol{\eta}_{\text{dev}} \right| \right)^2$ which is attained when $\boldsymbol{\varepsilon}_{\text{dev}}$ and $\boldsymbol{\eta}_{\text{dev}}$ are co-aligned (see \cite[Appendix C]{Vicentini_variational_2026}).
This substitution, originally derived from energy minimization in the quasi-static setting, remains valid in dynamics: the original $\psi$ (with the deviatoric term $\mu_0 \left| \boldsymbol{\varepsilon}_{\text{dev}} - \boldsymbol{\eta}_{\text{dev}} \right|^2$) is convex in $\boldsymbol{\eta}$ and the eigenstrain carries no inertia, so the point-wise stationarity condition of the action functional with respect to $\boldsymbol{\eta}$ coincides with the first-order minimality condition of the quasi-static problem -- and, by convexity, this stationary point is in fact a minimum.} \cite{Vicentini_variational_2026}
\begin{equation}\label{eq:psi_cohesive}
    \psi (\boldsymbol{\varepsilon}, \boldsymbol{\eta}, \alpha) =  \frac{\kappa_0}{2} \left( \text{tr}(\boldsymbol{\varepsilon}) - \text{tr}(\boldsymbol{\eta}) \right)^2+ \mu_0 \left( \left| \boldsymbol{\varepsilon}_{\text{dev}} \right| - \left| \boldsymbol{\eta}_{\text{dev}} \right| \right)^2+ \pi (\boldsymbol{\eta}, \alpha) \quad\text{,}
\end{equation}
where the \textit{eigenstrain potential} $\pi(\boldsymbol{\eta},\alpha)$ is defined as
\begin{equation}\label{eq:eigenstrain_potential}
    \pi(\boldsymbol{\eta},\alpha) =
    \begin{cases}
        a(\alpha)\pi_0(\text{tr}(\boldsymbol{\eta}), \left|\boldsymbol{\eta}_{\text{dev}}\right|) \qquad&\text{if}\ \text{tr}(\boldsymbol{\eta}) \geq 0\\
        +\infty &\text{otherwise}
    \end{cases}\quad\text{.}
\end{equation}
Here, $\pi_0$ is the support function of the initial elastic domain of the material $ \mathcal{S}_0$, which is considered a material property.
The eigenstrain potential \eqref{eq:eigenstrain_potential} is the support function of the damaged elastic domain $ \mathcal{S} (\alpha)$, which is assumed to homothetically shrink with increasing $\alpha$ from $\mathcal{S}_0=\mathcal{S} (0)$ through the degradation function $a(\alpha)$.
This means that in this model $\alpha$ leads to the  degradation of the local strength rather than of the stiffness.
Further assumptions  involve $ \mathcal{S} (\alpha)$ being convex and bounded in any direction apart from the purely hydrostatic compressive stresses, i.e. for $\boldsymbol{\sigma}_{\text{dev}}=\boldsymbol{\sigma}-\tfrac{1}{d}\,\text{tr}(\boldsymbol{\sigma})\boldsymbol{I}=\boldsymbol{0}$ and $\text{tr}(\boldsymbol{\sigma})<0$ \cite{Vicentini_variational_2026}.
Apart from these properties, the selection of $\mathcal{S} (\alpha)$ is arbitrary and, for the case at hand, is specified in Section~\ref{sec:cohesive_compact_formulation} along with the choice of $a(\alpha)$.

Assuming sufficient temporal regularity and adopting the spatial regularity framework of the quasi-static setting \cite{Vicentini_variational_2026}, the state vector $\boldsymbol{z} = (\boldsymbol{u}, \boldsymbol{\eta}, \alpha)$ at each time instant is piecewise smooth with a singular part localized on a jump set $J(\boldsymbol{z}) \subset \Omega$ of co-dimension $1$. 
The displacement $\boldsymbol{u}$ is continuously differentiable on $\Omega \setminus J(\boldsymbol{z})$ and admits jumps only across $J(\boldsymbol{z})$, so the strain decomposes into a regular and a singular part as
\begin{equation}
    \boldsymbol{\varepsilon} = \boldsymbol{\varepsilon}_{\text{R}} + \boldsymbol{\varepsilon}_{\text{S}}
    \quad \text{, with} \quad
    \boldsymbol{\varepsilon}_{\text{R}} = \nabla_{\text{sym}} \boldsymbol{u}
    \quad \text{, and} \quad
    \boldsymbol{\varepsilon}_{\text{S}} = (\jump{\boldsymbol{u}} \otimes_{\text{sym}} \boldsymbol{m}) \, \delta_{J(\boldsymbol{z})} \qquad\text{,}
    \label{eq:CF_strain_jump}
\end{equation}
where $\delta_{J(\boldsymbol{z})}$ is the Dirac surface measure concentrated on $J(\boldsymbol{z})$, $\otimes_{\text{sym}}$ denotes the symmetrized outer product and $\jump{\boldsymbol{u}}(\boldsymbol{x},t) = \boldsymbol{u}^+(\boldsymbol{x},t) - \boldsymbol{u}^-(\boldsymbol{x},t)$ is the displacement jump across $J(\boldsymbol{z})$, with $\boldsymbol{u}^+(\boldsymbol{x},t)$ the limit of $\boldsymbol{u}(\boldsymbol{x},t)$ approaching $\boldsymbol{x}$ from the direction of the unit normal $\boldsymbol{m}$ to the jump set at any given $t$.
Analogously to \eqref{eq:CF_strain_jump}, we have $\boldsymbol{\eta} = \boldsymbol{\eta}_{\text{R}} + \boldsymbol{\eta}_{\text{S}}$ and, enforcing $\boldsymbol{\varepsilon} - \boldsymbol{\eta} \in L^2(\Omega)$ to guarantee a finite energy, we obtain 
\begin{equation}
    \boldsymbol{\eta}_{\text{S}} = \boldsymbol{\varepsilon}_{\text{S}}= \left( \jump{\boldsymbol{u}} \otimes_{\text{sym}} \boldsymbol{m} \right) \, \delta_{J(\boldsymbol{z})} \qquad\text{,}
    \label{eq:CF_eta_jump}
\end{equation}
which gives the link between the singular part of the eigenstrain and the displacement jump.
The condition $\text{tr}(\boldsymbol{\eta}) \geq 0$ in \eqref{eq:eigenstrain_potential} naturally enforces non-interpenetration at the crack lips, $\jump{\boldsymbol{u}} \cdot \boldsymbol{m} \geq 0$.
For further details, we refer to \cite{alessi2015gradient,bourdin_variational_2025,Vicentini_variational_2026}.

In contrast to the brittle model with stiffness and density degradation of Section~\ref{sec:density_degradation}, the cohesive model does not degrade the mass, so the bulk density remains $\rho_0$; the only question is whether the crack itself carries mass.
Writing the total mass as $m = \int_{\Omega \setminus J(\boldsymbol{z})} \rho_0 \, \mathrm{d}x + m_J$, where $m_J$ denotes any mass concentrated on the jump set, mass conservation between an undamaged configuration ($J(\boldsymbol{z}) = \emptyset$) at time $t_A$ and a cracked configuration at $t_B$ requires $m_J = 0$: the bulk integral is unchanged since the density is undegraded, so the added surface set $J(\boldsymbol{z})$, of co-dimension~1, must carry no mass.
Hence the density admits no singular part and the kinetic energy is
\begin{equation}\label{eq:kin_cohesive}
    \mathcal{K} (\dot{\boldsymbol{u}}) = \int_\Omega \frac{1}{2} \left|\dot{\boldsymbol{u}}\right|^2 \mathrm{d}m
    = \int_{\Omega\setminus J(\boldsymbol{z})} \frac{1}{2} \rho_0 \left|\dot{\boldsymbol{u}}\right|^2 \mathrm{d}\boldsymbol{x} \quad\text{.}
\end{equation}
The fracture energy remains identical to the brittle case of \eqref{eq:P}.

The governing equations are obtained as usual from the principle of stationary action with irreversibility and energy balance.
The momentum balance is obtained as
\begin{equation}\label{eq:u_strongform_cohesive}
    - \nabla \cdot \boldsymbol{\sigma} + \rho_0 \ddot{\boldsymbol{u}} = \boldsymbol{0} \qquad \forall (\boldsymbol{x},t) \in \Omega \times [0,T]
    \qquad\text{,}\quad
    \boldsymbol{\sigma} \cdot \boldsymbol{n} = \boldsymbol{f} \qquad \forall (\boldsymbol{x},t) \in \partial \Omega_{\text{N}} \times [0,T]
\end{equation}
with the cohesive stress tensor given by
\begin{equation}\label{eq:stress_cohesive_pressure_shear}
    \boldsymbol{\sigma} (\boldsymbol{\varepsilon}, \boldsymbol{\eta}, \alpha) =\frac{\partial \psi(\boldsymbol{\varepsilon}, \boldsymbol{\eta}, \alpha)}{\partial\boldsymbol{\varepsilon}}= p \boldsymbol{I} + \tau \frac{\boldsymbol{\varepsilon}_{\text{dev}}}{\left|\boldsymbol{\varepsilon}_{\text{dev}}\right|}
\end{equation}
where
\begin{equation}
    p = \frac{\partial \psi(\boldsymbol{\varepsilon}, \boldsymbol{\eta}, \alpha)}{\partial \text{tr}(\boldsymbol{\varepsilon})} = \kappa_0 \left( \text{tr}(\boldsymbol{\varepsilon}) - \text{tr}(\boldsymbol{\eta}) \right)
    \quad\text{,}\quad
    \tau = \frac{\partial \psi(\boldsymbol{\varepsilon}, \boldsymbol{\eta}, \alpha)}{\partial \left| \boldsymbol{\varepsilon}_{\text{dev}}\right|} = 2 \mu_0 \left( \left|\boldsymbol{\varepsilon}_{\text{dev}} \right| - \left| \boldsymbol{\eta}_{\text{dev}} \right| \right)
\end{equation}
are the hydrostatic and the shear stress, respectively.

The eigenstrain evolution criterion \cite{Vicentini_variational_2026} reads
\begin{equation}\label{eq:eigenstrain_evolution_criterion}
    p \zeta + \tau \xi \leq a(\alpha) \pi_0^\prime (\text{tr}(\boldsymbol{\eta}), \left|\boldsymbol{\eta}_{\text{dev}}\right|) (\zeta, \xi)
\end{equation}
for any admissible variations $\zeta$ and $\xi$.
Since $\nabla\alpha$ may be discontinuous across $J(\boldsymbol{z})$, for the phase field we obtain two sets of KKT conditions
\begin{subequations}
    \begin{equation}
        -Y (\boldsymbol{\eta}, \alpha) + \frac{G_{\text{c}}}{c_w}\!\left(\!\frac{w^\prime(\alpha)}{\ell}\!-\!2 \ell \Delta \alpha\!\right) \geq 0
        \text{,}\,
        \dot{\alpha} \geq 0
        \text{,}\,
        \left[-Y (\boldsymbol{\eta}, \alpha) + \frac{G_{\text{c}}}{c_w}\!\left(\!\frac{w^\prime(\alpha)}{\ell}\!-\!2 \ell \Delta \alpha\!\right) \right] \dot{\alpha} = 0
        \quad \forall (\boldsymbol{x},t) \in \left( \Omega \setminus J(\boldsymbol{z}) \right) \times [0,T]
        \,\text{,}
    \end{equation}
    \begin{equation}
        -Y (\boldsymbol{\eta}, \alpha) - 2 \ell \frac{G_{\text{c}}}{c_w} \jump{\nabla \alpha} \cdot \boldsymbol{n} \geq 0
        \quad\text{,}\quad
        \dot{\alpha} \geq 0
        \quad\text{,}\quad
        \left[-Y (\boldsymbol{\eta}, \alpha) - 2 \ell \frac{G_{\text{c}}}{c_w} \jump{\nabla \alpha} \cdot \boldsymbol{n} \right] \dot{\alpha} = 0
        \quad \forall (\boldsymbol{x},t) \in J(\boldsymbol{z})\times [0,T]
        \,\text{,}
    \end{equation}
\end{subequations}
in addition to \eqref{KKT_boundary} at the boundary.
The bulk conditions mirror those of the brittle model~\eqref{eq:alpha_strongform} with an energy release rate
\begin{equation}
    Y (\boldsymbol{\eta}, \alpha) = -\frac{\partial \psi(\boldsymbol{\varepsilon}, \boldsymbol{\eta}, \alpha)}{\partial \alpha}= -\frac{\partial \pi(\boldsymbol{\eta}, \alpha)}{\partial \alpha}=-a^\prime(\alpha) \pi_0 (\text{tr}(\boldsymbol{\eta}), \left|\boldsymbol{\eta}_{\text{dev}}\right|) \qquad\text{.}
\end{equation}

\subsection{Adopted model and compact reformulation}\label{sec:cohesive_compact_formulation}
We adopt $w(\alpha)=\alpha^2$ and $c_w=2$ as in \cite{Vicentini_variational_2026}, and the linear degradation function $a(\alpha) = 1-\alpha$ from \cite{bourdin_variational_2025}.
Note that no residual value is necessary for this model, i.e. we can have $a(\alpha=1) = 0$.
For $\pi_0$ we adopt the $2$-norm model in \cite{Vicentini_variational_2026} which reads
\begin{equation}\label{eq:strength_pot}
    \pi_0(\text{tr}(\boldsymbol{\eta}), \left|\boldsymbol{\eta}_{\text{dev}}\right|) = \sqrt{p_{\text{c}}^2 \text{tr}(\boldsymbol{\eta})^2 + \tau_{\text{c}}^2 \left| \boldsymbol{\eta}_{\text{dev}} \right|^2}\,,
\end{equation}
where $p_{\text{c}}$ and $\tau_{\text{c}}$ are the critical pressure and shear strength, respectively.
As clarified later in Section~\ref{sec:model_comparison}, this leads to the same initial elastic limit as for the brittle case with the volumetric-deviatoric split \eqref{eq:vol_dev_decomposition}.

Before analyzing the model behavior, we introduce a more compact reformulation for the special case $p_{\text{c}}^2 / \tau_{\text{c}}^2 = \kappa_0 / (2 \mu_0)$.
For the considered special case, the strength potential reads
\begin{equation}\label{eq:cohesive_strength_potential}
    \pi_0 (\text{tr}(\boldsymbol{\eta}), \left|\boldsymbol{\eta}_{\text{dev}}\right|) = \sqrt{w_{\text{c}}} \sqrt{\kappa_0 \text{tr}(\boldsymbol{\eta})^2 + 2 \mu_0 \left| \boldsymbol{\eta}_{\text{dev}} \right|^2} \quad \text{with} \quad w_{\text{c}} = \frac{\tau_{\text{c}}^2}{2\mu_0} = \frac{p_{\text{c}}^2}{\kappa_0} \quad\text{,}
\end{equation}
hence it is $\pi(\boldsymbol{\eta}, \alpha) = a(\alpha) \sqrt{w_{\text{c}}} \sqrt{ \mathbb{C}_0 \boldsymbol{\eta} \cdot \boldsymbol{\eta}}$, with $\mathbb{C}_0$ as the undamaged stiffness tensor and the parameter $w_{\text{c}}$ controlling the size of the initial elastic domain $\mathcal{S}_0$.
Substituting \eqref{eq:strength_pot} in \eqref{eq:psi_cohesive}, it is possible to optimize the action functional with respect to the eigenstrain, yielding a strain energy density depending only on the displacement and the phase field.
This introduces a piece-wise defined elastic strain energy density but eliminates the need to solve for $\boldsymbol{\eta}$ as done in \cite{Vicentini_variational_2026}. Instead, the eigenstrain is evaluated at the Gauss points during assembly, analogously as with a return mapping algorithm in plasticity.
Although additional nonlinear iterations are introduced whenever the eigenstrain is non-zero, the resulting problem for the displacement is unconstrained, thus improving the efficiency of the numerical computations.
This approach shares some similarities with the one in \cite{Hageman_2026}; however, the present case only requires the evaluation of an algebraic expression rather than the solution of a system of equations at each Gauss point.

Stationarity with respect to $\boldsymbol{\eta}$ under the constraint $\text{tr}(\boldsymbol{\eta})\geq 0$ yields
\begin{equation}\label{eq:psi_cohesive_condensed}
    \hat{\psi} (\boldsymbol{\varepsilon}, \alpha) =
    \begin{cases}
        a(\alpha) \sqrt{w_{\text{c}}} \sqrt{ 2\psi_0 (\boldsymbol{\varepsilon}) } - \frac{1}{2} a(\alpha)^2 w_{\text{c}} + \epsilon \psi_0 (\boldsymbol{\varepsilon}) &\text{if}\,\,\, \text{tr}(\boldsymbol{\varepsilon}) \geq 0, \psi_0 (\boldsymbol{\varepsilon}) \geq \frac{1}{2} a(\alpha)^2 w_{\text{c}}\\
        \frac{\kappa_0}{2} \left( \text{tr}(\boldsymbol{\varepsilon}) \right)^2 - \frac{1}{2} a(\alpha)^2 w_{\text{c}} + a(\alpha) \sqrt{2 \mu_0 w_{\text{c}}} \left| \boldsymbol{\varepsilon}_{\text{dev}} \right| + \epsilon \psi_0 (\boldsymbol{\varepsilon}) &\text{if}\,\,\, \text{tr}(\boldsymbol{\varepsilon}) < 0, \left| \boldsymbol{\varepsilon}_{\text{dev}} \right| \geq a(\alpha) \sqrt{\frac{w_{\text{c}}}{2 \mu_0}}\\
        (1+\epsilon) \psi_0(\boldsymbol{\varepsilon}) &\text{else}\\
    \end{cases} \quad \text{.}
\end{equation}
The term $\epsilon \psi_0$ with $\epsilon = 10^{-7}$ is introduced, analogously to the brittle case, as a residual energy density to avoid numerical issues which would result from the linearity of $\hat{\psi}$ in $\boldsymbol{\varepsilon}$ in the first two branches (in the second one linearity emerges only for purely deviatoric deformations).
We add this term in all branches for numerical stability.
Having $a(\alpha=1)>0$ as in the brittle models would not fix the rank-deficiency of the stiffness matrix with this compact reformulation.
The detailed derivation as well as the optima for the eigenstrain and the condensed form of the stress are given in Appendix~\ref{app:compact_reformulation}.

As in the quasi-static case, we require that the strain-hardening condition be fulfilled, which yields \cite{Vicentini_variational_2026}
\begin{equation}\label{eq:strain_hardening}
    \ell \leq \ell_{\text{ch}}
    \qquad\text{with}\qquad
    \ell_{\text{ch}} = \min_{\boldsymbol{\sigma}_{\text{c}} \in \partial \mathcal{S}_0} \frac{G_{\text{c}}}{\mathbb{S}_0 \boldsymbol{\sigma}_{\text{c}} \cdot \boldsymbol{\sigma}_{\text{c}}}
\end{equation}
where $\ell_{\text{ch}}$ is the cohesive (or Irwin) length of the material and $\mathbb{S}_0 = \mathbb{C}_0^{-1}$ is the undamaged compliance tensor.
In the present case, this condition simplifies to $\ell \leq \ell_{\text{ch}} = G_{\text{c}} / w_{\text{c}}$.

\subsection{1D cohesive model}\label{sec:cohesive_1D_behavior}
\begin{figure}[t!]
    \centering
    \includegraphics{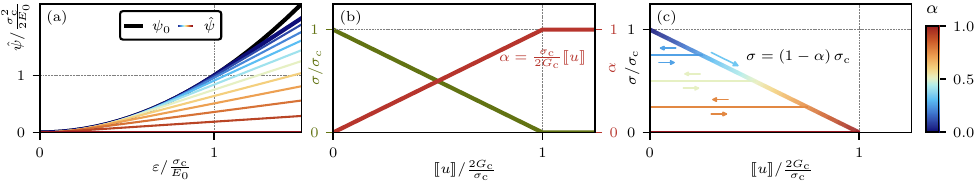}
    \caption{Visualization of the model behavior in the 1D setting, in terms of reduced elastic energy density (a), cohesive law (b), and loading-unloading behavior (c).}
    \label{fig:cohesive_model_1D}
\end{figure}

Following the same procedure as for the brittle models, we now analyze the cohesive model in the 1D setting.
The 1D elastic domain is $\mathcal{S}_0 = \{ \sigma \in \mathbb{R} : \sigma \leq \sigma_{\text{c}} \}$, where $\sigma_{\text{c}}>0$ denotes the tensile strength, so that the eigenstrain potential reads
\begin{equation}
    \pi (\eta, \alpha) =
    \begin{cases}
        a(\alpha) \sigma_{\text{c}} \eta &\text{if } \, \eta \geq 0\\
        +\infty &\text{else}
    \end{cases} \quad\text{.}
\end{equation}
In this case, \eqref{eq:CF_strain_jump} and \eqref{eq:CF_eta_jump} simplify to
\begin{equation}
    \varepsilon = u' + \jump{u} \, \delta_{J(\boldsymbol{z})}, \hspace{3mm} \eta = \eta_R + \jump{u} \, \delta_{J(\boldsymbol{z})},
    \label{eq:1Djump}
\end{equation}
where we see that at the location of a cohesive crack the eigenstrain coincides with the displacement jump (which we also refer to as crack opening in the following).
The eigenstrain evolution criterion~\eqref{eq:eigenstrain_evolution_criterion} takes the explicit form
\begin{equation}\label{eq:strength_criterion_1D}
    \sigma (\varepsilon, \eta) \leq a(\alpha) \sigma_{\text{c}} \qquad \eta\geq0 \qquad \left(\sigma (\varepsilon, \eta) - a(\alpha) \sigma_{\text{c}}\right) \eta = 0 \qquad \forall (x,t) \in \Omega \times [0, T]
\end{equation}
with $\sigma = E_0\,(\varepsilon-\eta)$.
Stationarity of $\psi$ with respect to $\eta \geq 0$ yields the optimal eigenstrain
\begin{equation}
    \eta^\star (\varepsilon, \alpha) = \arg\,\text{stat}_{\eta \geq 0} \psi (\varepsilon, \eta, \alpha) =
    \begin{cases}
        0 &\varepsilon < (1-\alpha) \frac{\sigma_{\text{c}}}{E_0}\\
        \varepsilon - (1-\alpha) \frac{\sigma_{\text{c}}}{E_0} &\text{else}\\
    \end{cases}\,.
\end{equation}
The resulting reduced form of the elastic strain energy density is represented in Fig.~\ref{fig:cohesive_model_1D}a and reads 
\begin{equation}
    \hat{\psi} (\varepsilon, \alpha) = \psi (\varepsilon, \eta^\star(\varepsilon, \alpha), \alpha) =
    \begin{cases}
        \psi_0 (\varepsilon) + \epsilon \psi_0 (\varepsilon) &\varepsilon < (1-\alpha) \frac{\sigma_{\text{c}}}{E_0}\\
        (1-\alpha) \sigma_{\text{c}} \varepsilon - \frac{1}{2} (1-\alpha)^2 \frac{\sigma_{\text{c}}^2}{E_0} + \epsilon \psi_0 (\varepsilon) &\text{else}\\
    \end{cases}\qquad\text{.}
\end{equation}
The KKT conditions for the phase field on the jump set read
\begin{equation}\label{eq:phase_field_KKT_1D}
    a^\prime(\alpha) \sigma_{\text{c}} \jump{u} - \frac{2 G_{\text{c}} \ell}{c_w} \jump{\alpha^\prime} \geq 0
    \qquad
    \dot{\alpha} \geq 0
    \qquad
    \left[ a^\prime(\alpha) \sigma_{\text{c}} \jump{u} - \frac{2 G_{\text{c}} \ell}{c_w} \jump{\alpha^\prime} \right] \dot{\alpha} = 0
    \qquad
    \forall (x,t) \in J(\boldsymbol{z})\times[0,T] \qquad\text{.}
\end{equation}
The complementarity condition in case of evolving damage ($\dot{\alpha} > 0$) and using the optimal \texttt{AT2} profile, for which $\jump{\alpha^\prime} = -2\alpha/\ell$, yields
\begin{equation}\label{eq:lindeg_alpha_jump}
    \jump{u} = \frac{2G_{\text{c}}}{\sigma_{\text{c}}} \alpha\quad\text{.}
\end{equation}
Combined with the complementarity condition of~\eqref{eq:strength_criterion_1D} for $\eta > 0$, this gives the cohesive law
\begin{equation}\label{eq:cohesive_stress_jump}
    \sigma = \left(1 - \frac{\sigma_{\text{c}}}{2G_{\text{c}}} \jump{u} \right) \sigma_{\text{c}} \qquad\text{,}
\end{equation}
hence the model exhibits linear softening (Fig.~\ref{fig:cohesive_model_1D}b) \cite{bourdin_variational_2025}.
For a displacement jump value $\jump{u}_{\text{ult}}={2G_{\text{c}}}/{\sigma_{\text{c}}}$, complete failure is reached.
In Fig.~\ref{fig:cohesive_model_1D}c we observe that, upon unloading from the softening branch, the displacement jump gradually closes at constant stress $\sigma = (1-\alpha) \sigma_{\text{c}}$. Once the jump is fully closed, the response follows the initial linear elastic branch until  complete unloading.
The initial phase of reloading follows again the initial elastic branch up to $\sigma = (1-\alpha) \sigma_{\text{c}}$, after which the jump reopens at constant stress until the maximum value reached previously. Then, the phase field continues to evolve with the response governed by the softening branch.

\subsection{Interaction of a stress wave with a pre-existing fully developed cohesive crack in a 1D bar}\label{sec:interaction_wave_cohesive_crack_fully}
We now investigate the interaction of an elastic wave with a pre-existing crack.
Note that a cohesive crack can transmit cohesive forces if it is not fully developed, i.e. if the displacement jump across its faces has never reached $\jump{u}_{\text{ult}}$ (and the maximum value of the phase field has never reached $1$).
In the following, we begin with the simpler case of a fully developed crack and will treat the case of a partially developed crack in the next subsection.

Thus, at the center of our bar we place a crack that has been previously opened beyond $\jump{u}_{\text{ult}}$, and whose faces have been then brought back in contact, i.e. $\alpha=1$ at $x=0$ and $\eta = 0$ everywhere.
From now on, we denote the value of the phase-field at the central crack as $\breve{\alpha}(t)$ (here, $\breve{\alpha}(0)=1$).
The initial crack is introduced by imposing the optimal \texttt{AT2} profile $\alpha(x,0) = \exp(-|x|/\ell)$; here, however, we directly allow for the phase field to evolve.

Apart from these conditions, the setup is identical to the brittle case (Section~\ref{sec:interaction_wave_brittle_crack}, Fig.~\ref{fig:setup_1D_brittle}) with a bar of length $L=1000$~mm and material parameters $E_0 = 30\,000$~MPa, $G_{\text{c}} = 0.1$~N/mm, and $\rho_0 = 2\,400$~kg/m$^3$.
As the optimal profile of the \texttt{AT2} model is wider, we now set $\ell = 20$~mm.
The strength is set to $\sigma_{\text{c}} = 5$~MPa, matching one of the values of $\hat{\sigma}_{\text{c}}(0)$ for the brittle phase-field model \eqref{eq:elastic_domain_brittle_1D}.
We again impose a half-sine pulse at the left boundary via~\eqref{eq:1D_pulse_loading}, either compressive with $\ell/\tilde{\lambda} = 0.05$, or tensile with $\ell/\tilde{\lambda} = 0.05$ or $0.15$.
The pulse amplitude is again chosen such that $\tilde{\sigma} / \sigma_{\text{c}} = 0.9$.
The FEM discretization uses $\ell / \Delta x \approx 20$ and $c_0\Delta t / \Delta x = 0.1$.

\begin{figure}[t!]
    \centering
    \includegraphics{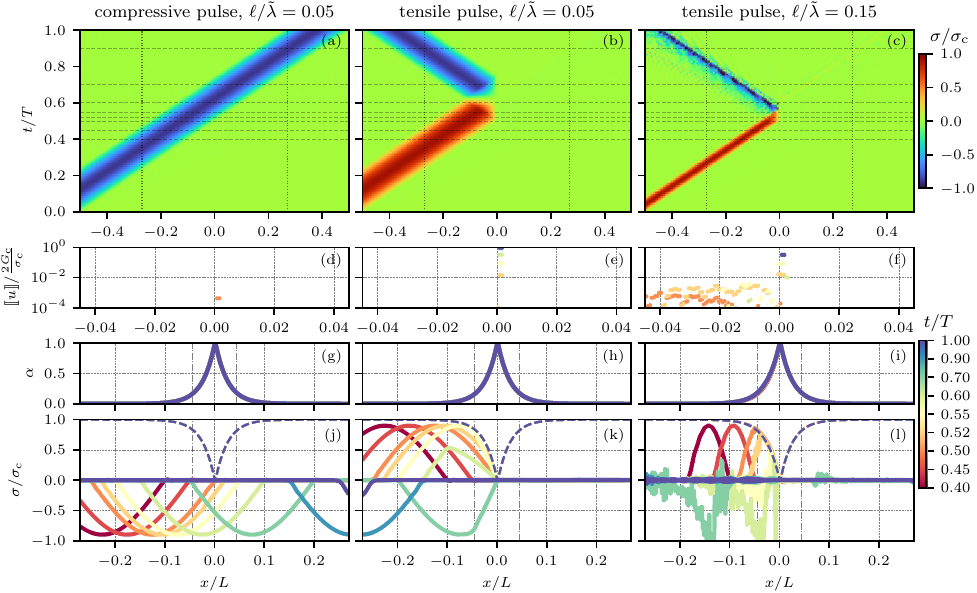}
    \caption{Interaction of a sinusoidal stress wave with  an \textit{evolving} fully developed cohesive crack in terms of $x$-$t$ diagram (a,b,c) and snapshots of the displacement jump (d,e,f), phase field (g,h,i) and stress field (j,k,l).
    To visualize also smaller non-zero values, the displacement jump is plotted in a logarithmic scale and only on a restricted region of the domain (highlighted by dashed-dotted lines in the lower two rows).
    The dashed horizontal lines in the first row represent the time instants for the snapshots of the plots in the rows below.}
    \label{fig:interaction_cohesive}
\end{figure}

The results are summarized in Fig.~\ref{fig:interaction_cohesive}, where the four rows show the stress field in $x$-$t$ space, the evolution of the displacement jump, the phase-field profile, and the stress along with the local strength at selected time instants.
Since the stiffness is not degraded, both the wave speed and the acoustic impedance within the support of the phase field remain constant and equal to that of the pristine material, while no energy reflection takes place.
The original wave equation is fully recovered, hence, the compressive wave (Fig.~\ref{fig:interaction_cohesive}a,d,g,j) is completely transmitted across the crack, with the condition $\eta \geq 0$ preventing interpenetration.
Since the stress is always negative, the eigenstrain evolution criterion~\eqref{eq:strength_criterion_1D} remains inactive and the phase field does not evolve.
This holds independently of the wave shape and frequency, as long as the wave is fully compressive.
The negligible evolution of the eigenstrain up to values still below $10^{-3}$ visible in Fig.~\ref{fig:interaction_cohesive}d is due to numerical artifacts.

For the tensile case with $\ell/\tilde{\lambda} = 0.05$ (Fig.~\ref{fig:interaction_cohesive}b,e,h,k), the wave behaves as expected in the sharp-crack case.
The results in Fig.~\ref{fig:interaction_cohesive}e can be explained using \eqref{eq:strength_criterion_1D}.
In this case the stress remains always lower than the local strength of the material, i.e. $\sigma(x,t)=E_0 u^\prime (x,t) <  \left(1-\alpha(x,t)\right) \sigma_{\text{c}}$ in all points and at all times, therefore neither the eigenstrain nor the damage evolve and the behavior remains linear elastic.
Also, since the fully developed cohesive crack can no longer transmit tensile cohesive stresses, it acts as a free end, hence the incoming wave is completely reflected without distortion but with reversed sign (i.e., as a compressive wave).
The jump opens up exclusively at the two Gauss points of the element with $\alpha = 1$, while neighboring points with $\alpha < 1$ remain unaffected.

For the tensile wave with the shorter wavelength $\ell/\tilde{\lambda} = 0.15$ (Fig.~\ref{fig:interaction_cohesive}c,f,i,l) the behavior is different. 
Here, the stress gradient of the incoming wave is steeper than the spatial variation of the material strength $a(\alpha(x,t))\sigma_{\text{c}}$ (reported as a dashed line in Fig.~\ref{fig:interaction_cohesive}l), leading to the evolution of the eigenstrain and, hence, of the  damage at several locations in the neighborhood of the crack center.
Although much less severely than in Fig.~\ref{fig:widening_standard}, this triggers the widening of the damage band and an increase of the dissipated energy.
Each of the jumps reflects and transmits a portion of the wave, and the subsequent closure of the jumps introduces stress discontinuities in time or \textit{shock waves} (better illustrated in Section~\ref{sec:interaction_wave_cohesive_crack_developing}), causing the observed oscillations.

The above results suggest that the interaction of stress waves with a fully developed cohesive crack depends on the $\ell/\tilde{\lambda}$ ratio, with smaller ratios leading to the expected sharp-crack response and larger  ratios yielding diffuse jumps accompanied by shock waves and high-frequency oscillations.
To achieve a better quantitative understanding of this phenomenon, in the following we formulate a general condition for the emergence of diffuse jumps.

\begin{figure}[t!]
    \centering
    \includegraphics{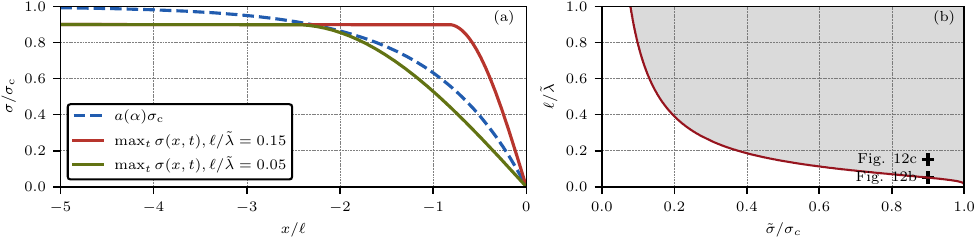}
    \caption{Comparison between maximum stress of an elastic wave and locally degraded strength (a), and numerically obtained limits for the emergence of diffuse jumps in dependence of $\tilde{\sigma}/\sigma_{\text{c}}$ and $\ell/\lambda$ for a harmonic wave (b).
    Note that the red envelope in (a) would not be observable as it violates~\eqref{eq:stress_strengthprofile}.}
    \label{fig:strength_intersection}
\end{figure}

Preventing the occurrence of diffuse jumps requires that the local strength of the material is reached only at the center of the crack, i.e. where $\alpha=1$ and the strength vanishes.
At any other position, the stress must be strictly below the strength profile at all times, namely
\begin{equation}\label{eq:stress_strengthprofile}
    \sigma (x, t) < a(\alpha (x,t)) \sigma_{\text{c}} \qquad \forall x \in \Omega \setminus \{0\}\quad\text{,}\quad\forall t \in [0,T]\text{.}
\end{equation}
To check whether this condition is fulfilled, we trace the maximum stress reached during the reflection process at each material point and compare it to the strength profile.
For an incoming wave \eqref{eq:1D_pulse_loading}, the temporal maximum at each spatial point reads
\begin{equation}\label{eq:max_stress}
    \max_{t} \sigma (x, t) =
    \begin{cases}
        \tilde{\sigma} \sin (4 \pi \tfrac{|x|}{\tilde{\lambda}}) &\text{for} \, |x| \leq \frac{\tilde{\lambda}}{8}\\
        \tilde{\sigma} &\text{else}\\
    \end{cases}\quad\text{with} \quad \tilde\sigma\le\sigma_c\quad\text{.}
\end{equation}
The strength profile is given by the optimal \texttt{AT2} profile
\begin{equation}\label{eq:opt_profile_AT2}
    a(\alpha(x,t))\sigma_{\text{c}} = \left(1 - \alpha(x,t)\right) \sigma_{\text{c}} \qquad\text{with}\qquad \alpha (x,t) = \breve{\alpha}(t) \exp \left( - \frac{|x|}{\ell} \right) \qquad\text{.}
\end{equation}
Both \eqref{eq:max_stress} and \eqref{eq:opt_profile_AT2} are illustrated in Fig.~\ref{fig:strength_intersection}a for the case studied here, i.e. with $\breve{\alpha}(0)=1$ and $\ell/\tilde{\lambda} = 0.05$ or $0.15$.
Using \eqref{eq:max_stress} and \eqref{eq:opt_profile_AT2} in \eqref{eq:stress_strengthprofile} gives the condition
\begin{equation}\label{eq:nodiffusecrack_cond}
    \frac{\tilde{\sigma}}{\sigma_{\text{c}}} \sin \left(4 \pi \frac{\ell}{\tilde{\lambda}} \frac{|x|}{\ell}\right) < \left(1 - \breve{\alpha}(0) \exp \left(-\frac{|x|}{\ell}\right)\right)  \quad \tilde\sigma\le\sigma_c\qquad \forall x \in [-L/2, L/2] \setminus \{0\}\quad\text{,}
\end{equation}
which depends on the $\tilde{\sigma}/\sigma_{\text{c}}$ and ${\ell}/\tilde{\lambda}$ ratios.
Evaluating \eqref{eq:nodiffusecrack_cond} with an equal sign gives the limit which separates the cases where no diffuse jumps are expected from those where the boundary of the elastic domain is reached, triggering the evolution of eigenstrain and damage.
The resulting domains for the case at hand are illustrated in Fig.~\ref{fig:strength_intersection}b. The red line represents the limit case; for $\tilde{\sigma}/\sigma_{\text{c}}$ and ${\ell}/\tilde{\lambda}$ ratios falling within the lower white area, we expect the correct wave-crack interaction, i.e. a sharp-crack behavior, whereas $\tilde{\sigma}/\sigma_{\text{c}}$ and ${\ell}/\tilde{\lambda}$ ratios within the upper gray-shaded area are expected to lead to diffuse jumps accompanied by shock waves and high-frequency oscillations.
The tensile cases illustrated in Fig.~\ref{fig:interaction_cohesive} are marked and, as expected, the waves with  $\ell/\tilde{\lambda} = 0.05$ and $0.15$ lie respectively below and above the red limit line, further confirming the obtained results.
Although the criterion \eqref{eq:nodiffusecrack_cond} is derived for harmonic waves, we can draw some general observations.
For sufficiently small stress wave amplitudes compared to the initial material strength ($\tilde{\sigma}/\sigma_{\text{c}} \rightarrow 0$), the whole range of $\ell/\tilde{\lambda}$ ratios recovers the expected sharp-crack behavior.
The same occurs for the whole range of amplitudes $\tilde{\sigma}/\sigma_{\text{c}}$ if the regularization length is sufficiently small compared to the wavelength ($\ell/\tilde{\lambda} \rightarrow 0$).
In this context, an advantage of the cohesive model lies in the interpretation of the regularization length $\ell$ as a pure numerical parameter, which can be adapted to the expected wavelength.
We remark that for the special case of the applied wave, these results do not only hold for the half-wave of width $\tilde{\lambda}$, but also for the full wave of wavelength $\lambda$.
Note that additionally the strain-hardening condition \eqref{eq:strain_hardening} must be satisfied, independently of $\ell/\lambda$.
However, in case of high-frequency waves ($\lambda\to0$) or shock waves, avoiding diffuse jumps calls for $\ell \to 0$.

Note that, although the dependence on $\ell/\lambda$ resembles that of the brittle model with stiffness degradation of Section~\ref{sec:interaction_wave_brittle_crack} (both recovering the sharp crack for small ratios), the cohesive model admits two additional degrees of freedom to reach it: the sharp-crack behavior is restored for any $\ell/\lambda$ at small enough $\tilde{\sigma}/\sigma_{\text{c}}$, and $\ell$ -- being decoupled from the strength unlike in the brittle case -- can be freely reduced to accommodate high-frequency waves.

\subsection{Interaction of a stress wave with a pre-existing partially developed cohesive crack in a 1D bar}\label{sec:interaction_wave_cohesive_crack_developing}
We now place at the center of our bar a cohesive crack that has experienced a maximum damage $\breve{\alpha} = 0.5$, hence a maximum displacement jump $2 G_{\text{c}} \breve{\alpha} / \sigma_{\text{c}}$ (see \eqref{eq:lindeg_alpha_jump}).
This crack can still transmit a cohesive stress $a(\breve{\alpha})\sigma_{\text{c}}$ across its faces, and we further assume that its faces have been then brought back in contact, i.e. $\eta = 0$ everywhere. 
We adopt the same setup as before, but we set $\ell = 10$~mm.
Moreover, we choose $\ell/\Delta x = 20$, $c_0 \Delta t / \Delta x = 0.1$, $\ell / \tilde\lambda = 0.015$ and $\tilde{\sigma} / \sigma_{\text{c}} = 0.9$.
To test different cohesive responses, we vary the fracture toughness as $G_{\text{c}} \in \{0.01, 0.025, 0.1\}$~N/mm, yielding different Irwin lengths $\ell_{\text{ch}} = E_0 G_{\text{c}} / \sigma_{\text{c}}^2$ following \eqref{eq:strain_hardening}.

\begin{figure}[t!]
    \centering
    \includegraphics{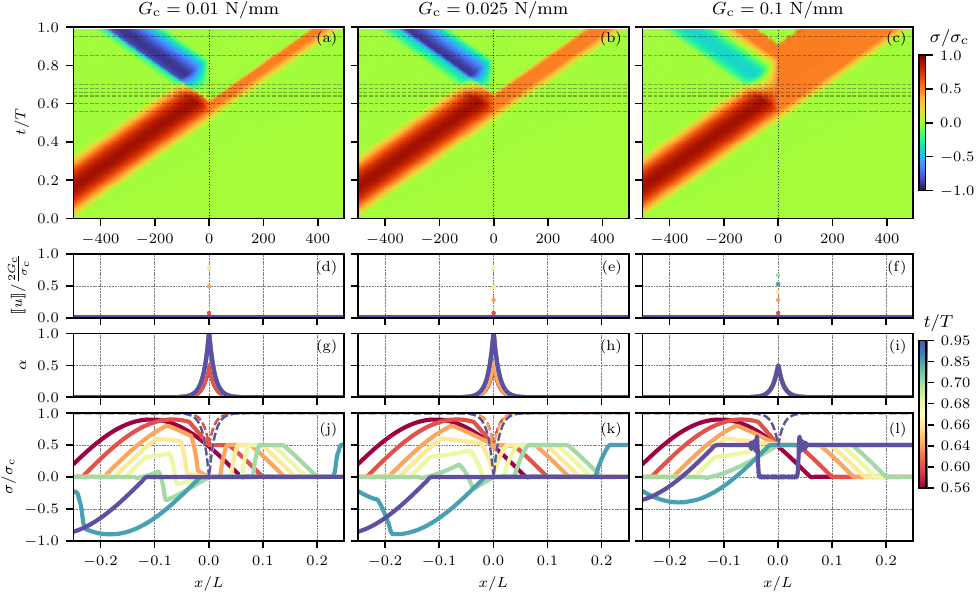}
    \caption{Interaction of a sinusoidal stress wave with an \textit{evolving} partially developed cohesive crack with $\breve{\alpha} = 0.5$ in terms of $x$-$t$ diagram (a,b,c) and snapshots of the displacement jump (d,e,f), phase field (g,h,i) and stress field (j,k,l).
    The dashed horizontal lines in the first row represent the time instants for the snapshots of the plots in the rows below.}
    \label{fig:interaction_cohesive_evolution}
\end{figure}

\begin{figure}[t!]
    \centering
    \includegraphics{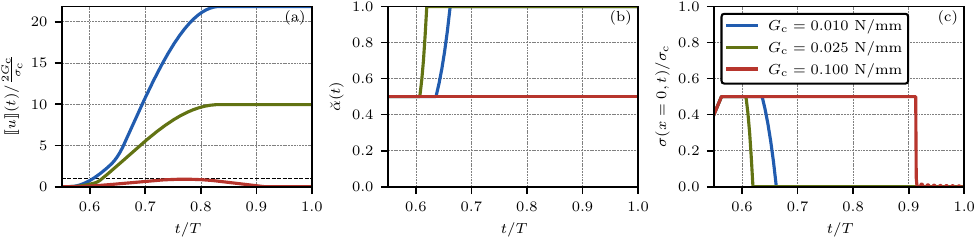}
    \caption{Time evolution of the jump (a), the phase field (b) and the stress (c) at the crack for the test in Fig.~\ref{fig:interaction_cohesive_evolution} for various values of the fracture toughness $G_{\text{c}}$.}
    \label{fig:interaction_cohesive_evolution_crack}
\end{figure}

The numerical results are shown in Fig.~\ref{fig:interaction_cohesive_evolution}, while the time evolution of the displacement jump $\jump{u}(t)$, the phase-field value at the crack $\breve{\alpha}(t)$, and the cohesive stress $\sigma(0,t)$ at the center of the crack are illustrated in Fig.~\ref{fig:interaction_cohesive_evolution_crack}.
Once the wave reaches the crack, the response shows four different phases: 
\begin{enumerate}[I]
    \item As long as the strength criterion at the crack is not met, i.e. $\sigma(0,t) < a(\breve{\alpha}(0))\sigma_{\text{c}}$, the wave is fully transmitted and $\eta = 0$, $\dot{\eta} = 0$, $\dot{\alpha} = 0$.
    \item When at time $t_{I}$ the cohesive strength is reached, i.e. $\sigma(0,t_{I}) = a(\breve{\alpha}(0))\sigma_{\text{c}}$, the crack starts reopening, $\dot{\eta} > 0$, with a transmitted stress equal to $a(\breve{\alpha}(0))\sigma_{\text{c}}$ while the damage remains constant and equal to $\breve{\alpha}(0)$. This phase continues until the displacement jump reaches the maximum value  attained in the past, namely $2 G_{\text{c}} \breve{\alpha}(0) / \sigma_{\text{c}}$, at time $t_{II}$. This corresponds to the horizontal branch of the reloading path in Fig.~\ref{fig:cohesive_model_1D}c. 
    \item Once this value of the opening is reached, the damage evolution criterion is fulfilled and the phase field starts to evolve along with the eigenstrain, i.e. $\dot{\eta} > 0$ and $\dot{\alpha} > 0$ (softening branch  in Fig.~\ref{fig:cohesive_model_1D}c).
    Provided that the incoming wave supplies sufficient energy, this phase continues until at the crack center $\alpha = 1$ and accordingly $\jump{u}_{\text{ult}}=2 G_{\text{c}} / \sigma_{\text{c}}$ is reached at time $t_{III}$.
    \item From this point on, no cohesive stresses can be transmitted. Therefore, the crack behaves as a free end and any further portion of wave is reflected while the displacement jump keeps increasing.
\end{enumerate}

As Fig.~\ref{fig:interaction_cohesive_evolution}i shows, for the case with $G_{\text{c}} = 0.1$~N/mm the energy carried by the wave is not enough to  reach $\jump{u}=2 G_{\text{c}} \breve{\alpha}(0) / \sigma_{\text{c}}$ needed to activate the damage evolution criterion.
Therefore, after the initial linear elastic reloading phase, the transmitted stress remains constant and the phase field does not evolve.
The displacement jump first increases and, when the energy carried by the incoming wave starts decreasing, it begins decreasing again following the horizontal unloading branch in Fig.~\ref{fig:cohesive_model_1D}c.
Once the stress of the incoming wave drops below the cohesive strength, the jump closes completely, triggering a shock wave propagating through the domain (Fig.~\ref{fig:interaction_cohesive_evolution}l).
This behavior can be explained considering that the constant stress during the jump closure forces also the velocities of the crack faces to be constant due to the balance of linear momentum.
When the jump finally closes, the faces velocities drop abruptly from a constant non-zero value to zero, producing a discontinuity in velocity, hence, a shock wave.
As discussed in Section~\ref{sec:interaction_wave_cohesive_crack_fully}, this creates high-frequency oscillations as visible in Fig.~\ref{fig:interaction_cohesive}l and also, albeit to a more limited extent, in Fig.~\ref{fig:interaction_cohesive_evolution}l.
This explains the scatter visible in Fig.~\ref{fig:interaction_cohesive}c: since the strength criterion is reached at multiple points, multiple cracks open and close again subsequently, which triggers multiple smaller shock waves.
With the emergence of shock waves, the assumption of 'sufficiently regular' temporal derivatives made in the model derivation (Section~\ref{sec:cohesive_model_formulation}) is no longer satisfied.
The behavior, directly stemming from the reversibility of the eigenstrain, might be reduced or avoided  by modifying the reversibility condition so as to involve a different unloading/reloading behavior as also noted in \cite[Section 3.7]{Vicentini_variational_2026} and \cite[Section 5.2]{bourdin_2008}.
This is, however, outside of the scope of the present work.

From Fig.~\ref{fig:interaction_cohesive_evolution_crack} we note that a smaller value of  $G_{\text{c}}$ leads to a faster rate of change of the displacement jump.
To make this dependence quantitative, we analytically derive the  evolution in time of the cohesive crack opening from the kinematic conditions, stress continuity, and the complementarity conditions for the eigenstrain and the phase field. The full derivation is given in Appendix~\ref{app:dynamic_cohesive_law}, while here we report only the essential results.
Under the condition that the strength criterion is met at a single point only (i.e., for sufficiently small $\ell/\lambda$), the opening evolution is governed by the ordinary differential equation (ODE)
\begin{equation}\label{eq:opening_evolution_ODE_mainbody}
    \dot{v}_L^{\text{ref}}(0^-,t) - \frac{c_0}{\ell_{\text{ch}}} v_L^{\text{ref}}(0^-,t) = \dot{v}_L^{\text{inc}}(0^-,t) \qquad \forall t_{\text{II}} \leq t \leq t_{\text{III}} \qquad\text{,}
\end{equation}
where $v_L^{\text{ref}}$, $v_L^{\text{inc}}$ are respectively the reflected and the incoming velocities of the wave traveling toward the left at the crack position.
The ODE admits the closed-form solution
\begin{equation}
    v_L^{\text{ref}}(0^-,t) = \exp \left( \tfrac{c_0}{\ell_{\text{ch}}} (t-t_{\text{II}}) \right) v_L^{\text{ref}}(0^-,t_{\text{II}}) + \int_{t_{\text{II}}}^{t} \exp \left( \tfrac{c_0}{\ell_{\text{ch}}} (t-\tau) \right) \dot{v}_L^{\text{inc}}(0^-,\tau) \mathrm{d}\tau \qquad \forall t_{\text{II}} \leq t \leq t_{\text{III}} \quad\text{,}
\end{equation}
from which the crack opening follows as
\begin{equation}\label{eq:dynamic_opening_evolution}
    \jump{u} (t) = - 2 \tfrac{\ell_{\text{ch}}}{c_0}  \Bigl[ v_L^{\text{ref}}(0^-,t) - v_L^{\text{ref}}(0^-,t_{\text{II}}) - \left(v_L^{\text{inc}}(0^-,t) - v_L^{\text{inc}}(0^-,t_{\text{II}})\right) \Bigr] - 2  u_L^{\text{ref}}(0^-,t_{\text{II}}) \quad\text{.}
\end{equation}
From \eqref{eq:dynamic_opening_evolution} we deduce that the evolution is independent of $\ell$ and governed only by the  $c_0/\ell_{\text{ch}}$ ratio and by the incoming wave, which determines the initial condition at $t_{\text{II}}$.
An interactive visualization of the theoretical solution is available at \url{https://github.com/jonas-heinzmann/phase_field_dynamics} in the folder \texttt{/math/crack\-opening.py}.
In Appendix~\ref{app:dynamic_cohesive_law_pulse}, we also evaluate these expressions for the specific case of the half-sine pulse used in the numerical computations.
There, we also show that the theoretical results and the FEM results from Fig.~\ref{fig:interaction_cohesive_evolution_crack}a coincide.

\begin{figure}[t!]
    \centering
    \includegraphics{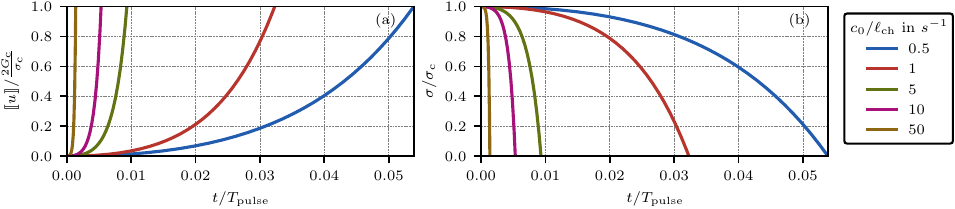}
    \caption{Visualization of the dynamic cohesive law for different $c_0/\ell_{\text{ch}}$ ratios with a harmonic incoming wave, in terms of the jump (a), as well as the stress (b).}
    \label{fig:cohesive_opening}
\end{figure}

Analyzing \eqref{eq:opening_evolution_ODE_mainbody} the following two limit cases can be obtained:
\begin{itemize}
    \item For $c_0/\ell_{\text{ch}} \rightarrow 0$ (small wave speed or very large Irwin length), the second term on the left-hand side vanishes and $\dot{v}_L^{\text{ref}}(0^-,t) = \dot{v}_L^{\text{inc}}(0^-,t)$, the reflected and incoming wave velocities can differ only by a constant offset, while $\breve{\alpha}(t)$ does not evolve and a constant cohesive stress $a(\breve{\alpha}(t))\sigma_{\text{c}}$ is transmitted.
    \item For $c_0/\ell_{\text{ch}} \rightarrow \infty$, the ODE becomes singularly perturbed and the phase field jumps instantaneously to $\alpha = 1$ since the time range $t_{III}-t_{II}\to0$.
        This corresponds to the brittle limit $G_{\text{c}}/\sigma_{\text{c}} \rightarrow 0$ (or  $\ell_{\text{ch}} \rightarrow 0$), for which no cohesive softening is present.
\end{itemize}
Fig.~\ref{fig:cohesive_opening}, which illustrates the time evolution of $\jump{u}(t)$ and $\sigma/\sigma_c$ for different  $c_0/\ell_{\text{ch}}$ ratios for the harmonic wave used here, confirms these observations.
The derivation of the curves in Fig.~\ref{fig:cohesive_opening} is reported in Appendix~\ref{app:dynamic_cohesive_law_pulse}.

\subsection{Summary of the 1D analyses: Stiffness and stiffness+density degradation vs. strength degradation}
\begin{figure}[t!]
    \centering
    \includegraphics{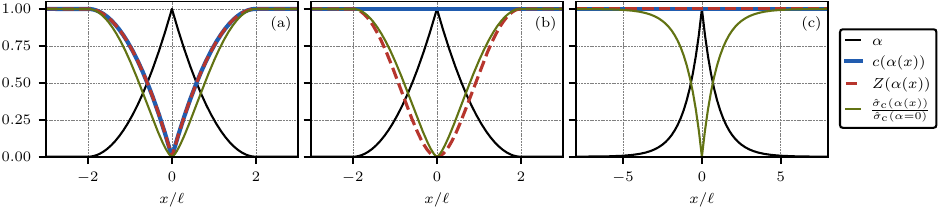}
    \caption{Phase-field profile and its effect on local wave speed, acoustic impedance, and material strength in the brittle model with stiffness degradation (a), brittle model with stiffness+density degradation (b), and cohesive model with strength degradation (c).}
    \label{fig:model_comparison}
\end{figure}

We close the 1D analyses by comparing in Fig.~\ref{fig:model_comparison} how the phase-field regularization affects the local material properties in the three formulations and briefly summarizing the results.
The figure uses \texttt{AT1} for the brittle models and \texttt{AT2} for the cohesive model, consistently with the dissipation functions adopted in Sections~\ref{sec:phase_field_model_brittle} and \ref{sec:phase_field_model_cohesive}; the qualitative comparison is independent of this choice.
In the brittle model with stiffness degradation (Fig.~\ref{fig:model_comparison}a), the wave speed and the acoustic impedance are reduced within the phase-field support, and elastic waves are partially reflected and partially transmitted at the regularized crack.
Combined with the locally reduced critical stress $\hat{\sigma}_{\text{c}}(\alpha)$, this drives the widening of the phase-field profile (Fig.~\ref{fig:widening_standard}). Compressive waves are transmitted correctly.
Degrading the density alongside the stiffness restores a constant wave speed and leads to an even more strongly reduced acoustic impedance for tensile loading, which yields full reflection (Fig.~\ref{fig:model_comparison}b). However, mass balance is not fulfilled; moreover, in compression the stiffness remains undegraded while the density is degraded, leading to partial reflection and transmission.
Both brittle variants recover the sharp-crack response -- including undistorted waveforms and correct reflection or transmission -- only in the  limit $\ell/\lambda \rightarrow 0$ (Section~\ref{sec:interaction_wave_brittle_crack}).

In the cohesive model (Fig.~\ref{fig:model_comparison}c), neither the stiffness nor the density is degraded, so both the bulk wave speed and the acoustic impedance are preserved while fulfilling mass balance.
The phase-field regularization reduces the local strength in the vicinity of the crack, while the sharp crack itself is represented by the eigenstrain at the jump set.
No strain energy decomposition is needed since tension-compression asymmetry follows from the eigenstrain constraint $\text{tr}(\boldsymbol{\eta}) \geq 0$.
The only restriction is that elastic waves must reach the strength criterion exclusively on the jump set and not in its vicinity, leading to limits in terms of $\ell/\lambda$ and  $\tilde \sigma/\sigma_c$ ratios for which the sharp-crack behavior is fully recovered (Section~\ref{sec:interaction_wave_cohesive_crack_fully}).

\FloatBarrier
\section{Comparison of the models in the multi-dimensional setting}\label{sec:model_comparison}
We now compare the three formulations on a benchmark problem in 2D plane-strain conditions ($d=3$).
Preliminarily, we compare the formulations of the elastic domain for the three models and introduce a closeness measure to the strength surface.

\subsection{Elastic domains and a closeness measure to the strength surface for the brittle and cohesive models}\label{sec:strength_surfaces}
The equations of motion in 3D for the brittle models with stiffness and stiffness+density degradation and for the cohesive model are compared in Table~\ref{tab:eq_motion_comparison}, while the expressions for the stress and energy release rate are presented in Table~\ref{tab:sigma_Y_comparison}.
We remark that, in contrast to pure linear elastodynamics, the dilational and deviatoric waves are no longer decoupled as soon as the phase field is non-zero for the brittle models in the multidimensional setting.
This comes from the degradation of the elastic properties, and is expected for heterogeneous materials~\cite{mueller_2007}.

\begin{table}[b!]
    \caption{Comparison of the equations of motion obtained by inserting the constitutive law and small-strain kinematics into the balance of linear momentum.}
    \label{tab:eq_motion_comparison}
    \footnotesize
    \begin{tabularx}{\textwidth}{cc}
        \toprule
        model & equations of motion\\
        \midrule
        brittle
        &$\begin{aligned}
            &g^\prime(\alpha) \nabla \alpha \left[ \kappa_0 \langle\nabla\cdot\boldsymbol{u}\rangle_+ \boldsymbol{I} + \mu_0 \left( (\nabla\boldsymbol{u} + \nabla\boldsymbol{u}^\intercal) - \frac{2}{3} \nabla\cdot\boldsymbol{u} \boldsymbol{I} \right) \right] + \left[ g(\alpha) \kappa_0 H(\nabla \cdot \boldsymbol{u}) + \kappa_0 H(-\nabla \cdot \boldsymbol{u}) + \frac{4}{3} g(\alpha) \mu_0 \right] \nabla (\nabla \cdot \boldsymbol{u})\\
            & - g(\alpha) \mu_0 \nabla \times (\nabla \times \boldsymbol{u}) = \begin{cases}
                \rho_0 \ddot{\boldsymbol{u}} &\text{stiffness degr.}\\
                h^\prime(\alpha) \dot{\alpha} \rho_0 \dot{\boldsymbol{u}} + h(\alpha) \rho_0 \ddot{\boldsymbol{u}} &\text{stiff.+dens. degr.}\\
            \end{cases} \qquad \forall (\boldsymbol{x},t) \in \Omega \times [0,T]
        \end{aligned}$\\[0.5cm]
        \midrule
        cohesive
        &$(\lambda_0 + 2\mu_0) \nabla (\nabla \cdot \boldsymbol{u}) - \mu_0 \nabla \times (\nabla \times \boldsymbol{u}) = \rho_0 \ddot{\boldsymbol{u}} \qquad \forall (\boldsymbol{x},t) \in (\Omega \setminus J) \times [0,T]$\\
        \bottomrule
    \end{tabularx}
\end{table}

\begin{table}[b!]
    \caption{Comparison of the stress and the damage energy release rate for the various models.}
    \label{tab:sigma_Y_comparison}
    \footnotesize
    \begin{tabularx}{\textwidth}{ccc}
        \toprule
        model &stress $\boldsymbol{\sigma} = \partial \psi / \partial \boldsymbol{\varepsilon}$ &damage energy release rate $Y = -\partial \psi / \partial \alpha$\\
        \midrule
        brittle
        &$g(\alpha) \left[ \kappa_0 \langle \text{tr} (\boldsymbol{\varepsilon}) \rangle_+ \boldsymbol{I} + 2 \mu_0 \boldsymbol{\varepsilon}_{\text{dev}} \right] + \kappa_0 \langle \text{tr} (\boldsymbol{\varepsilon}) \rangle_- \boldsymbol{I}$
        &$\begin{cases}
            - g^\prime(\alpha) \left( \frac{\kappa_0}{2} \langle \text{tr} (\boldsymbol{\varepsilon}) \rangle_+^2 + \mu_0 \left|\boldsymbol{\varepsilon}_{\text{dev}}\right|^2 \right) &\text{stiffness degr.}\\
            - g^\prime(\alpha) \left( \frac{\kappa_0}{2} \langle \text{tr} (\boldsymbol{\varepsilon}) \rangle_+^2 + \mu_0 \left|\boldsymbol{\varepsilon}_{\text{dev}}\right|^2 \right) + \frac{\rho_0}{2} h^\prime(\alpha) \left|\dot{\boldsymbol{u}}\right|^2 &\text{stiff.+dens. degr.}\\
        \end{cases}$\\[0.5cm]
        \midrule
        cohesive
        &$\kappa_0 \left( \text{tr}(\boldsymbol{\varepsilon}) - \text{tr}(\boldsymbol{\eta}) \right) \boldsymbol{I} + 2 \mu_0 \left( \left| \boldsymbol{\varepsilon}_{\text{dev}} \right| - \left| \boldsymbol{\eta}_{\text{dev}} \right| \right) \frac{\boldsymbol{\varepsilon}_{\text{dev}}}{\left|\boldsymbol{\varepsilon}_{\text{dev}}\right|}$
        &$-a^\prime(\alpha) \sqrt{p_{\text{c}}^2 \text{tr}(\boldsymbol{\eta})^2 + \tau_{\text{c}}^2 \left| \boldsymbol{\eta}_{\text{dev}} \right|^2}$\\
        \bottomrule
    \end{tabularx}
\end{table}

We analyze now the elastic domains obtained for the brittle and cohesive local models (i.e., for $\nabla\alpha \equiv \boldsymbol{0}$). 
For the brittle models the elastic domain is a consequence of the adopted strain energy decomposition~\cite{Vicentini_variational_2026} and of the selected fracture toughness, regularization length and elastic parameters \cite{Tanne_2018,Vicentini_energy_2024}.
It is obtained from the KKT conditions~\eqref{eq:alpha_strongform}; for the \texttt{AT1} dissipation function, quadratic degradation function, and volumetric-deviatoric split \eqref{eq:vol_dev_decomposition} it reads~\cite{Vicentini_energy_2024}
\begin{equation}\label{eq:elastic_domain_brittle}
    \frac{\langle \text{tr}(\boldsymbol{\sigma}) \rangle_+^2}{d^2 \kappa_0} + \frac{\left|\boldsymbol{\sigma}_{\text{dev}}\right|^2}{2 \mu_0}
    \leq \begin{cases}
        \frac{3 (1-\alpha)^3 G_{\text{c}}}{8\ell} &\text{stiffness degradation}\\
        \frac{3 (1-\alpha)^3 G_{\text{c}}}{8\ell} - h^\prime (\alpha) (1-\alpha)^3 \frac{\rho_0}{2}\left|\dot{\boldsymbol{u}}\right|^2 &\text{stiffness+density degradation}\\
    \end{cases}
\end{equation}
where the residual stiffness is neglected and $d$ is the spatial dimension.
Although the elastic domain is affected by the damage gradient, looking at the elastic domains for the local models is useful to compare their behavior.
Moreover, the initial elastic domain obtained for zero damage, i.e. for an intact domain, is exact and its boundary provides the elastic limit for the pristine material. 
Evaluating~\eqref{eq:elastic_domain_brittle} along the pure hydrostatic pressure ($\tau=0$) and pure shear ($p=0$) directions yields the damage-dependent critical pressure $\hat{p}_{\text{c}}(\alpha)$ and shear $\hat{\tau}_{\text{c}}(\alpha)$ given in Table~\ref{tab:strength_comparison}.
As observed in Section~\ref{sec:density_degradation} for the 1D case, also in 3D the model with stiffness+density degradation leads to a strength surface that depends on the local velocity $\dot{\boldsymbol{u}}$.
For both models the initial elastic domain is an ellipse for positive pressure and an unbounded cylinder in the negative pressure direction and it shrinks homothetically with damage, so the elastic domain $\mathcal{S}(\alpha)$ can be written as
\begin{equation}\label{eq:elastic_domain}
    \begin{cases}
        \tau \leq \hat{\tau}_{\text{c}} (\alpha) &p < 0\\
        \left( \frac{p}{\hat{p}_{\text{c}} (\alpha)} \right)^2 + \left( \frac{\tau}{\hat{\tau}_{\text{c}} (\alpha)} \right)^2 \leq 1 &p\geq0\\
    \end{cases} \quad\text{.}
\end{equation}

For the cohesive model, the elastic domain is governed by the strength potential $\pi_0(\boldsymbol{\eta})$ and by the degradation function $a(\alpha)$, which homothetically scales $\pi_0(\boldsymbol{\eta})$.
In particular, the strength potential is independent of $\ell$ and $\nabla\alpha$.
Also, it is parametrized through $p_{\text{c}}$ and $\tau_{\text{c}}$, which are independent input parameters.  
Although its choice is flexible \cite{Vicentini_variational_2026}, with the adopted strength potential \eqref{eq:cohesive_strength_potential} the elastic domain can still be written as \eqref{eq:elastic_domain}, with the damaged critical pressure and shear given in Table~\ref{tab:strength_comparison} for $a(\alpha)=1-\alpha$.

\begin{table}[b!]
    \caption{Comparison of the damaged critical pressure and shear stress for the different models.}
    \label{tab:strength_comparison}
    \footnotesize
    \begin{tabularx}{\textwidth}{ccc}
        \toprule
        model &damaged critical pressure $\hat{p}_{\text{c}} (\alpha)$ &damaged critical shear $\hat{\tau}_{\text{c}} (\alpha)$\\
        \midrule
        brittle
        &$\begin{cases}
            \sqrt{(1-\alpha)^3 \frac{3 \kappa_0 G_{\text{c}}}{8 \ell}} &\text{stiffness degr.}\\
            \sqrt{(1-\alpha)^3 \kappa_0 \left( \frac{3 G_{\text{c}}}{8 \ell} - h^\prime (\alpha) \frac{\rho_0}{2} \left|\dot{\boldsymbol{u}}\right|^2 \right)} &\text{stiff.+dens. degr.}\\
        \end{cases}$
        &$\begin{cases}
            \sqrt{(1-\alpha)^3 \frac{3 \mu_0 G_{\text{c}}}{4 \ell}} &\text{stiffness degr.}\\
            \sqrt{(1-\alpha)^3 \mu_0 \left( \frac{3 G_{\text{c}}}{2 \ell} - h^\prime (\alpha) \frac{\rho_0}{2} \left|\dot{\boldsymbol{u}}\right|^2 \right)} &\text{stiff.+dens. degr.}\\
        \end{cases}$\\[0.5cm]
        \midrule
        cohesive
        &$(1-\alpha) p_{\text{c}}$
        &$(1-\alpha) \tau_{\text{c}}$\\
        \bottomrule
    \end{tabularx}
\end{table}

\begin{figure}[t!]
    \centering
    \includegraphics{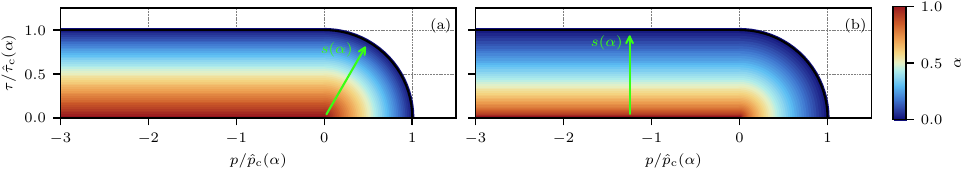}
    \caption{Strength surfaces $\mathcal{S} (\alpha)$ in the volumetric-deviatoric stress space for the cohesive (a) and the brittle model with stiffness degradation (b).
    The 'closeness' of a stress state to the damaged strength surface $s(\boldsymbol{x})$ is depicted with green arrows for an example with positive and negative pressure.}
    \label{fig:elastic_domain}
\end{figure}

From Fig.~\ref{fig:elastic_domain}, where the strength surfaces of the brittle model with stiffness degradation and of the  cohesive model are compared, we can observe that the initial elastic domain is the same for both models.
However, with our choice of degradation functions, the brittle elastic domain shrinks at a faster rate compared to the cohesive one.
The brittle model with stiffness+density degradation is not reported since it leads to an elastic domain that depends on the velocity, whereas it gives the same elastic domain of the brittle model with stiffness degradation in the quasi-static case.
Exploiting~\eqref{eq:elastic_domain} and Table~\ref{tab:strength_comparison}, we finally define a closeness measure $s \in [0,1]$ of a stress state to the damaged strength surface as follows:
\begin{equation}
    s(p, \tau, \alpha) =
    \begin{cases}
        \frac{\tau}{\hat{\tau}_{\text{c}} (\alpha)} &p<0\\
        \sqrt{\left( \frac{p}{\hat{p}_{\text{c}} (\alpha)} \right)^2 + \left( \frac{\tau}{\hat{\tau}_{\text{c}} (\alpha)} \right)^2} &p\geq0\\
    \end{cases}
\end{equation}
illustrated with green arrows in Fig.~\ref{fig:elastic_domain}. The measure is devised so that when $s=1$ a stress state lies at the boundary of the elastic domain.
We recall that, for the cohesive model, $s$ is exact by virtue of the explicit strength criterion, while for the brittle models it is exact only under homogeneous damage conditions.

\subsection{Comparison in the brittle limit}\label{sec:2D_brittle}
\begin{figure}[t!]
    \centering
    \includegraphics{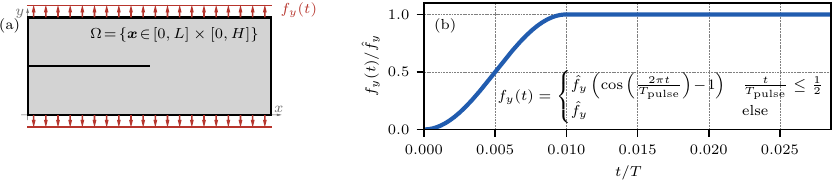}
    \caption{Setup for the pre-notched plate (a), as well as the loading ramp (b) for the traction on the upper and lower edges.}
    \label{fig:setup_prenotchedplate}
\end{figure}

To compare the various models, let us consider the benchmark problem of a brittle notched plate under tension, adopting the setup from \cite[Sec. 4.2]{Borden_2012_phase} and \cite[Sec. 5.2]{Song_comparative_2008}. The setup is presented in Fig.~\ref{fig:setup_prenotchedplate}a and involves a rectangular 2D domain of length $L=100$~mm and height $H=40$~mm with an initial crack of length $a_0 = 50$~mm starting from the left edge and placed at $y=H/2$.
We describe this initial crack through an initial phase field, enabling us to evaluate the interaction of the waves with the initial phase-field crack.
Following \cite{Borden_2012_phase}, we consider plane-strain conditions, Young's modulus $E_0 = 32\,000$~MPa, Poisson's ratio $\nu_0 = 0.2$, mass density $\rho_0 = 2\,450$~kg/m$^3$, fracture toughness $G_{\text{c}} = 0.003$~N/mm, and regularization length $\ell = 0.25$~mm.
According to Table~\ref{tab:strength_comparison}, the resulting initial critical pressure and shear for the brittle model with stiffness degradation are $p_{\text{c}} = 8.94$~MPa and $\tau_{\text{c}} = 10.95$~MPa.
The same values are also adopted for the cohesive model to allow for a fair comparison.
The maximum of the ultimate displacement jumps in case of pure hydrostatic or shear loading, i.e. $2G_{\text{c}} / \min(p_{\text{c}}, \tau_{\text{c}}) = 0.6\ \mu\text{m}$, is five orders of magnitude smaller than the smallest structural dimension, which indicates that we are close to the brittle limit.
Following Section~\ref{sec:interaction_wave_cohesive_crack_developing}, we expect the ratios of the wave speeds to the cohesive length to govern the dynamic crack opening evolution also in the multi-dimensional case.
Given the value of $3.5 \cdot 10^6~\text{s}^{-1}$ obtained here, we expect that the period during which the crack transmits cohesive forces (phase III in Section~\ref{sec:interaction_wave_cohesive_crack_developing}) is close to zero.

Unlike in \cite{Borden_2012_phase} where the tensile loading is applied abruptly, we impose a traction on the top and bottom edges via a smooth ramp with $T_{\text{pulse}}=1.6$~$\mu$s (Fig.~\ref{fig:setup_prenotchedplate}b) to avoid introducing a shock wave.
We consider a traction per unit thickness of final magnitude $\hat{f}_y \in \{1, 2\}$~N/mm to promote more or less dominant crack branching.
The time domain is $t \in [0, T]$ with $T=80$~$\mu$s and $\Delta t = 0.01$~$\mu$s.
The spatial discretization employs a structured mesh of linear quadrilateral elements with $\ell / \Delta x = \ell / \Delta y \approx 5$, yielding $1\,604\,802$ nodes and $1\,602\,000$ elements.
This gives $c_{\text{S}_0} \Delta t / \Delta x = 0.47$, with $c_{\text{S}_0}$ denoting the shear wave speed (lower than the compressive wave speed).

\subsubsection{Single branching case} \label{sec:single_branch}
\begin{figure}[t!]
    \centering
    \includegraphics{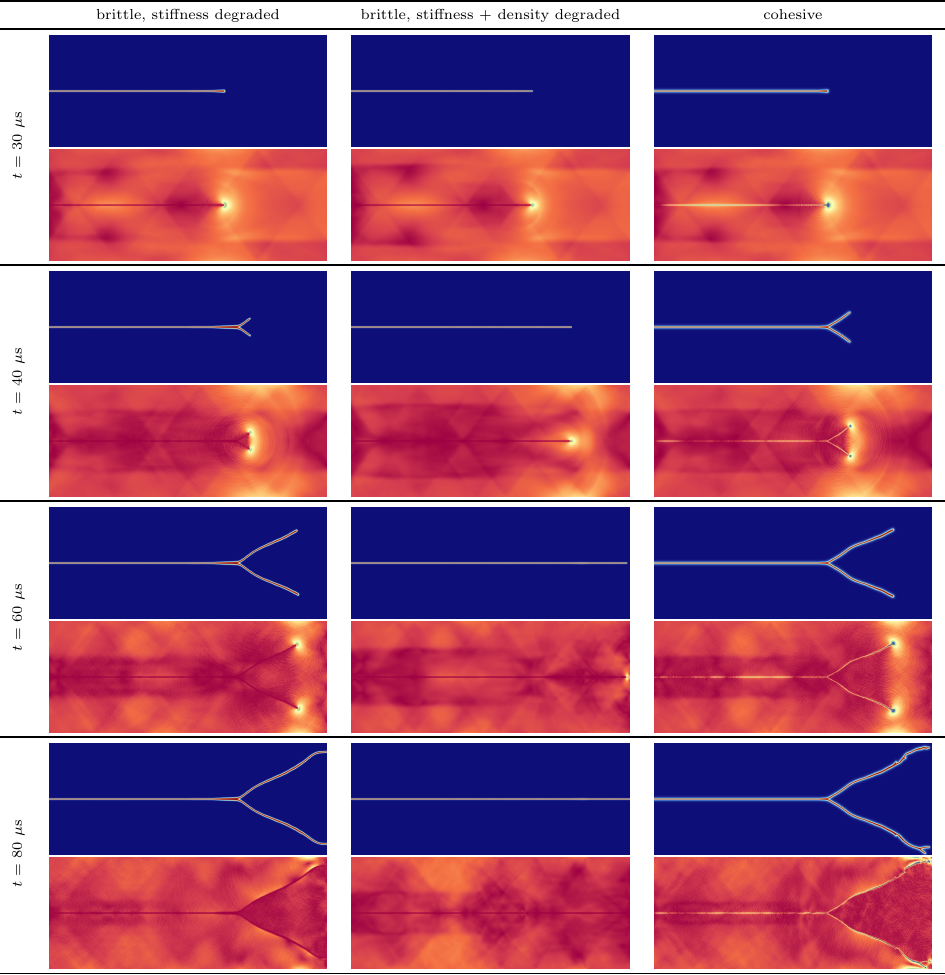}
    \caption{Phase field and $s(\boldsymbol{x})$ at various time instants for the pre-notched plate with $\hat{f}_y = 1$~N$/$mm.}
    \label{fig:2D_fields_table_1_0}
\end{figure}

We start with $\hat{f}_y = 1$~N/mm, for which we observe a single crack branching event.
Fig.~\ref{fig:2D_fields_table_1_0} shows the phase field before first branching at $t=30\ \mu\text{s}$, shortly after the branching at $t=40\ \mu\text{s}$, with fully developed branches  at $t=60\ \mu\text{s}$, and at the final time $t=80\ \mu\text{s}$ for the three models.

\begin{figure}[t!]
    \centering
    \includegraphics{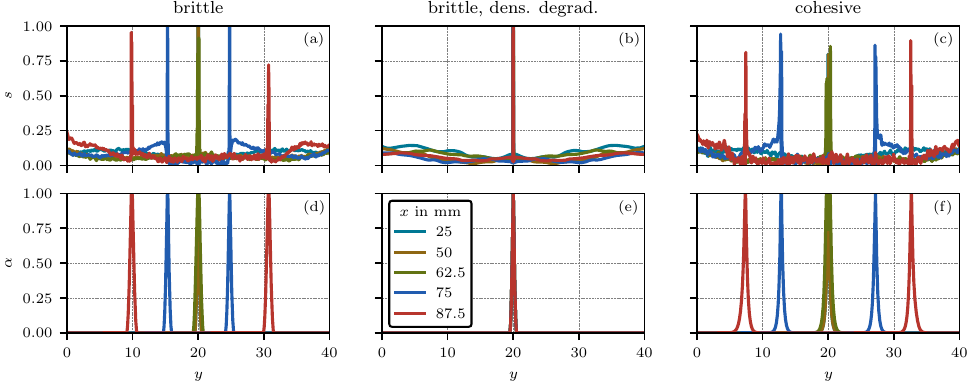}
    \caption{Slices of $s(y)$ (a-c) and $\alpha(y)$ (d-f) at various $x$-coordinates for the pre-notched plate with $\hat{f}_y = 1$~N$/$mm at the final time step ($t=80$~$\mu$s).}
    \label{fig:2D_fields_slices}
\end{figure}

While the brittle model with stiffness degradation and the cohesive model both branch, the brittle model with stiffness and density degradation produces straight crack propagation up to complete separation of the domain in two parts.
In \cite{Tian_dynamic_2020} the crack tip splitting for the brittle case is attributed to the wave speed reduction within the damaged regions, which causes an accumulation of energy at the crack tip.
With this interpretation, the different behavior of the brittle model with stiffness+density degradation is likely due to the constant wave speed under tensile loading which avoids any energy accumulation at the crack tip.
Consistently with the closeness to the brittle limit and with the large  wave speeds to cohesive length ratios of the adopted cohesive model, in the cohesive results the phase field directly evolves to $\alpha=1$ during propagation.
The plots of $s(\boldsymbol{x})$ further reveal that for the brittle model with stiffness degradation and for the cohesive model the crack tip emits ripples into the domain alongside the elastic energy release.
For the brittle model with stiffness+density degradation no ripples are observed, likely due to the additional velocity-dependent term which decreases the value of the energy release rate.
The obtained results  are further confirmed in Fig.~\ref{fig:2D_fields_slices}, where vertical slices of $s$ and $\alpha$ are shown.
High-frequency oscillations of $s$ related to the ripples are visible for the brittle model with stiffness degradation and for the cohesive model but not for the brittle model with stiffness+density degradation. 
For the brittle model with stiffness degradation, the oscillations stem from the interaction of the elastic waves with the regularized crack leading to partial reflection and transmission and from the widening of the phase-field support discussed in Section~\ref{sec:interaction_wave_brittle_crack}.
Instead, for the cohesive model they correspond to the occurrence of diffuse jumps due to the strength criterion being met in the vicinity of the crack (Section~\ref{sec:interaction_wave_cohesive_crack_fully}).

The brittle model with stiffness degradation and the cohesive model produce very similar crack patterns, which we attribute to their identical initial elastic domain.
Also, the different rates of homothetic shrinkage with damage evolution (Fig.~\ref{fig:elastic_domain}) do not play a big role since $\alpha$ rapidly reaches $1$.
The cohesive model has a slightly higher tendency to branch, especially as the cracks approach the boundaries.
We ascribe this to the brittle model dissipating additional energy by spuriously widening the phase-field profile -- as a result, as discussed earlier, the energy per unit crack extension increases over $G_{\text{c}}$ -- whereas in the cohesive model this excess energy is available for branching.

\begin{figure}[t!]
    \centering
    \includegraphics{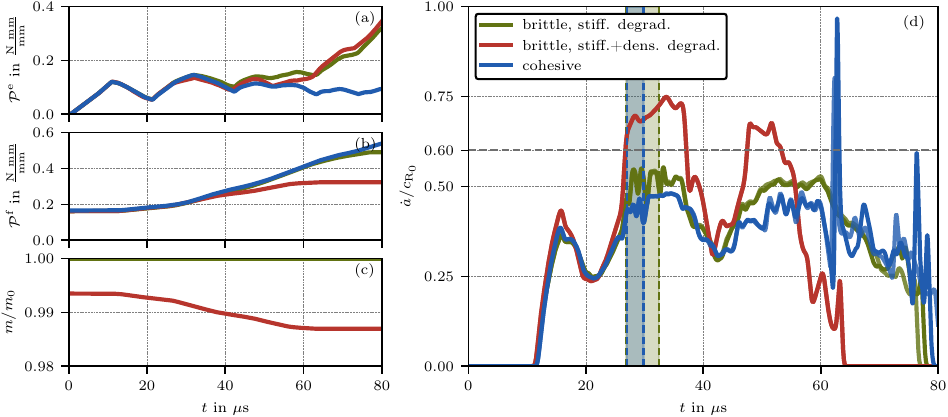}
    \caption{Energy contributions (a,b), mass (c) and crack tip speed (d) monitoring for the pre-notched plate with $\hat{f}_y = 1$~N$/$mm.
    Top and bottom branches are drawn in different shades of the color corresponding to each model.
    The shaded areas, colored according to the models, in (d) correspond to the periods during which widening of the phase-field profile occurs, where the dashed lines mark the registered start and end thereof.}
    \label{fig:2D_monitoring_1_0}
\end{figure}

Further information comes from the elastic and fracture energy contributions (Fig.~\ref{fig:2D_monitoring_1_0}a,b).
The brittle model with stiffness degradation and the cohesive model show very similar fracture energy evolutions, mirroring their similar crack patterns.
Their elastic energies also agree up to roughly $t=40$~$\mu$s, after which both brittle models show a higher elastic energy storage (Fig.~\ref{fig:2D_monitoring_1_0}b).
Such difference can be traced back to the distinct representations of a fully developed crack.
In the brittle models, the crack opening is smeared as an elastic strain over the localization band; since the degradation is complete only on the centerline of the phase-field profile, the elastic energy density $g(\alpha) \psi_{\text{D}}(\boldsymbol{\varepsilon})$ does not vanish in the band, where small values of $g(\alpha)$ meet very large strains.
As the crack faces continue to separate under the sustained loading, the strains in the band — and with them this stored energy — keep growing, which explains the increasing surplus of elastic energy after $t\approx40$~$\mu$s.
In the cohesive model, by contrast, the opening is accommodated by the eigenstrain $\boldsymbol{\eta}$, and beyond the ultimate displacement jump the stored energy is insensitive to further opening; the elastic energy therefore reflects only the bulk response and saturates accordingly.
The system mass $m=\int_\Omega h(\alpha) \rho_0\, \mathrm{d}\boldsymbol{x}$, normalized by the undegraded mass $m_0 = \int_\Omega \rho_0 \mathrm{d}\boldsymbol{x}$, is shown in Fig.~\ref{fig:2D_monitoring_1_0}c.
The brittle model with stiffness+density degradation exhibits a mass loss of roughly $1.2$~\% during crack propagation.

The crack tip speeds $\dot{a}$ (Fig.~\ref{fig:2D_monitoring_1_0}d) are computed assuming the presence of, at most, two crack tips as the moving averages of the temporal derivative of the crack tip positions.
Namely, we track the $x$-coordinates corresponding to minimum and maximum $y$-coordinate where $\alpha \geq 0.95$.
When no branching occurs, both of them coincide, while in the event of branching or widening, their vertical distance increases.
In particular, we identify the start of widening as the instant at which upper and lower tip coordinates are separated by more than $\Delta y$ for the first time.
The instant at which the interpolation of the phase-field variable on the line connecting the two tips shows values below $0.95$ for the first time marks the end of the widening, and hence the occurrence of branching.

As expected for stable dynamic crack propagation \cite{Ravi_experimental_1984a,Ravi_experimental_1984b}, the brittle model with stiffness degradation and the cohesive model both remain below $60$~\% of the Rayleigh wave speed $c_{\text{R}_0}$ before branching, after which the individual branches slow down and then re-accelerate.
The cohesive model attains slightly slower speeds and yields  an additional branching event around $t=65$~$\mu$s accompanied by spikes in the measured crack tip speeds.
In the final part of the test, the crack tip velocity decreases for all models, due to the reflection of the tensile wave at the boundary that introduces compressive stresses.
For $t \lesssim 25$~$\mu$s, the brittle model with stiffness+density degradation shows a similar behavior to the other two; at later times, the crack tip speed reaches values above $60$~\% of $c_{\text{R}_0}$ without branching. It then decreases around $t \approx 35$~$\mu$s due to the boundary wave reflections.

\begin{figure}[t!]
    \centering
    \includegraphics{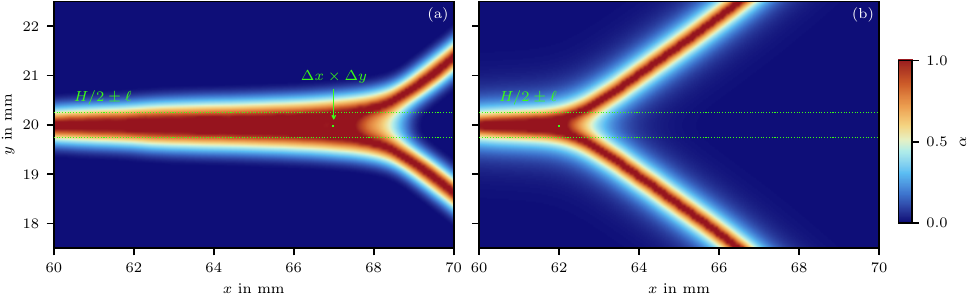}
    \caption{The region around the branching point $\boldsymbol{x} \in [60, 70]\,\text{mm} \times [17.5, 22.5]\,\text{mm}$ at the final time step $t=80$~$\mu$s for the brittle model with stiffness degradation (a), and the cohesive model (b).
    For reference, the outline of a single element is drawn in green, while the green lines mark $H/2\pm \ell$.}
    \label{fig:2D_fields_clipped_branching}
\end{figure}

In Fig.~\ref{fig:2D_monitoring_1_0}d we have also marked with dashed lines the time instants at which widening of the phase-field profile starts and ends.
Accordingly, the second dashed line of a respective color marks the branching event.
Compared to the cohesive model, the brittle model with stiffness degradation exhibits a longer region of widening prior to branching, as well as a more extensive widening resulting in a thicker damage band (Fig.~\ref{fig:2D_fields_clipped_branching}). This behavior is consistent with the observations made in Section~\ref{sec:interaction_wave_cohesive_crack_fully} for the 1D model.
The separated branches themselves do not show a significantly widened phase-field profile in either model.

\subsubsection{Multiple branching case}
\begin{figure}
    \centering
    \includegraphics{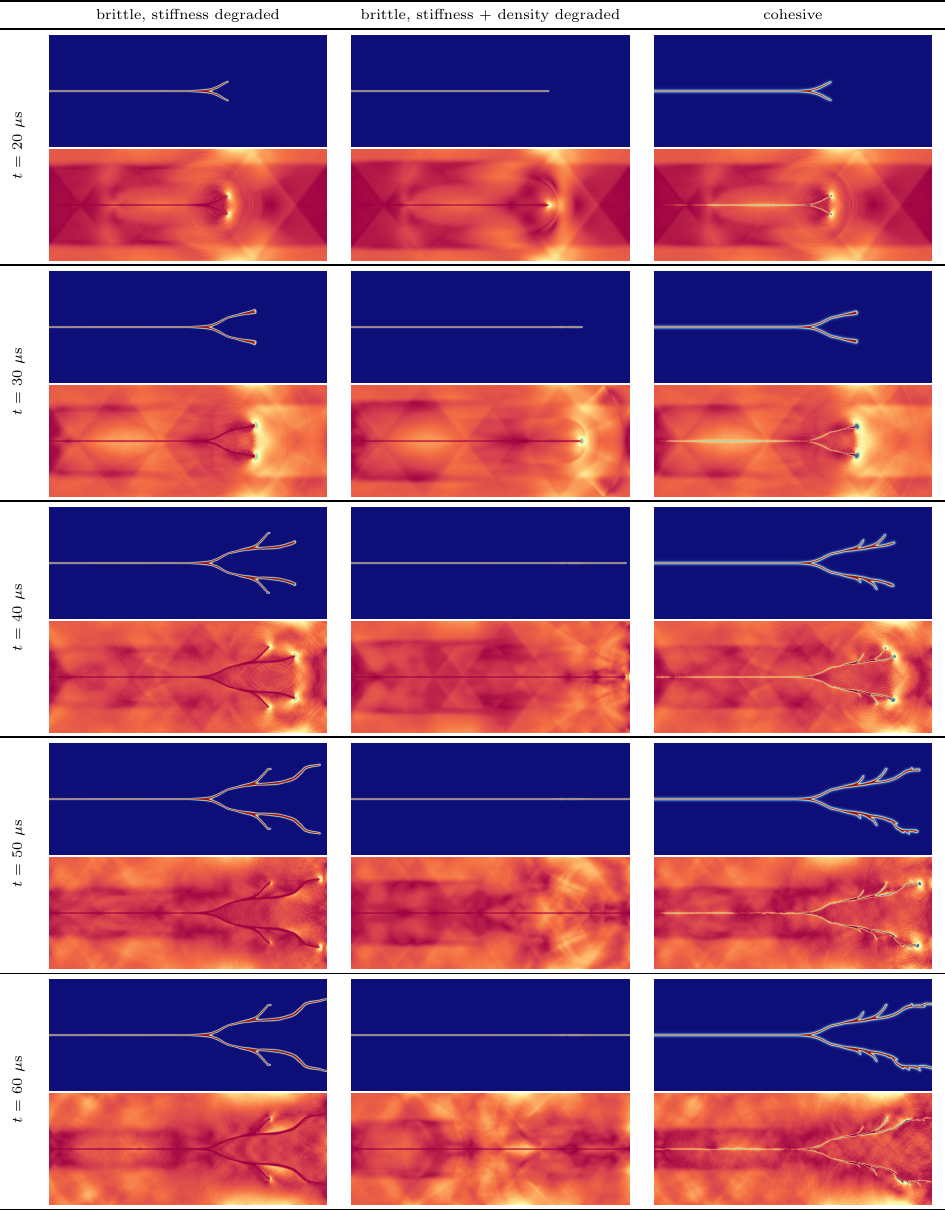}
    \caption{Phase field and $s(\boldsymbol{x})$ at various time instants for the pre-notched plate with $\hat{f}_y = 2$~N$/$mm.}
    \label{fig:2D_fields_table_2_0}
\end{figure}

Fig.~\ref{fig:2D_fields_table_2_0} shows the phase field and $s(\boldsymbol{x})$ for $\hat{f}_y = 2$~N/mm.
The brittle model with stiffness+density degradation again does not branch, producing a single straight crack up to the right edge.
Conversely, the other two models branch significantly more compared to the case in Section~\ref{sec:single_branch}, with the cohesive model showing a higher number of tip splitting events.
This is consistent with the observation in Fig.~\ref{fig:2D_fields_clipped_branching}. The brittle model with stiffness degradation triggers a longer widening period of the phase-field profile, which spuriously dissipates more energy; this energy is no longer available for further tip splittings.
In addition, in this case the crack branches generated after a crack tip splitting event display a wider phase-field profile (Fig.~\ref{fig:2D_fields_table_2_0}).
The cohesive model both branches more and yields a less symmetric pattern, features we attribute to its higher sensitivity to small perturbations.

\begin{figure}[t]
    \centering
    \includegraphics{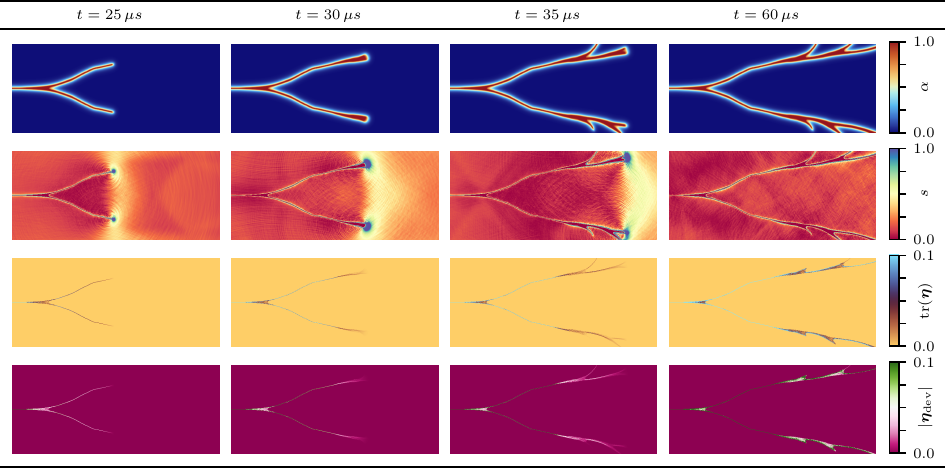}
    \caption{Phase field, $s(\boldsymbol{x})$, and eigenstrain obtained with the cohesive model at various time instants for the pre-notched plate with $\hat{f}_y = 2$~N$/$mm, zoomed in the region $\boldsymbol{x} \in [50, 85]\,\text{mm} \times [12.5, 27.5]\,\text{mm}$ around the branching points.}
    \label{fig:2D_fields_table_clipped}
\end{figure}

To examine the branching mechanism of the cohesive model in more detail, Fig.~\ref{fig:2D_fields_table_clipped} shows a close-up of $\alpha(\boldsymbol{x})$, $s(\boldsymbol{x})$, and the volumetric and deviatoric parts of the eigenstrain, $\text{tr}(\boldsymbol{\eta})$ and $\left|\boldsymbol{\eta}_{\text{dev}}\right|$ at different instants.
At a branching event, the eigenstrain is not localized in a single element but is instead diffuse over a region close to the crack: the smooth transition from a single crack to two branches necessarily widens the damaged band, and with it the eigenstrain.
A diffuse jump band persists, which we attribute to the interactions between the ripples emitted by the crack tip, as well as the interaction with the boundaries.
The resulting opening and closing of these diffuse jumps then generate shock waves and with them high-frequency oscillations, which in turn foster the further widening of the crack band.

\section{Conclusions}\label{sec:conclusions}
In this work, we analyzed the influence of phase-field regularization on the dynamic extensions of both brittle and cohesive phase-field fracture formulations.
The following main insights were obtained. 

For the brittle models, the interaction of a phase-field crack with an incoming elastic wave is dictated by the profiles of the wave speed and of the acoustic impedance, which depend on the choice of the degradation functions for the stiffness and for the mass density. For the model with stiffness degradation, we demonstrated the occurrence of spurious partial reflection and transmission of tensile elastic waves.
The overall wave-crack interaction is controlled by the ratio $\ell/\lambda$ of the phase-field regularization length to the elastic wavelength, and the sharp-crack response is recovered fully only in the asymptotic limit where $\ell/\lambda \rightarrow 0$.
Incidentally, this limit is not compatible with the interpretation of $\ell$ as a material parameter, which is needed with this model to calibrate the desired nucleation behavior under quasi-static conditions.
Furthermore, wave-crack interactions directly trigger a widening of the phase-field profile, causing the energy dissipated per unit crack extension to spuriously increase over the material's critical energy release rate $G_{\text{c}}$.  However, with this model a phase-field crack correctly transmits a compressive wave if combined with a standard energy decomposition.
The brittle model with stiffness and density degradation, while leading to a constant wave speed and to full reflection of a tensile wave if both degradation functions coincide, inherently violates mass conservation and cannot correctly transmit a compressive wave with a standard strain energy decomposition.

To attempt to resolve the intrinsic issues of brittle regularization in the dynamic setting, we extended our recently proposed cohesive phase-field fracture model with strength degradation to elastodynamics.
Since it does not modify the stiffness nor the density as a result of damage, this model leaves the bulk wave equation unchanged with respect to the one of the pure linearly elastic problem. Moreover, with this model a phase-field crack correctly transmits a compressive wave with no need for an energy decomposition; it also correctly (i.e. completely) reflects a tensile wave under the condition that this wave does not exceed the locally degraded strength adjacent to the crack, which is controlled by the combination of the $\ell/\lambda$ ratio and of the ratio $\tilde{\sigma}/\sigma_c$ between the amplitude of the stress wave and the tensile strength of the material. Incidentally, with this model the value of $\ell$ is completely independent from the material strength and can be chosen based on purely numerical considerations. 
For the one-dimensional case, we further derived  an analytical crack-opening evolution law and demonstrated that, for a combination of $\ell/\lambda$ and $\tilde{\sigma}/\sigma_c$ leading to the full reflection of a tensile wave, the dynamic response in time is fundamentally governed by the ratio $c_0/\ell_{\text{ch}}$ of the bulk wave speed to the Irwin cohesive characteristic length. 

Finally, a comparative study on a 2D pre-notched plate under tensile loading highlighted critical differences in the predictions of the three models.
While the brittle model with stiffness degradation and the cohesive model yielded remarkably similar crack branching patterns, the brittle model with stiffness and density degradation failed to capture crack-tip branching instabilities, resulting in purely straight crack propagation. 

Taken together, these results identify the cohesive formulation as a promising starting point for further work on dynamic phase-field fracture.

\section*{Acknowledgements}
We gratefully acknowledge funding from the Swiss National Science Foundation through Grant No. 200021-219407 'Phase-field modeling of fracture and fatigue: from rigorous theory to fast predictive simulations'.

\section*{Data and code availability}
Data will be made available upon request.
The implementations used in this work are publicly available at \url{https://github.com/jonas-heinzmann/phase_field_dynamics}. Animations of all numerical results can be found at \url{https://github.com/jonas-heinzmann/phase_field_dynamics} in the folder \texttt{animations}.

\section*{Declaration of generative AI in the writing process}
During the preparation of this work, the authors used Anthropic Claude in order to improve the readability of the manuscript.
After using this tool, the authors reviewed and edited the content as needed and take full responsibility for the content of the publication.

\FloatBarrier
\bibliographystyle{elsarticle-num}
\bibliography{references}

\addcontentsline{toc}{section}{Appendices}
\renewcommand{\thesubsection}{\Alph{subsection}}
\section*{Appendices}

\subsection{Scattering analysis of the 1D heterogeneous wave equation}\label{scatt}
In this section, we analyze the frequency-dependent reflection and transmission of an elastic wave in a 1D medium with smoothly varying Young's modulus $E(x) = g(x)E_0$ and density $\rho(x) = h(x)\rho_0$, as obtained from both brittle models for a fixed phase field (whereby the model with stiffness degradation takes $h(x)\equiv 1$). We first
derive an exact reformulation of the wave equation that separates propagation from reflection \cite{bremmer_1951}, from which we can obtain the adiabatic criterion \eqref{eq:WKB}; the transfer matrix method (TMM) then arises as its exact
solution for layerwise-constant properties.

Non-dimensionalizing the elastic wave equation by $\check{x}=x/\ell$ and $\check{t}=c_0 t/\ell$, as well as $\check{u} = u / u_0$ with some reference displacement magnitude $u_0$, yields
\begin{equation}
\label{wave}
    \frac{\partial}{\partial \check{x}}\!\left(g(\check{x})\,\frac{\partial \check{u}}{\partial \check{x}}\right)
    = h(\check{x})\,\frac{\partial^{2}\check{u}}{\partial \check{t}^{\,2}} \quad\text{,}
\end{equation}
where, with a little abuse of notation, we are denoting $g(\check{x})=g(\alpha(\check{x}))$ and $h(\check{x})=h(\alpha(\check{x}))$ as the compositions of the degradation functions for stiffness and density, respectively, with the phase-field profile \eqref{eq:profile_AT1}. Since, for a fixed phase field, the properties of the medium are time-independent, we work in the frequency domain: for a time-harmonic field $\check u (\check x, \check t)=\text{Re}[U(\check x)e^{-i\check \omega \check t}]$ at dimensionless angular frequency $\check\omega = \omega\ell/c_0$, and denoting with uppercase letters the complex amplitudes of the corresponding lower-case fields, we introduce the velocity and stress amplitudes
$V = -i\check\omega\,U$ and $\Sigma = g\,U'$, by which \eqref{wave} can be reformulated as the first-order system
\begin{equation}
  V' = -\frac{i\check k}{\check Z}\,\Sigma , \qquad
  \Sigma' = -i\check k\,\check Z\,V ,
  \qquad\text{with}\qquad
  \check k(\check x) = \check\omega\sqrt{\frac{h(\check x)}{g(\check x)}} , \quad
  \check Z(\check x) = \sqrt{g(\check x)\,h(\check x)}.
  \label{eq:B-VS}
\end{equation}
Note that this system is fully characterized by two acoustic profiles: the local wavenumber $\check k$ (or equivalently the local wave speed $\check c(\check x) = \sqrt{g(\check x)/h(\check x)}$) and the local impedance $\check Z$. In a homogeneous medium, right- and left-going waves satisfy $\Sigma = \mp\check Z V$; accordingly, we
define the local wave amplitudes
\begin{equation}
  a = \tfrac12\Bigl( \sqrt{\check Z}\, V
      - \Sigma/\sqrt{\check Z} \Bigr) , \qquad
  b = \tfrac12\Bigl( \sqrt{\check Z}\, V
      + \Sigma/\sqrt{\check Z} \Bigr) ,
  \label{eq:B-ab}
\end{equation}
normalized such that $|a|^2 - |b|^2$ equals (twice) the time-averaged rightward power flux. Differentiating \eqref{eq:B-ab} and using \eqref{eq:B-VS} yields the first-order system
\begin{equation}
  a' - i\check k\,a = \frac{\check Z'}{2\check Z}\,b , \qquad
  b' + i\check k\,b = \frac{\check Z'}{2\check Z}\,a .
  \label{eq:B-coupled}
\end{equation}
which is still exact and fully equivalent to \eqref{wave}.
In each equation of \eqref{eq:B-coupled}, the term proportional to $i\check k$ rotates the phase of the corresponding amplitude at the local rate $\check k(\check x)$ while conserving its modulus, and describes transmission; the right-hand sides, with coefficient $\check Z'/2\check Z$, couple the two amplitudes and are the only mechanism
converting right- into left-going waves, i.e., generating reflection. Moreover, $|a|^2 - |b|^2$ is exactly conserved. Since the coupling vanishes for $\check Z'=0$, a medium with constant impedance is reflectionless regardless of its wave-speed profile. 

The WKB (adiabatic) approximation consists in neglecting the coupling wherever it is slow compared to the phase rotation: factoring out the accumulated phase, the reflected contributions generated over one local wavelength then cancel by destructive interference \cite{brekhovskikh_1980}. The validity ratio of the two rates, $\bigl(|\check Z'(\check x)|/2\check Z(\check x)\bigr)/\check k(\check x)$, is (up to the factor of two) the quantity $\delta$ introduced in \eqref{eq:WKB}.

\subsection{Transfer matrix method}\label{app:tmm}
For quantitative computations we solve \eqref{eq:B-ab} exactly by the TMM, which approximates the heterogeneous medium with a stack of $n$ homogeneous layers with constant material properties $g_l$, $h_l$; for more details on the TMM applied to elastic waves see e.g. \cite{cretu_2004}.
Within each layer $l$ the coupling vanishes and the amplitudes only accumulate phase, so that the solution reads
\begin{equation}
    U_l(\check{x})=A_l\,e^{i\check{k}_l\check{x}}+B_l\,e^{-i\check{k}_l\check{x}},
    \qquad \check{k}_l=\check{\omega}\sqrt{h_l/g_l} \quad\text{,}
\end{equation}
where $A_l$ and $B_l$ are the amplitudes of the rightward and leftward waves, respectively.
By means of continuity of displacement and stress across the interface between layers $l$ and $l{+}1$, the \textit{discontinuity matrix}
\begin{equation}\label{discont_matrix}
    \mathbf{D}(\check{Z}_l,\check{Z}_{l+1})=\frac{1}{2}
    \begin{bmatrix}
        1+\check{Z}_{l+1}/\check{Z}_l & 1-\check{Z}_{l+1}/\check{Z}_l\\[2pt]
        1-\check{Z}_{l+1}/\check{Z}_l & 1+\check{Z}_{l+1}/\check{Z}_l
    \end{bmatrix}
\end{equation}
relates the amplitudes of the two layers, with the reflection and transmission at the interface being  governed by the acoustic impedances $\check{Z}_l=\sqrt{g_l h_l}$; for a small impedance contrast, its off-diagonal entries linearize to $\Delta\check Z/2\check Z$, recovering the coupling terms of \eqref{eq:B-coupled}. 
Propagation across a layer of thickness $\Delta\check{x}_l$ is described by the \textit{propagation matrix}
\begin{equation}\label{propag_matrix}
    \mathbf{P}(\check{k}_l,\Delta\check{x}_l)=
    \begin{bmatrix}
        e^{i\check{k}_l\Delta\check{x}_l} & 0\\
        0 & e^{-i\check{k}_l\Delta\check{x}_l}
    \end{bmatrix} \quad\text{,}
\end{equation}
which integrates the propagation terms of \eqref{eq:B-coupled} within a layer.
The global \textit{transfer matrix} is assembled as
\begin{equation}\label{global_transfer_matrix}
    \mathbf{T}(\check{\omega})=\mathbf{D}(\check{Z}_0,\check{Z}_1)\,
    \prod_{l=1}^{n-1}\mathbf{P}(\check{k}_l,\Delta\check{x}_l)\,
    \mathbf{D}(\check{Z}_l,\check{Z}_{l+1}) \quad\text{,}
\end{equation}
and links the wave amplitudes in the two semi-infinite homogeneous media bounding the layer stack -- the undamaged material on either side of the regularized crack -- via
\begin{equation}
    \begin{bmatrix}A_\text{inc}\\ A_\text{ref}\end{bmatrix}
    =\mathbf{T}(\check{\omega})\begin{bmatrix}A_\text{tra}\\ 0\end{bmatrix} \quad\text{,}
\end{equation}
with $A_\text{inc}$, $A_\text{ref}$ and $A_\text{tra}$ being the amplitudes of the incoming, reflected and transmitted waves, respectively.
Accounting for the impedance mismatch between the incident and transmitted media, the frequency-dependent power coefficients follow as
\begin{equation}
    t(\check{\omega})=\frac{\check{Z}_n}{\check{Z}_0}\left|\frac{1}{T_{11}(\check{\omega})}\right|^{2}
    \quad\text{, and}\quad
    r(\check{\omega})=\left|\frac{T_{21}(\check{\omega})}{T_{11}(\check{\omega})}\right|^{2} \quad\text{.}
\end{equation}
The results in Figs.~\ref{fig:tmm_standard} and~\ref{fig:tmm_degraded} were obtained with $n=10^{5}$ layers and $10^{3}$ frequencies, expressed through the dimensionless ratio $\ell/\lambda$.


\subsection{Finite element implementation}\label{app:fem}
We use \texttt{FEniCSx} v0.9.0 \cite{baratta_2023,scroggs_2022a,scroggs_2022b} and \texttt{PETSc} release 3.24.5 \cite{balay_2024a,balay_2024b,balay_1997} for the implementation of the models in this work.
The starting point is the respective action functional, with the temporal approximations according to either Newmark-$\beta$ or generalized-$\alpha$ inserted already.
We use \texttt{UFL} \cite{alnaes_2014} to automatically derive the residuals and their linearization, which are passed to \texttt{PETSc} through the \texttt{petsc4py} interface \cite{dalcin_2011}.

While some previous contributions use a monolithic solution scheme to solve the time-discrete equations of dynamic phase-field fracture in a fully coupled way \cite{Schlueter_phase_2014,Borden_2012_phase,Tian_dynamic_2020}, we adopt a staggered solution scheme as also done in \cite{Li_numerical_2016,Hofacker_continuum_2012,Steinke_comparative_2016}.
As described in \cite{heinzmann_exact_2026}, this means that we solve alternately the $\boldsymbol{u}$-problem while keeping $\alpha$ fixed, and the $\alpha$-problem while keeping the last converged $\boldsymbol{u}$ fixed.
We solve both subproblems with Newton's method until convergence, while we employ the reduced-space active set strategy based on Newton's method \cite{benson_flexible_2006} for the damage problem to enforce the constraints.
For the linear solve in each Newton iteration, we use the \texttt{MUMPS} package \cite{amestoy_2001,amestoy_2019} with the Cholesky factorization.
For the model with density degradation, we use the $LU$ factorization instead.
We determine convergence of both subproblems as $\|\mathbf{R}_{\mathbf{u}} (\mathbf{u}, \boldsymbol{\alpha}^{i-1})\|_2 \leq 10^{-7}$ and $\|\mathbf{R}_{\boldsymbol{\alpha}} (\mathbf{u}^i, \boldsymbol{\alpha})\|_2 \leq 10^{-7}$, while we check the convergence of the staggered scheme as $\|\mathbf{R}_{\mathbf{u}} (\mathbf{u}^i, \boldsymbol{\alpha}^{i})\|_2 \leq 10^{-5}$, where $i$ is the staggered iteration.
All our implementations are available at \url{https://github.com/jonas-heinzmann/phase_field_dynamics}.

\subsection{Compact reformulation of the cohesive strain energy density}\label{app:compact_reformulation}
To derive a compact reformulation for the cohesive model, we seek a stationary point of the action functional with respect to $\boldsymbol{\eta}$, respecting the constraint $\text{tr}(\boldsymbol{\eta}) \geq 0$.
Since $\boldsymbol{\eta}$ carries no inertia, we can optimize the strain energy functional directly.
By the normality rule \cite[Appendix A]{Vicentini_variational_2026}, the sign of $\text{tr}(\boldsymbol{\eta})$ coincides with the sign of the pressure $p = \tfrac{1}{d} \text{tr}(\boldsymbol{\sigma})$, allowing us to distinguish the following two cases.

\paragraph{Non-negative pressure}
For $p \geq 0$, stationarity of $\psi^{\text{c}}$ with respect to $\boldsymbol{\eta}$ yields
\begin{equation}\label{eq:CF_psi_voldev_postr}
    \frac{\partial}{\partial \boldsymbol{\eta}} \left( \frac{1}{2} \mathbb{C}_0 (\boldsymbol{\varepsilon} - \boldsymbol{\eta}^\star) \cdot (\boldsymbol{\varepsilon} - \boldsymbol{\eta}^\star) + a(\alpha) \sqrt{w_{\text{c}}} \sqrt{ \mathbb{C}_0 \boldsymbol{\eta}^\star \cdot \boldsymbol{\eta}^\star} \right) = \boldsymbol{0}
    \quad\Rightarrow\quad
    \boldsymbol{\varepsilon} - \boldsymbol{\eta}^\star = a(\alpha) \sqrt{w_{\text{c}}} \frac{\boldsymbol{\eta}^\star}{\sqrt{ \mathbb{C}_0 \boldsymbol{\eta}^\star \cdot \boldsymbol{\eta}^\star}} \quad\text{,}
\end{equation}
where $\boldsymbol{\eta}^\star$ is the optimum. Noting the coaxiality between $\boldsymbol{\eta}^\star$ and $\boldsymbol{\varepsilon}$, we can write $\boldsymbol{\eta}^\star = q(\alpha) \boldsymbol{\varepsilon}$ with
\begin{equation}
    \boldsymbol{\varepsilon} \left( 1 - q(\alpha) -  \frac{a(\alpha) \sqrt{w_{\text{c}}}}{\sqrt{ \mathbb{C}_0 \boldsymbol{\varepsilon} \cdot \boldsymbol{\varepsilon}}} \right) = \boldsymbol{0}
    \qquad\Rightarrow\qquad
    q(\alpha) = 1 - \frac{a(\alpha) \sqrt{w_{\text{c}}}}{\sqrt{ \mathbb{C}_0 \boldsymbol{\varepsilon} \cdot \boldsymbol{\varepsilon}}} \qquad\text{.}
\end{equation}
Requiring $q(\alpha) \geq 0$ yields $\psi_0 (\boldsymbol{\varepsilon}) \geq \frac{1}{2} a(\alpha)^2 w_{\text{c}}$. 
Substituting in the cohesive strain energy density gives
\begin{equation}
    \hat{\psi}^{\text{c}} (\boldsymbol{\varepsilon},\alpha) = a(\alpha) \sqrt{w_{\text{c}}} \sqrt{2\psi_0(\boldsymbol{\varepsilon})} - \frac{1}{2} a(\alpha)^2 w_{\text{c}} \qquad \text{if}\,\,\, \text{tr}(\boldsymbol{\varepsilon}) \geq 0, \psi_0 (\boldsymbol{\varepsilon}) \geq \frac{1}{2} a(\alpha)^2 w_{\text{c}}.
\end{equation}

\paragraph{Negative pressure}
For $p < 0$, the eigenstrain potential~\eqref{eq:eigenstrain_potential} enforces $\text{tr}(\boldsymbol{\eta}) = 0$, reducing~\eqref{eq:psi_cohesive} to
\begin{equation}\label{CF_psi_voldev_treta_0}
    \psi^{\text{c}} (\boldsymbol{\varepsilon}, \boldsymbol{\eta}, \alpha) |_{\text{tr}(\boldsymbol{\eta}) = 0} = \frac{\kappa_0}{2} \left( \text{tr}(\boldsymbol{\varepsilon}) \right)^2 + 2 \mu_0 \left( \left| \boldsymbol{\varepsilon}_{\text{dev}} \right| - \left| \boldsymbol{\eta}_{\text{dev}} \right| \right)^2 + a(\alpha) \sqrt{w_{\text{c}}} \sqrt{2 \mu_0} \left| \boldsymbol{\eta}_{\text{dev}} \right| \quad\text{.}
\end{equation}
Stationarity with respect to $\left| \boldsymbol{\eta}_{\text{dev}} \right|$ then gives the optimum
\begin{equation}\label{eq:CF_compact_trneg_opt}
    \left| \boldsymbol{\eta}_{\text{dev}} \right|^\star = \left| \boldsymbol{\varepsilon}_{\text{dev}} \right| - a(\alpha) \sqrt{\frac{w_{\text{c}}}{2\mu_0}} \quad\text{.}
\end{equation}
Non-negativity of the norm requires $\left| \boldsymbol{\varepsilon}_{\text{dev}} \right| \geq a(\alpha) \sqrt{w_{\text{c}}/(2\mu)}$ for this branch.
Combining the two cases and adding the residual energy density $\epsilon \psi_0$ to avert numerical issues yields the condensed form of the strain energy density reported in the main body, \eqref{eq:psi_cohesive_condensed}.

From it, the stress~\eqref{eq:stress_cohesive_pressure_shear} can now be evaluated directly from $\hat{\psi}^{\text{c}}$:
\begin{subequations}
    \begin{equation}
        p (\boldsymbol{\varepsilon}, \alpha) = \frac{\partial \hat{\psi}^{\text{c}} (\boldsymbol{\varepsilon}, \alpha)}{\partial \text{tr}(\boldsymbol{\varepsilon})} =
        \begin{cases}
            \kappa_0 \left( \frac{a(\alpha) \sqrt{w_{\text{c}}}}{\sqrt{2\psi_0 (\boldsymbol{\varepsilon})}} + \epsilon \right) \text{tr}(\boldsymbol{\varepsilon}) &\text{if}\,\,\, \text{tr}(\boldsymbol{\varepsilon}) \geq 0, \psi_0 (\boldsymbol{\varepsilon}) \geq \frac{1}{2} a(\alpha)^2 w_{\text{c}}\\
            \kappa_0 (1 + \epsilon) \text{tr}(\boldsymbol{\varepsilon}) &\text{else}
        \end{cases}
        \qquad\text{and}
    \end{equation}
    \begin{equation}
        \tau (\boldsymbol{\varepsilon}, \alpha) = \frac{\partial \hat{\psi}^{\text{c}} (\boldsymbol{\varepsilon}, \alpha)}{\partial \left| \boldsymbol{\varepsilon}_{\text{dev}} \right|} =
        \begin{cases}
            2\mu_0 \left( \frac{ a(\alpha) \sqrt{w_{\text{c}}}}{\sqrt{2\psi_0 (\boldsymbol{\varepsilon})}} + \epsilon \right) \left| \boldsymbol{\varepsilon}_{\text{dev}} \right| &\text{if}\,\,\, \text{tr}(\boldsymbol{\varepsilon}) \geq 0, \psi_0 (\boldsymbol{\varepsilon}) \geq \frac{1}{2} a(\alpha)^2 w_{\text{c}}\\
            a(\alpha) \sqrt{2 \mu_0 w_{\text{c}}} + 2 \epsilon \mu_0 \left| \boldsymbol{\varepsilon}_{\text{dev}} \right| &\text{if}\,\,\, \text{tr}(\boldsymbol{\varepsilon}) < 0, \left| \boldsymbol{\varepsilon}_{\text{dev}} \right| \geq a(\alpha) \sqrt{\frac{w_{\text{c}}}{2\mu_0}}\\
            2\mu_0 (1 + \epsilon) \left| \boldsymbol{\varepsilon}_{\text{dev}} \right| &\text{else}
        \end{cases} \qquad\text{.}
    \end{equation}
\end{subequations}
The eigenstrain can be post-processed from $\boldsymbol{\varepsilon}$ as
\begin{subequations}
    \begin{equation}
        \text{tr}(\boldsymbol{\eta}) =
        \begin{cases}
            q(\alpha) \text{tr}(\boldsymbol{\varepsilon}) &\text{if}\,\,\, \text{tr}(\boldsymbol{\varepsilon}) \geq 0, \psi_0 (\boldsymbol{\varepsilon}) \geq \frac{1}{2} a(\alpha)^2 w_{\text{c}}\\
            0 &\text{else}
        \end{cases}
        \qquad\text{and}\qquad
    \end{equation}
    \begin{equation}
        \left| \boldsymbol{\eta}_{\text{dev}} \right| =
        \begin{cases}
            q(\alpha) \left| \boldsymbol{\varepsilon}_{\text{dev}} \right| &\text{if}\,\,\, \text{tr}(\boldsymbol{\varepsilon}) \geq 0, \psi_0 (\boldsymbol{\varepsilon}) \geq \frac{1}{2} a(\alpha)^2 w_{\text{c}}\\
            \left| \boldsymbol{\varepsilon}_{\text{dev}} \right| - a(\alpha) \sqrt{\frac{w_{\text{c}}}{2\mu_0}} &\text{if}\,\,\, \text{tr}(\boldsymbol{\varepsilon}) < 0, \left| \boldsymbol{\varepsilon}_{\text{dev}} \right| \geq a(\alpha) \sqrt{\frac{w_{\text{c}}}{2\mu_0}}\\
            0 &\text{else}
        \end{cases}
    \end{equation}
\end{subequations}
where the residual energy density does not affect the definitions since the stationarity conditions with respect to $\boldsymbol{\eta}$, \eqref{eq:CF_psi_voldev_postr} and \eqref{eq:CF_compact_trneg_opt}, are not affected by $\epsilon \psi_0(\boldsymbol{\varepsilon})$.

\subsection{The dynamic cohesive law}\label{app:dynamic_cohesive_law}
In the following, we derive the crack opening evolution for the dynamic cohesive phase-field model.

\subsubsection{General solution}\label{app:dynamic_cohesive_law_general}
\begin{figure}[t!]
    \centering
    \includegraphics{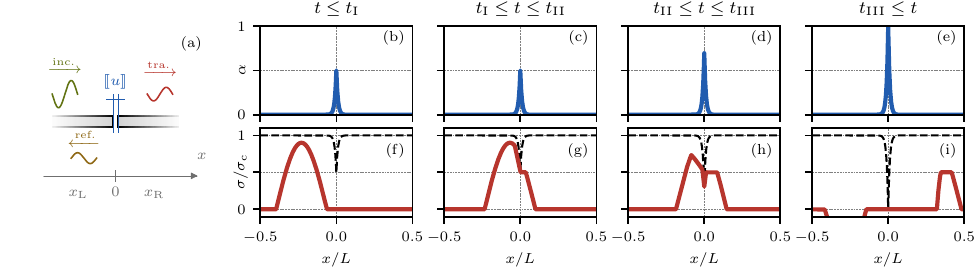}
    \caption{Setup for the study of the dynamic cohesive response (a), phase field and stress field during the four phases of the crack opening evolution (b).
    In the stress plots, the dashed line corresponds to the local strength of the material $(1-\alpha(x,t))\sigma_{\text{c}}$.}
    \label{fig:cohesive_opening_theory}
\end{figure}

We consider an isolated material point at $x=0$ at which a crack will open, adjacent to two semi-infinite half-lines representing the bulk on the left and right sides of the crack.
We assume that the bulk is linearly elastic ($\eta \equiv 0$ for all $x \neq 0$), i.e., the wave interacts only with a single crack, see Section~\ref{sec:interaction_wave_cohesive_crack_fully} and Fig.~\ref{fig:cohesive_opening_theory}a.
We allow for a pre-existing phase-field crack at $x=0$, with corresponding maximum phase-field value $\breve{\alpha}(t)$, while the jump is initially closed, $\jump{u}(t=0)=0$.
The tensile loading is assumed large enough that the stress at $x=0$ reaches the critical stress $\sigma_{\text{c}}$ at some time, and that it provides enough energy for the crack to fully open.
On the left side of the crack ($x_L = \{x : x < 0\}$) we consider an incoming, right-traveling wave $u_L^{\text{inc}}$ and a potentially reflected, left-traveling wave $u_L^{\text{ref}}$; on the right side of the crack ($x_R = \{x : x > 0\}$) we consider only a transmitted, right-traveling wave $u_R^{\text{tra}}$.

In both half-spaces, the standard d'Alembert solution holds along the characteristics $t \pm x/c_0 = \text{const}$:
\begin{equation}\label{eq:u_DAlambert}
    u_L(x_L,t) = \underbrace{f_L(t - \tfrac{x_L}{c_0})}_{u_L^{\text{inc}}(x_L,t)} + \underbrace{g_L(t + \tfrac{x_L}{c_0})}_{u_L^{\text{ref}}(x_L,t)} \qquad\text{and}\qquad u_R(x_R,t) = \underbrace{f_R(t - \tfrac{x_R}{c_0})}_{u_R^{\text{tra}}(x_R,t)} \qquad\text{,}
\end{equation}
with the arbitrary perturbations $f_L,\,g_L,\,f_R$ determined by the initial conditions.
The stress follows directly from the velocity,
\begin{equation}\label{eq:DAlambert_stess_velocity}
    \sigma_L(x_L,t) = Z_0 \left( - v_L^{\text{inc}}(x_L,t) + v_L^{\text{ref}}(x_L,t) \right) \qquad\text{and}\qquad \sigma_R(x_R,t) = - Z_0\, v_R^{\text{tra}}(x_R,t)\quad\text{,}
\end{equation}
with the (constant) bulk acoustic impedance $Z_0 = E_0/c_0 = \rho_0 c_0 = \sqrt{E_0 \rho_0}$.

At the crack, the kinematic compatibility on the jump and the stress continuity (with the cohesive law~\eqref{eq:cohesive_stress_jump}) read
\begin{align}
    \jump{u}(t) &= u_R^{\text{tra}}(0^+,t) - u_L^{\text{inc}}(0^-,t) - u_L^{\text{ref}}(0^-,t) \quad\text{,}\label{eq:crack_kinematic_condition}\\
    \sigma_L(0^-,t) &= \sigma_R(0^+,t) = \left(1 - \breve{\alpha}(t)\right)\sigma_{\text{c}} = \left(1 - \frac{\sigma_{\text{c}}}{2G_{\text{c}}} \jump{u}(t)\right)\sigma_{\text{c}} \quad\text{.}\label{eq:crack_stress_continuity}
\end{align}
Using~\eqref{eq:DAlambert_stess_velocity} in the stress continuity gives $v_R^{\text{tra}}(0^+,t) = v_L^{\text{inc}}(0^-,t) - v_L^{\text{ref}}(0^-,t)$.
Inserting this into the time-integrated form of~\eqref{eq:crack_kinematic_condition} eliminates the incident and transmitted contributions and yields
\begin{equation}\label{eq:cohesive_response_raw_jump}
    \jump{u}(t) = -2 \int_{0}^{t} v_L^{\text{ref}}(0^-,\tau)\,\mathrm{d}\tau \quad\text{.}
\end{equation}
Owing to the reversibility of the eigenstrain, we have to distinguish the four phases of Section~\ref{sec:interaction_wave_cohesive_crack_developing}, illustrated in Fig.~\ref{fig:cohesive_opening_theory}b-i.

\paragraph{Phase I: $\dot{\eta} = 0$, $\dot{\alpha} = 0$}
For $0 \leq t \leq t_{\text{I}}$, the stress at $x=0$ remains below $\sigma_{\text{c}}$, so the incoming wave is fully transmitted with no reflection:
\begin{equation}
    u_L^{\text{ref}}(x_L,t) = 0 \quad\text{,}\quad \jump{u}(t) = 0 \quad\text{,}\quad u_R^{\text{tra}}(0^+,t) = u_L^{\text{inc}}(0^-,t)\quad\text{,}
\end{equation}
with the maximum phase field still at its initial value, $\breve{\alpha}(t_{\text{I}}) = \breve{\alpha}(0)$.

\paragraph{Phase II: $\dot{\eta} > 0$, $\dot{\alpha} = 0$}
For $t_{\text{I}} \leq t \leq t_{\text{II}}$, the jump reopens to the value associated with the (irreversible) phase-field value $\breve{\alpha}(t_{\text{I}})$.
The strength criterion thus holds with $\sigma_L(0^-,t) = (1 - \breve{\alpha}(t_{\text{I}}))\sigma_{\text{c}}$, and~\eqref{eq:DAlambert_stess_velocity} together with the initial condition $u_L^{\text{ref}}(0^-,t_{\text{I}}) = 0$ yields
\begin{align}
    v_L^{\text{ref}}(0^-,t) &= v_L^{\text{inc}}(0^-,t) + \frac{1 - \breve{\alpha}(t_{\text{I}})}{Z_0}\,\sigma_{\text{c}}\quad\text{,} \\
    u_L^{\text{ref}}(0^-,t) &= u_L^{\text{inc}}(0^-,t) - u_L^{\text{inc}}(0^-,t_{\text{I}}) + \frac{1 - \breve{\alpha}(t_{\text{I}})}{Z_0}\,\sigma_{\text{c}}\,(t - t_{\text{I}})\quad\text{.}
\end{align}
The jump and the transmitted wave then follow from~\eqref{eq:cohesive_response_raw_jump} and~\eqref{eq:crack_kinematic_condition} as
\begin{align}
    \jump{u}(t) &= -2\,\frac{1 - \breve{\alpha}(t_{\text{I}})}{Z_0}\,\sigma_{\text{c}}\,(t - t_{\text{I}}) - 2 u_L^{\text{inc}}(0^-,t) + 2 u_L^{\text{inc}}(0^-,t_{\text{I}})\quad\text{,} \\
    u_R^{\text{tra}}(0^+,t) &= -\frac{1 - \breve{\alpha}(t_{\text{I}})}{Z_0}\,\sigma_{\text{c}}\,(t - t_{\text{I}}) + u_L^{\text{inc}}(0^-,t_{\text{I}})\quad\text{.}
\end{align}
Phase II ends at $t_{\text{II}}$, defined by the jump reaching the value at which the phase field is about to evolve, $\jump{u}(t_{\text{II}}) = \tfrac{2G_{\text{c}}}{\sigma_{\text{c}}}\breve{\alpha}(t_{\text{I}})$.

\paragraph{Phase III: $\dot{\eta} > 0$, $\dot{\alpha} > 0$}
For $t_{\text{II}} \leq t \leq t_{\text{III}}$, the cohesive stress decreases as the jump grows.
Substituting~\eqref{eq:cohesive_response_raw_jump} into~\eqref{eq:crack_stress_continuity} and differentiating in time yields, after introducing the Irwin length $\ell_{\text{ch}} = E_0 G_{\text{c}}/\sigma_{\text{c}}^2$ via $\sigma_{\text{c}}^2/(Z_0 G_{\text{c}}) = c_0/\ell_{\text{ch}}$, the linear ODE
\begin{equation}\label{eq:opening_evolution_ODE}
    \dot{v}_L^{\text{ref}}(0^-,t) - \frac{c_0}{\ell_{\text{ch}}}\,v_L^{\text{ref}}(0^-,t) = \dot{v}_L^{\text{inc}}(0^-,t) \quad\text{,}
\end{equation}
which reveals a dependence of the solution on the ratio of wave speed to Irwin length and on the incoming wave.
With the initial condition at $t_{\text{II}}$, its solution is
\begin{equation}\label{eq:cohesive_response_ODE_solution}
    v_L^{\text{ref}}(0^-,t) = \exp\!\left(\tfrac{c_0}{\ell_{\text{ch}}}(t-t_{\text{II}})\right) v_L^{\text{ref}}(0^-,t_{\text{II}}) + \int_{t_{\text{II}}}^{t} \exp\!\left(\tfrac{c_0}{\ell_{\text{ch}}}(t-\tau)\right) \dot{v}_L^{\text{inc}}(0^-,\tau)\,\mathrm{d}\tau\quad\text{.}
\end{equation}
Phase III ends at $t_{\text{III}}$ when $\breve{\alpha}$ reaches one, marking full opening.
All other quantities follow from~\eqref{eq:cohesive_response_ODE_solution}.
To avoid the explicit evaluation of the integral, we directly integrate the ODE~\eqref{eq:opening_evolution_ODE} between $t_{\text{II}}$ and $t$, which together with~\eqref{eq:cohesive_response_raw_jump} yields
\begin{equation}
    \begin{aligned}
        u_L^{\text{ref}}(0^-,t) &= u_L^{\text{ref}}(0^-,t_{\text{II}}) + \tfrac{\ell_{\text{ch}}}{c_0}\Bigl[v_L^{\text{ref}}(0^-,t) - v_L^{\text{ref}}(0^-,t_{\text{II}}) - \bigl(v_L^{\text{inc}}(0^-,t) - v_L^{\text{inc}}(0^-,t_{\text{II}})\bigr)\Bigr]\quad\text{,} \\
        \jump{u}(t) &= -2 u_L^{\text{ref}}(0^-,t_{\text{II}}) - 2\tfrac{\ell_{\text{ch}}}{c_0}\Bigl[v_L^{\text{ref}}(0^-,t) - v_L^{\text{ref}}(0^-,t_{\text{II}}) - \bigl(v_L^{\text{inc}}(0^-,t) - v_L^{\text{inc}}(0^-,t_{\text{II}})\bigr)\Bigr]\quad\text{,}
    \end{aligned}
\end{equation}
the latter being the expression~\eqref{eq:dynamic_opening_evolution} reported in the main body.

\paragraph{Phase IV: $\dot{\eta} \gtreqless 0$, $\dot{\alpha} = 0$}
For $t \geq t_{\text{III}}$, the cohesive stress is zero and the crack behaves as a free surface,
\begin{equation}
    v_L^{\text{ref}}(x_L,t) = v_L^{\text{inc}}(x_L,t)\quad\text{,}\qquad u_L^{\text{ref}}(x_L,t) = u_L^{\text{ref}}(x_L,t_{\text{III}}) + u_L^{\text{inc}}(x_L,t) - u_L^{\text{inc}}(x_L,t_{\text{III}})\quad\text{.}
\end{equation}

In all phases, the reflected and transmitted wave solutions in the bulk are obtained from these crack-located expressions by propagation along the characteristics.

\subsubsection{Solution for a harmonic half-sine wave}\label{app:dynamic_cohesive_law_pulse}
We finally evaluate the above for the sinusoidal half-pulse~\eqref{eq:1D_pulse_loading} used in our numerical examples.
With $\tilde{T} = \lambda/c_0$, the incoming wave and its velocity are
\begin{equation}
    u_L^{\text{inc}}(x_L,t) =
    \begin{cases}
        0 & t - \tfrac{x_L}{c_0} < 0\\
        \frac{\tilde{\sigma}}{Z_0}\frac{\tilde{T}}{2\pi}\!\left(\cos\!\left(2\pi \frac{t - x_L/c_0}{\tilde{T}}\right) - 1\right) & t - \tfrac{x_L}{c_0} \in [0,\tfrac{\tilde{T}}{2}]\\
        -\frac{\tilde{\sigma}}{Z_0}\frac{\tilde{T}}{\pi} & t - \tfrac{x_L}{c_0} > \tfrac{\tilde{T}}{2}
    \end{cases}
    \text{,}\quad
    v_L^{\text{inc}}(x_L,t) =
    \begin{cases}
        -\frac{\tilde{\sigma}}{Z_0}\sin\!\left(2\pi \frac{t - x_L/c_0}{\tilde{T}}\right) & t - \tfrac{x_L}{c_0} \in [0,\tfrac{\tilde{T}}{2}]\\
        0 & \text{else}
    \end{cases}\quad\text{.}
\end{equation}
The reflected wave velocity and displacement at any $x_L$ then follow from the general derivation as
\begin{equation}
    v_L^{\text{ref}}(x_L,t) =
    \begin{cases}
        0 & t < t_{\text{I}}\\
        v_L^{\text{inc}}\!\left(0,t+\tfrac{x_L}{c_0}\right) + \bigl(1-\breve{\alpha}(t_{\text{I}})\bigr)\frac{\sigma_{\text{c}}}{Z_0} & t_{\text{I}} \leq t < t_{\text{II}}\\
        \exp\!\left(\tfrac{c_0}{\ell_{\text{ch}}}\bigl(t+\tfrac{x_L}{c_0}-t_{\text{II}}\bigr)\right)\left[v_L^{\text{ref}}(0,t_{\text{II}}) - \tilde{v}^{\text{ref}}(t_{\text{II}})\right] + \tilde{v}^{\text{ref}}\!\left(t+\tfrac{x_L}{c_0}\right) & t_{\text{II}} \leq t < t_{\text{III}}\\
        v_L^{\text{inc}}\!\left(0,t+\tfrac{x_L}{c_0}\right) & t_{\text{III}} \leq t
    \end{cases}
\end{equation}
with
\begin{equation}
    \tilde{v}^{\text{ref}}(\tau) = \frac{\dfrac{2\pi \tilde{\sigma}}{\tilde{T} Z_0}}{\left(\dfrac{c_0}{\ell_{\text{ch}}}\right)^{2} + \left(\dfrac{2\pi}{\tilde{T}}\right)^{2}}\left[\frac{c_0}{\ell_{\text{ch}}}\cos\!\left(2\pi\tfrac{\tau}{\tilde{T}}\right) - \frac{2\pi}{\tilde{T}}\sin\!\left(2\pi\tfrac{\tau}{\tilde{T}}\right)\right]\quad\text{,}
\end{equation}
and
\begin{equation}
    u_L^{\text{ref}}(x_L,t) =
    \begin{cases}
        0 & t < t_{\text{I}}\\
        u_L^{\text{inc}}(0,t+\tfrac{x_L}{c_0}) - u_L^{\text{inc}}(0,t_{\text{I}}) + (1-\breve{\alpha}(t_{\text{I}}))\frac{\sigma_{\text{c}}}{Z_0}\bigl(t+\tfrac{x_L}{c_0}-t_{\text{I}}\bigr) & t_{\text{I}} \leq t < t_{\text{II}}\\
        u_L^{\text{ref}}(0,t_{\text{II}}) + \frac{\ell_{\text{ch}}}{c_0}\Bigl[v_L^{\text{ref}}(0,t+\tfrac{x_L}{c_0}) - v_L^{\text{ref}}(0,t_{\text{II}}) - \bigl(v_L^{\text{inc}}(0,t+\tfrac{x_L}{c_0}) - v_L^{\text{inc}}(0,t_{\text{II}})\bigr)\Bigr] & t_{\text{II}} \leq t < t_{\text{III}}\\
        u_L^{\text{ref}}(0,t_{\text{III}}) + u_L^{\text{inc}}(0,t+\tfrac{x_L}{c_0}) - u_L^{\text{inc}}(0,t_{\text{III}}) & t_{\text{III}} \leq t
    \end{cases}
\end{equation}
from which all remaining quantities of interest can be obtained.
The end of Phase I is given in closed form by $t_{\text{I}} = \tfrac{\tilde{T}}{2\pi}\sin^{-1}\!\left((1-\breve{\alpha}(0))\,\sigma_{\text{c}}/\tilde{\sigma}\right)$, while $t_{\text{II}}$ and $t_{\text{III}}$ must be obtained numerically.

\begin{figure}[t!]
    \centering
    \includegraphics{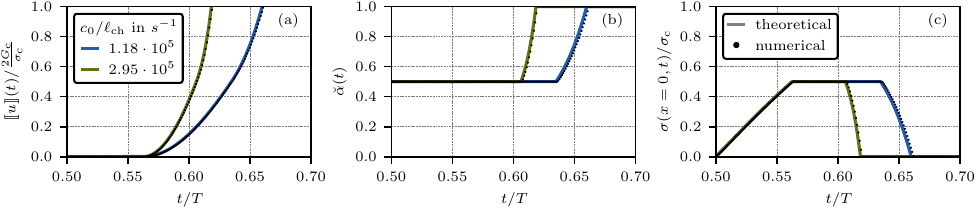}
    \caption{Comparison of numerical and theoretical results on the cohesive crack evolution, in terms of the jump (a), the maximum value of the phase field (b), and the stress (c).}
    \label{fig:cohesive_opening_fem_theoy}
\end{figure}

In Fig.~\ref{fig:cohesive_opening_fem_theoy}, we show a comparison between these theoretical results and the FEM results from Fig.~\ref{fig:interaction_cohesive_evolution_crack}a:
the chosen fracture toughnesses yield $c_0/\ell_{\text{ch}} = 2.95 \cdot 10^5$~s$^{-1}$ for $G_{\text{c}} = 0.01$~N/mm and $c_0/\ell_{\text{ch}} = 1.18 \cdot 10^5$~s$^{-1}$ for $G_{\text{c}} = 0.025$~N/mm.
Clearly, analytical and FEM results coincide; the case producing a shock wave is excluded from this comparison.

\end{document}